\documentclass[twocolumn,amsfonts,showpacs,superscriptaddress,amsmath,amssymb,aps,
longbibliography,prx]{revtex4-2}
\usepackage{dutchcal}
\usepackage{graphicx}% Include figure files
\usepackage{bm}% bold math
 \usepackage{mathtools}
\usepackage{color}
\usepackage{alphabeta}
\usepackage{comment}
\usepackage[english]{babel}
\usepackage{dsfont}
\usepackage[caption=false]{subfig}
\usepackage{float}
\usepackage{xcolor} 

\usepackage{epstopdf}
\usepackage{leftidx}
\newcommand{\bc}{\begin{center}}
\newcommand{\ec}{\end{center}}
\def\ba#1{\begin{array}{#1}\displaystyle}
\newcommand{\ea}{\end{array}}

\newcommand{\beq}{\begin{equation}}
\newcommand{\eeq}{\end{equation}}
\newcommand{\beqa}{\begin{eqnarray}}
\newcommand{\eeqa}{\end{eqnarray}}

\newcommand{\n}{\nonumber\\}
\newcommand{\bi}{\begin{itemize}}
\newcommand{\ei}{\end{itemize}}
\def\b#1{\bar{#1}}

\newcommand{\bra}{\langle}
\newcommand{\ket}{\rangle}

\newcommand{\Tr}{{\rm Tr}}

\newcommand{\trmt}{t_\mathrm{ramp}}

\newcommand{\ii}{{\rm i}}

\newcommand{\dd}{{\rm d}}

\usepackage[colorlinks, linkcolor=red, citecolor=blue]{hyperref}

\begin{document}

\title{Operator dynamics in Floquet many-body systems}
\author{Takato Yoshimura}
\thanks{T.Y. and S.J.G. contributed equally to this work.}
\affiliation{All Souls College, Oxford OX1 4AL, UK}
\affiliation{Rudolf Peierls Centre for Theoretical Physics, University of Oxford, 1 Keble Road, Oxford OX1 3NP, UK}
\author{Samuel J. Garratt}
\thanks{T.Y. and S.J.G. contributed equally to this work.}
\affiliation{Department of Physics, University of California, Berkeley, California 94720, USA}
\author{J. T. Chalker}
\affiliation{Rudolf Peierls Centre for Theoretical Physics, University of Oxford, 1 Keble Road, Oxford OX1 3NP, UK}

\date{\today}
\begin{abstract}
We study operator dynamics in many-body quantum systems, focusing on generic features of systems that are ergodic, spatially extended, and lack conserved densities. Quantum circuits of various types provide simple models for such systems.
%and important insights have come from studying behaviour averaged over ensembles of random quantum circuits in which successive time-steps are statistically independent. Here w
We focus on Floquet quantum circuits, comparing their behaviour with what has been found previously for circuits that are random in time. Floquet circuits, which have discrete time-translation symmetry, represent an intermediate case between circuits that are random in time and lack any symmetry, and systems with a time-independent Hamiltonian and continuous time-translation invariance. By making this comparison, one of our aims is to identify signatures of time-translation symmetry in Floquet operator dynamics. To characterise behaviour we examine a variety of quantities in solvable models and numerically: operator autocorrelation functions; the partial spectral form factor; the out-of-time-order correlator (OTOC); and the paths in operator space that make the dominant contributions to the ensemble-averaged autocorrelation functions. Our most striking result is that ensemble-averaged autocorrelation functions show behaviour that is distinctively different in Floquet systems compared to systems in which successive time-steps are independent. Specifically, while average autocorrelation functions decay on a microscopic timescale for circuits that are random in time, in Floquet systems they have a late-time tail with a duration that grows parametrically with the size of the operator support. In the simplest models this tail is separated from the initial decay by a minimum, so that the average autocorrelation function has an intermediate-time peak. The existence of these tails provides a way to understand deviations of the spectral form factor from random matrix behaviour at times shorter than the Thouless time. In contrast to this feature in autocorrelation functions, we find no new aspects to the behaviour of OTOCs for Floquet models compared to random-in-time circuits. We show that this difference between averaged autocorrelation functions and OTOCs can be understood in terms of the paths in operator space that contribute to the two quantities: paths for the former retain a limited support at late times, while paths for the latter are dominated by operator spreading.
\end{abstract}

\maketitle

\section{Introduction}

The essential phenomenology of dynamics in chaotic many-body quantum systems describes the evolution of operators 
and of state vectors. Operators that are initially simple evolve into increasingly complicated operators, and states that initially have low entanglement %(in a basis that is a product over sites) 
develop increasing entanglement under time evolution. These long-standing ideas \cite{DAlessio_From_2016,Maldacena_Bound_2016} have recently been illustrated by calculations using random unitary circuits (RUCs) \cite{Nahum_2017_Quantum,Nahum_Operator_2018,Keyserlingk_Operator_2018,Fisher_Random_2023}, in which the dynamics of a spin chain consists of a series of discrete steps and at each step pairs of spins are coupled by randomly chosen gates. 

In comparison with evolution under a time-independent Hamiltonian, RUCs embody two related but distinct simplifications. Energy is eliminated as a conserved quantity, and time-translation symmetry is broken by taking the gates at every step to be statistically independent. The consequences of each of these simplifications can be investigated separately. First, one can  construct RUCs that have an internal symmetry and an associated conserved density but no time-translation symmetry \cite{Rakovszky_Diffusive_2018,Khemani_Operator_2018}. Second, one can study random Floquet circuits (RFCs) that have discrete time-translation symmetry but no conserved densities \cite{Chan_Solution_2018,Bertini_Exact_2018}. Both alternatives show dynamical features that are absent in the simplest RUCs. These features appear either in correlation functions or in the spectral form factor (SFF) \cite{Mehta_Random_2004}, obtained from the trace of the evolution operator. For RUCs with an internal symmetry, the existence of a density that spreads diffusively is imprinted on the correlation functions of observables to which it couples \cite{Rakovszky_Diffusive_2018,Khemani_Operator_2018}, but the trace of the time evolution operator is typically small and time-independent for $t>0$. For RFCs, previous results for correlations functions are limited but it is known that time-translation symmetry gives rise to a distinctive ramp and plateau in the SFF \cite{Mehta_Random_2004}.

A natural question, and the subject of this paper, is what consequences time-translation symmetry has for operator dynamics in chaotic systems. Strikingly, we find that autocorrelation functions of observables --- the simplest quantities that characterise such dynamics --- show a generic feature that we believe has not previously been identified and that arises from this symmetry. In contrast, operator spreading, as characterised by the out-of-time order correlator (OTOC) \cite{Maldacena_Bound_2016}, has the same behaviour in the RFCs we study as in RUCs examined previously. While the calculations presented in this paper are restricted to Floquet models, we expect similar behaviour in Hamiltonian systems for observables that do not couple to conserved densities.  

\begin{figure}[htb]
\hspace*{-0cm}
\centering
\includegraphics[width=0.4\textwidth]{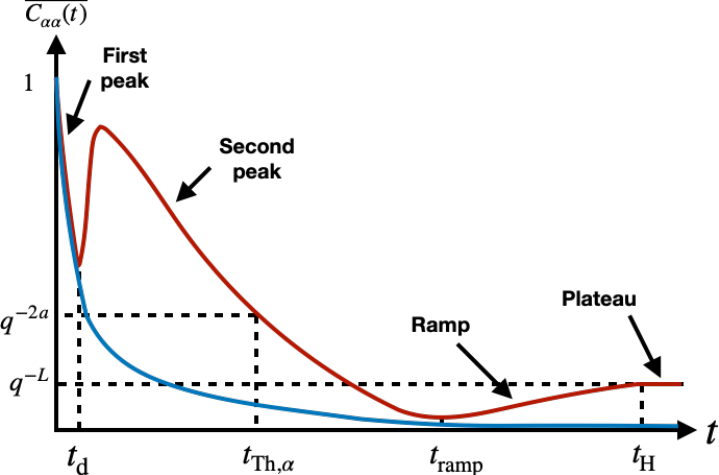}
\caption{Comparison of schematic behaviour of the average autocorrelation function $\overline{C_{\alpha\alpha}(t)}$ ($P_\alpha\neq I$) for (i) a spatially local Floquet many-body system without conservation laws (red curve) and (ii) a random unitary circuit (blue curve). For the Floquet system three successive features appear as time increases. The earliest is a peak at $t=0$. The second is a peak (or in some models, simply a tail) at intermediate times. The third is a ramp and plateau at late times, which have an amplitude that decreases exponentially with increasing system size. The random unitary circuit shows only the first feature.
%(Here we take the lengths $L$ of the system and $a$ of the operator support to satisfy $a<L/2$.)
}
\label{fig:schematic}
\end{figure}

In order to summarise our results more precisely, we introduce some notation. Consider a spin chain of $L$ sites with local Hilbert space dimension $q$ and hence full Hilbert space dimension $D=q^L$. The ($D \times D$) unitary evolution operator is defined for integer times $t$ as $W(t)\equiv W^t$, with $W$ the Floquet operator. Now take a complete orthonormal basis of $q^2$ Hermitian operators at each site, with the identity as one member of the basis, and construct from them as direct products the $D^2$ observables $P_\alpha$ satisfying $D^{-1}{\rm Tr}[P_\alpha P_\beta] = \delta_{\alpha \beta}$. The correlation function of the observables $P_\alpha$ and $P_\beta$ is
\begin{equation}\label{eq:def_corr}
    C_{\alpha\beta}(t) = \langle P_\alpha(t) P_\beta \rangle,\quad P_\alpha(t)=W^\dagger(t)P_\alpha W(t)
\end{equation}  
with $\langle \,\cdots \rangle \equiv D^{-1}{\rm Tr}[\,\cdots]$, the infinite-temperature average appropriate for a Floquet system. Autocorrelation functions correspond to the case $\alpha=\beta$. We denote the average of the correlation function over an ensemble of similar systems by $\overline{C_{\alpha\beta}(t)}$. 
 
To give an overview of behaviour, it is useful to have a simple picture of the dynamics in mind. An autocorrelation function takes its maximum value at time zero and is expected to decay on a short timescale, typically set by the period in Floquet systems and by the inverse of coupling energies in Hamiltonian models. After this initial decay, it may exhibit irregular fluctuations, for example because of spin precession in evolving local exchange fields. One expects such fluctuations to be a specific fingerprint of the particular observable and system. Averaging over an ensemble of similar systems, a natural possibility is that fluctuations would be reduced to zero. We find this is not the case: instead, a tail in the autocorrelation function emerges from the average. The duration of this tail grows indefinitely with the size of the support of the operator concerned, and it is therefore a parametrically distinct feature for `mesoscopic' operators, which have large but finite support. In addition, for a class of models, the autocorrelation function displays a minimum at early times, promoting the tail to a peak in the autocorrelation function at intermediate times. This class of models includes ensembles of circuits that are either left or right invariant (or both) under independent unitary rotations at each site, as is the case with Haar-distributed gates.

In more detail, the behaviour of the ensemble-averaged autocorrelation function $\overline{C_{\alpha\alpha}(t)}$ is shown schematically in Fig.~\ref{fig:schematic}. It exhibits three generic features as a function of time. The first and most prominent one is a peak at $t=0$ with amplitude $\overline{C_{\alpha\alpha}(0)}=1$ fixed by our choice of normalisation for the observable. In a system without locally conserved densities (or for observables that do not couple to conserved densities) the autocorrelation function
relaxes from this initial value over timescale $t_{\rm d}$, which we refer to as the decay time. 
%Of the three features, we believe that until recently only this first one has been recognised. By contrast, our concern in this paper is with the two additional generic features in the autocorrelation function which appear at longer timescales. 

The second generic feature is a tail which extends from a time of order $t_{\rm d}$ up to a timescale  that we call the Thouless time $t_{\text{Th},\alpha}$ of the operator $P_{\alpha}$ (we comment below on this terminology). As noted above, for some models this feature is separated from the initial peak by a minimum at $t_d$, and becomes an asymmetric peak in the autocorrelator as a function of time. For an operator having 
%nontrivial 
support on $a$ sites, we will see below that a natural scale for $\overline{C_{\alpha\alpha}(t)}$ is $q^{-2a}$, and we define $t_{\text{Th},\alpha}$ as the time at which the tail in $\overline{C_{\alpha\alpha}(t)}$ drops to a value of order $q^{-2a}$. For example, if the sites on which $P_{\alpha}$ has 
%non-trivial 
support are contiguous, 
%the peak in the autocorrelation grows exponentially with $a$, and
we find that $t_{\text{Th},\alpha} \sim \ln a$. Roughly, $t_{\text{Th},\alpha}$ can be identified as the time beyond which the dynamics of $P_{\alpha}$ resembles that with an unstructured $D\times D$ random unitary matrix as Floquet operator. 
%It is important to bear in mind that, in many-body systems, this dynamical time scale \cite{Serbyn_Thouless_2017,Sonner_Thouless_2021} is distinct from the `spectral' Thouless time $t_{\text{Th}}$ \cite{Bertrand_Anomalous_2016,Chan_Spectral_2018,Garratt_Local_2021,Sierant_Thouless_2020}, beyond which the SFF coincides with random matrix behaviour. A key result of this work is to elucidate the relation between these different time scales.

At the latest times we find the third generic feature: a linear growth and eventual plateau of the autocorrelation function, with an onset time $t_{\rm ramp} \propto L-L_A$ where $L_A$ is the support of the operator $P_\alpha$. The amplitude of this growth and the plateau value are exponentially small in system size, and this feature is therefore completely suppressed in the thermodynamic limit. This late-time feature in autocorrelation functions is the counterpart of the ramp and plateau in the SFF as a representation of random matrix correlations between eigenvalues of the Floquet operator or Hamiltonian for a system \cite{Mehta_Random_2004}. Similar behaviour has been discussed previously in several contexts. A ramp and plateau feature in autocorrelation functions of chaotic many-body quantum systems has been identified for Hamiltonian models in Ref.~\cite{Schiulaz_19}, and for Floquet circuits it has been analysed \cite{Garratt_Local_2021,Joshi_Probing_2022} in the context of the partial spectral form factor (PSFF). In addition, the same structure is present in the late-time return probability (the probability of a system to remain in a generic initial state), and has been examined for single-particle models of disordered conductors \cite{Prigodin_94} and for chaotic many-body systems \cite{Schiulaz_19,Santos_19}. 

We arrive at the picture outlined in the preceding paragraphs using a combination of analytical and numerical calculations in RFCs with short-range couplings. Our analytical results are obtained for the random phase model (RPM) introduced in Ref.~\cite{Chan_Spectral_2018}, which is exactly solvable for $q\to \infty$, while for numerical calculations we use the kicked random field Heisenberg model studied previously in Ref.~\cite{Garratt_ManyBody_2021}. In addition, we study numerically several models with a variety of choices of probability distribution for gates, comparing Floquet systems with their counterparts in which gates are chosen independently at every time-step. The results demonstrate that the intermediate-time tail or asymmetric peak is absent for RUCs but is generic in Floquet systems. In this sense, it is a consequence of time-translation symmetry. However, the existence of a minimum in the autocorrelation function at intermediate times, determining whether this function has an intermediate-time peak or simply a tail, is model specific. The minimum is enhanced by fast local scrambling combined with weak intersite coupling.

Both the tail or asymmetric peak in autocorrelation functions, and the ramp-and-plateau structure, are features visible only after a suitable average. One possibility is to average autocorrelation functions over an ensemble of systems. 
%In an individual system, one can also 
An alternative is to average by summing the autocorrelation functions over all choices of the initial operator with a given support. This generates the PSFF \cite{Garratt_Local_2021,Joshi_Probing_2022} and for large subregions it is a sufficiently effective average to reveal the asymmetric peak in an individual system without an ensemble average. The ramp and plateau, on the other hand, are obscured by large temporal fluctuations in individual systems, and are not visible without an explicit average over different systems or over time. The absence in this sense of self-averaging has long been recognised in the case of the SFF \cite{Prange_Spectral_1997,Kunz_Probability_1999}.

%Our analysis of dynamics is rooted in the deconstruction of 
%A useful perspective on our results is provided by expressing the SFF as a sum over autocorrelation functions of a complete set of observables \cite{Cotler_Chaos_2017,Gharibyan_Onset_2018}. 
%A related sum over observables with support on a specified subsystem yields a quantity known as the partial spectral form factor (PSFF) \cite{Joshi_Probing_2022}. 
%Under this correspondence the tail or 
%second feature in the autocorrelation function (the 
%asymmetric peak in autocorrelation functions at intermediate times is the source of deviations of the SFF from random matrix behaviour, identified recently in ergodic, spatially extended many-body systems \cite{Chan_Spectral_2018,Braun_Transition_2020,Garratt_Local_2021}. These deviations are absent in dual-unitary circuits \cite{Bertini_Exact_2018,Flack_Statistics_2020,Bertini_Random_2021}, where many autocorrelation functions also vanish. 
%We note that the behaviour of OTOCs \cite{Claeys_Maximum_2020} has been studied in dual unitary circuits, and more recently in systems where this restriction is weakly violated \cite{Rampp_From_2023}.

Our use of the term {Thouless time} deserves discussion, since it and the term {Thouless energy} for its reciprocal have been employed in a variety of contexts, and with definitions and interpretations that are not in all cases equivalent. As originally introduced in discussions of single-particle models for disordered conductors \cite{Thouless_77}, the terms have both a dynamical significance (as the time for a particle to diffuse over a distance set by the system size) and a spectral significance (as the energy scale below which eigenvalue correlations match the predictions of random matrix theory \cite{Altshuler_86}). In many-body systems the terms have also been used in both in a spectral sense \cite{Bertrand_Anomalous_2016,Chan_Spectral_2018,Friedman_Spectral_2019,Schiulaz_19,Santos_19,Garratt_ManyBody_2021,Sierant_Thouless_2020,Sonner_Thouless_2021} and a dynamical sense
\cite{Serbyn_Thouless_2017,Santos_19,Sonner_Thouless_2021}
but we believe that the link between the two has been unclear. A key result of the current work is to elucidate the connection between these scales. We will identify a Thouless time in three different settings: the dynamics of individual operators; the behaviour of the PSFF; and the behaviour of the SFF. These are related by the fact that the PSFF and SFF can be obtained as sums over autocorrelation functions of operators \cite{Cotler_Chaos_2017,Gharibyan_Onset_2018}. In all three settings the Thouless time has a dynamical significance, while in the third it also has a spectral significance. Under this relation between correlation functions and the SFF, the tail or 
%second feature in the autocorrelation function (the 
asymmetric peak in autocorrelation functions at intermediate times is the source of deviations of the SFF from random matrix behaviour, identified recently in ergodic, spatially extended many-body systems \cite{Chan_Spectral_2018,Friedman_Spectral_2019,Braun_Transition_2020,Garratt_Local_2021}. 

%Some comments on the relation between our definition of Thouless time and previous work are appropriate at this point. First, in dual-unitary Floquet circuits, deviations of the SFF from random-matrix behaviour are essentially absent  \cite{Bertini_Exact_2018,Flack_Statistics_2020,Bertini_Random_2021}. Consistency with the decomposition of the SFF as a sum over autocorrelation functions requires many autocorrelation functions to be zero in dual-unitary Floquet circuits, as is known to be the case. 
We note that the spectral timescale that we term the Thouless time is different from (and much shorter than) the one discussed for chaotic many-body Hamiltonians in Refs.~\cite{Schiulaz_19,Santos_19}. The latter arises from an interplay between the disconnected part of the SFF and the ramp that appears in the connected part of the SFF. It is estimated \cite{Schiulaz_19} by using the known form of the many-body density of states to find the disconnected part of the SFF, and by assuming that the ramp reflects only random-matrix level correlations. By contrast, the Thouless time identified here and in Refs.~\cite{Chan_Spectral_2018,Friedman_Spectral_2019,Garratt_Local_2021} characterises deviations from a random-matrix ramp in the connected part of the SFF, and therefore deviations of two-point level correlations from random matrix theory. The distinction between these two routes to identification of a Thouless time in many-body systems is particularly clear in the Floquet case, since in large many-body Floquet systems the density of eigenphases is uniform. As a consequence, in Floquet systems the disconnected part of the SFF is non-zero only for time zero. At later integer times, the SFF is given in full by the connected contribution and the arguments of \cite{Schiulaz_19} do not apply. We suspect that the origin of the Thouless time observed in numerical studies of many-body chaotic Hamiltonian systems \cite{Bertrand_Anomalous_2016,Serbyn_Thouless_2017,Sierant_Thouless_2020} lies in the mechanism discussed in the present paper and in Refs.~\cite{Chan_Spectral_2018,Friedman_Spectral_2019,Garratt_Local_2021}, and not from the interplay between disconnected and connected contributions to the SFF treated in \cite{Schiulaz_19}.

\begin{figure}[h!]
\subfloat[\label{fig:path_dominant}]{\includegraphics[width=0.2\textwidth]{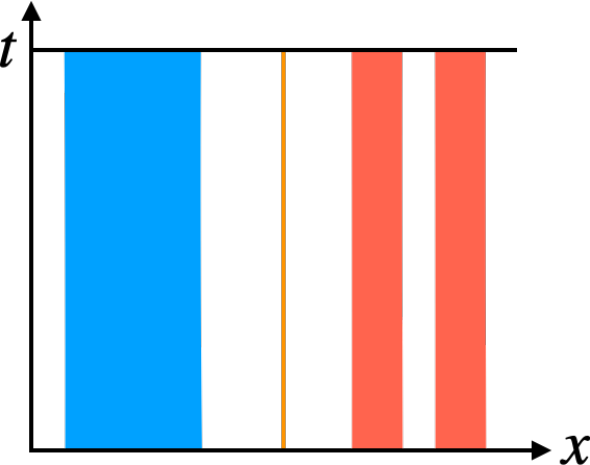}}
\hfill
\subfloat[\label{fig:path_subdominant}]{\includegraphics[width=0.2\textwidth]{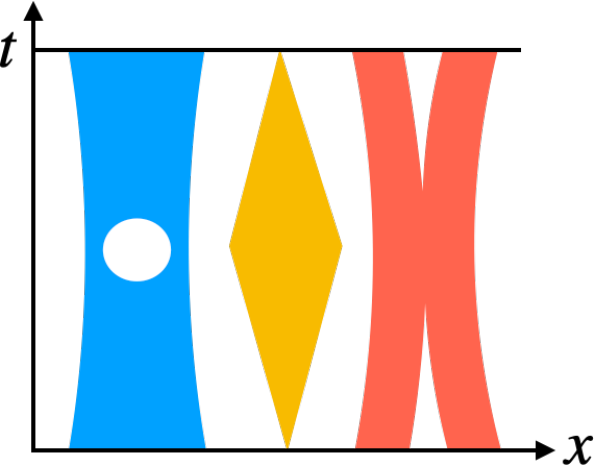}} 
\caption{Operator paths contributing to ensemble-averaged autocorrelation functions $\overline{C_{\alpha\alpha}(t)}$. Coloured regions represent regions of space where the operator acts non-trivially (i.e. not as the identity). (a) Examples of the dominant paths for various choices of $\overline{C_{\alpha\alpha}(t)}$ distinguished by colour. (b) Examples of paths that are sub-dominant for the same set of $\overline{C_{\alpha\alpha}(t)}$.}
\label{fig:paths}
\end{figure}

%The effects of time-translation symmetry can be understood in a unified way 
A complementary perspective on our results comes from interpreting an autocorrelation function as a sum of amplitudes associated with paths in the space of operators. More specifically, and in analogy with the Feynman formulation of quantum mechanics, the amplitude for evolution between given initial and final operators can be expressed as a sum over paths, with each path consisting of a sequence of operators at each intermediate time-step. Each path has an associated amplitude, which for Hermitian operators is real but not necessarily positive. Full specification of a path in terms of the operators at each step requires a great deal of information, but partial specification can be provided by simply identifying the operator support as a function of time: we refer to this support as the operator trajectory. We identify a subset of paths with positive amplitudes that make the dominant contribution to the averaged autocorrelation functions.  Interestingly, these paths have rather narrow trajectories in the full space of operators. For example, in the RPM at large $q$, the boundaries of operator paths form straight vertical lines in the space-time plane, as illustrated in Fig.~\ref{fig:paths}. This picture is to be contrasted with that for the averages of squared autocorrelation functions, and also of OTOCs (both of which are nonzero even in RUCs). In the first case the spatial widths of the dominant operator paths grow as $t^{1/2}$ \cite{Nahum_Real_2022}, whereas in the second they grow ballistically \cite{Nahum_Operator_2018,Keyserlingk_Operator_2018}. In systems with conservation laws, we note that the operator paths relevant to dynamics have recently been studied in Refs.~\cite{Rakovszky_Dissipation_2022,Keyserlingk_Operator_2022}.

We believe that there are encouraging prospects for experimental observation of the behaviour that we have identified in autocorrelation functions. The asymmetric peak is striking in some models even for operators supported on a few sites, as is demonstrated by our numerical results. This feature is therefore straightforward to observe in experiments involving only conventional dynamics and a number of samples that is independent of the system size. We should compare this situation with that for OTOCs and the SFF. For example, one method for determining OTOCs involves an explicit time-reversal of many-body dynamics \cite{Swingle_Measuring_2016}, while randomised measurement approaches for extracting either the OTOC \cite{Vermersch_Probing_2019} or the SFF \cite{Joshi_Probing_2022} suffer from a sample complexities that scale exponentially with the system size.

The remainder of this paper is organised as follows. In Sec.~\ref{sec:overview} we provide an overview of our results, and in Sec.~\ref{sect:rpm_derivation} we present calculations of averaged autocorrelation functions and PSFFs. Following this, in Sec.~\ref{sec:otoc_derivation_nospin}, we calculate the OTOC in the RPM. In Sec.~\ref{sec:paths} we provide an interpretation of our results in the language of operator paths. Finally, in Sec.~\ref{sec:fluctuations}, we discuss statistical fluctuations relevant to dynamics. We summarise and discuss our results in Sec.~\ref{sec:summary}.

\section{Overview}\label{sec:overview}
This section is organised as follows. In Sec.~\ref{sect:definitions} we introduce our notation and define the basic physical quantities of interest. Next, in Sec.~\ref{sect:models}, we describe the models we use for analytical as well as numerical calculations. In Secs.~\ref{sec:overview_corr} and \ref{sect:overview_psff_rpm} we then outline the behaviour of autocorrelation functions and PSFFs, respectively, before discussing OTOCs in Sec.~\ref{sect:RPM-OTOC}.

\subsection{Definitions}\label{sect:definitions}

In this work we study chaotic many-body dynamics in interacting spin chains. Our focus will be on the effects of locality in systems with fixed evolution operators, where there is time-translation symmetry in the dynamics. Our minimal models for this kind of dynamics are RFCs without conservation laws. As an orthonormal basis for the Hilbert space of many-body quantum states we choose the tensor products $|a\ket=|a_1\ldots a_L\ket$, where $a_x=0,\cdots,q-1$, and $a=0,\cdots,D-1$. Since our focus is primarily on the dynamics of operators, it is convenient to choose a Hermitian operator basis $P_{\alpha}$; here the label $\alpha=0,\ldots,(q^{2L}-1)$. Explicitly, $P_{\alpha}=P_{\alpha_1}\otimes\cdots\otimes P_{\alpha_L}$ where $P_{\alpha_x}$, with $\alpha_x=0,\ldots,(q^{2}-1)$, acts on site $x$. For example, for $q=2$, we can choose $P_{\alpha_x}$ to be Pauli matrices.
We reserve the label $\alpha_x=0$ to denote the $q \times q$ identity at site $x$, and $\alpha=0$ to denote the $D \times D$ identity. Note that in our convention $q^{-1}\text{Tr}[P_{\alpha_x}P_{\beta_x}]=\delta_{\alpha_x \beta_x}$, where the trace is over states at $x$, and $D^{-1}\text{Tr}[P_{\alpha}P_{\beta}]=\delta_{\alpha \beta}$, where the trace is over the full Hilbert space of states. This condition implies that all $P_{\alpha \neq 0}$ are traceless. 

We will make multiple uses of the completeness relation for our operator basis. In terms of the matrix elements $\bra a |P_{\alpha}|b\ket$ it has the form
\begin{align}\label{eq:res_identity}
    D^{-1}\sum_{\alpha=0}^{D^2-1} \bra a| P_{\alpha} |b\ket \bra c |P_{\alpha}|d\ket = \delta_{ad}\delta_{bc}.
\end{align}

Our probes of the dynamics are (infinite-temperature) correlation functions defined by  Eq.~\eqref{eq:def_corr}. Note that $C_{\alpha \beta}(t)$ is real number, and that the normalisation condition $D^{-1}\text{Tr}P_{\alpha}^2(t)=1$ implies $\sum_{P_\beta\in\mathcal{P}} C^2_{\alpha \beta}(t)=1$ for $\alpha \not= 0$, where $\mathcal{P}$ is the set of non-identity $P_\alpha$. This means that $C_{\alpha\beta}(t)$ is an orthogonal matrix. 
Using Eq.~\eqref{eq:res_identity} the autocorrelation function $C_{\alpha \beta}(t)$ can be recast as
\begin{align}
    C_{\alpha \beta}(t) &= \sum_{P_{\alpha(1)},\cdots,P_{\alpha(t-1)}\in\mathcal{P}}C_{\alpha \alpha(1)}\cdots C_{\alpha(t-1) \beta} , \label{eq:paths}
\end{align}
where in the correlation function for a single time step we omit the time argument, $C_{\alpha \beta}\equiv C_{\alpha \beta}(1)$. The expansion given in Eq.~\eqref{eq:paths} can be thought of as a sum of the amplitudes of paths $(\alpha, \alpha(1), \ldots, \beta)$. The normalisation condition $D^{-1}\text{Tr}P_{\alpha}^2(t)=1$ can now be rewritten as
\begin{align}
   \sum_{P_{\alpha(1)} \cdots P_{\alpha(t)}\in\mathcal{P}} C^2_{\alpha \alpha(1)}\ldots C^2_{\alpha(t-1)\alpha(t)} = 1, \label{eq:pathnorm}
\end{align}
and so we can interpret the squared amplitudes $C^2_{\alpha \alpha(1)}\ldots C^2_{\alpha(t-1)\alpha(t)}$ as probabilities for the various operator paths. 

Our focus in Sec.~\ref{sect:rpm_derivation} will be on autocorrelation functions $C_{\alpha \alpha}(t)$, the SFF
 \begin{equation}\label{eq:defSFF2}
     K(t) = |{\rm Tr} W(t)|^2
 \end{equation}
 and the PSFF
\begin{equation}\label{eq:defPSFF2}
    K_A(t)=D_AD^{-1}\mathrm{Tr}_{\bar{A}}\left[(\mathrm{Tr}_A W(t)^\dagger)(\mathrm{Tr}_A W(t))\right],
\end{equation}
where $A$ denotes a subsystem and $\overline{A}$ its complement, with $D_A$ the Hilbert space dimension of $A$. 

The significance of the SFF and the PSFF can be made clear by introducing the eigenvectors $|m\rangle$ and eigenphases $\theta_m$ of the Floquet operator, satisfying $W|m\rangle = e^{i\theta_m}|m\rangle$. The SFF contains information only on the spectrum, and is given by
\begin{equation}\label{eq:Kspec}
    K(t) = \sum_{mn} e^{it(\theta_m - \theta_n)}\,.
\end{equation}
The ensemble-averaged SFF, $\overline{K(t)}$, has a ramp and plateau structure as a function of time. For example, in the case that $W$ is a Haar-distributed random unitary matrix, one has \cite{Mehta_Random_2004}
\begin{equation}\label{eq:SFF_RMT}
    \overline{K(t)}=\begin{dcases*}
    D^2 & $t=0$\\
    t & $1\leq t< D$ \\
    D & $D\leq t$.
    \end{dcases*}
\end{equation}
From Eq.~\eqref{eq:Kspec}, the value of $\overline{K(t)}$ at $t=0$ follows immediately, and the plateau at large $t$ from assuming off-diagonal terms in the double sum vanish on averaging; the ramp for $0<t\leq D$ reflects spectral rigidity.

The PSFF, on the other hand, contains information from both the spectrum and the eigenvectors of the Floquet operator. We define the density matrix $\rho(m) =  |m\rangle \langle m|$ formed from an eigenvector, and the corresponding reduced density matrix for subsystem $\overline{A}$,
\begin{equation}
    \rho_{\overline{A}}(m) = {\rm Tr}_{A} \rho(m).
\end{equation}
Then 
\begin{equation}
    K_A(t) = D_AD^{-1}\sum_{mn}  {\rm Tr}_{\bar{A}}\rho_{\bar{A}}(m) \rho_{\bar{A}}(n)e^{it(\theta_m - \theta_n)}\,.
\end{equation}

Using the operator resolution of the identity, Eq.~\eqref{eq:res_identity}, the SFF can be expressed in terms of correlations functions as
\begin{align}\label{eq:sff1}
    K(t) = 1 + \sum_{P_\alpha\in\mathcal{P}} C_{\alpha \alpha}(t).
\end{align}
and the PSFF as
\begin{align}
    K_A(t) = 1 + \sum_{P_\alpha \in \mathcal{P}_A} C_{\alpha \alpha}(t), \label{eq:defPSFF}
\end{align}
where $\mathcal{P}_A$ is the set of $P_{\alpha}$ with non-trivial support in the region $A$ (and which act as the identity in its complement $\bar{A}$).

In Sec.~\ref{sec:otoc_derivation_nospin} we study the OTOC,
\begin{align}\label{eq:otoc}
     \mathcal{C}(x,t)&=-\frac{1}{2}\langle[
     \mathcal{O}(0,t),\mathcal{O}(x,0)]^2\rangle \\ 
     &=1-\sum_{P_\alpha,P_\beta\in\mathcal{P}}\langle P_\alpha \mathcal{O}(x,0)P_\beta \mathcal{O}(x,0)\rangle C_{\mathcal{O}\alpha}(t)C_{\mathcal{O}\beta}(t),\notag
\end{align}
where $\mathcal{O}(x,0)$ is a local operator supported at site $x$, and $C_{\mathcal{O}\alpha}(t):=\langle \mathcal{O}(0,t)P_\alpha\rangle$. One expects the support of $\mathcal{O}(0,t)$ to grow with time and the OTOC is designed to probe this spreading: at early times, the support does not extend as far as $x$ and $\mathcal{C}(x,t)=0$, while at late times the correlator on the right of the second line of Eq.~\eqref{eq:otoc} falls to zero and $\mathcal{C}(x,t)=1$.

From the perspective highlighted in this work, the crucial difference between correlation functions and OTOCs is that the former involve sums over amplitudes of paths whereas, in both RUCs and RFCs, the dominant contribution to the latter is a sum over probabilities. Meanwhile, a difference between RUCs and RFCs with Haar-random local unitary operations is that, in the former, ensemble averages of operator path amplitudes necessarily vanish. Their sums, from autocorrelations functions to the SFF, are therefore trivial. This is not the case in RFCs.
%because the amplitudes of different steps of operator paths are correlated.

%%%%%%%%%%%%%%%%%%%%%%%%%%%%%%%%%%%%%%%%%%%%%%%%%%%%%%%%%%
\subsection{Models}\label{sect:models}%%%%%%%%%%%%%%%%
%%%%%%%%%%%%%%%%%%%%%%%%%%%%%%%%%%%%%%%%%%%%%%%%%%%%%%%%%%

\begin{figure}[h!]
\centering
\includegraphics[width=8cm]{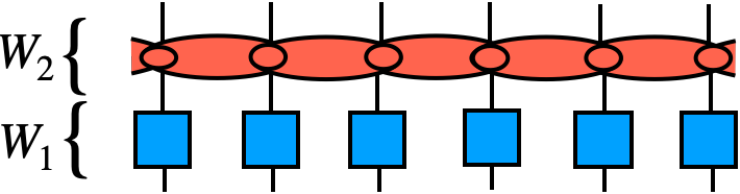}
\caption{Floquet operator in the RPM. The blue squares represent the local unitary operations $U_x$ [Eq.~\eqref{eq:RPM_W1}], and the red ovals represent the interactions in Eq.~\eqref{eq:RPM_W2}.} 
\label{fig:floquet_operator}
\end{figure}

We perform analytical calculations for the RPM \cite{Chan_Spectral_2018}, a minimal model for Floquet dynamics without conservation laws which is tractable in the limit of large local Hilbert space dimension $q$. It has the Floquet operator $W = W_2 W_1$, with 
\begin{align}
    W_1=\bigotimes_{x=1}^LU_x \label{eq:RPM_W1}
\end{align}
a tensor product of single-site ($q \times q$) Haar-random unitary operators, and $W_2$ a product of diagonal two-site unitary operators which couple neighbouring sites,
\begin{align}
W_2|a\rangle=\exp\left(\ii\sum_x\varphi_{x;a_x,a_{x+1}}\right)|a\rangle. \label{eq:RPM_W2}
\end{align}
Each $\varphi_{x;a_x,a_{x+1}}$ is an independent Gaussian random variable with mean zero and variance $\varepsilon$. The structure of the Floquet operator is illustrated in Fig.~\ref{fig:floquet_operator}. Previously it has been shown that, in this model and in the limit of large $q$, the average SFF behaves as in RMT, $\overline{K(t)} \simeq t$, beyond the spectral Thouless time $t_{\text{Th}} \sim \log L /\varepsilon$ \cite{Chan_Spectral_2018}. Due to the local Haar-random operations, the relaxation time $t_\mathrm{d}=1$, but this is non-generic; we explore a generalisation of the RPM which is not locally Haar random in  Appendix~\ref{sec:spinful}.

For numerical calculation we do not use the RPM because, for $q=2$, it is known \cite{Chan_Spectral_2018} to be in or close to a many-body localised regime for all $\varepsilon$, while setting $q > 2$ limits the system sizes that are accessible. Instead, we study the kicked random field Heisenberg model of Ref.~\cite{Garratt_ManyBody_2021} in its ergodic phase. This model involves single-site Haar-random unitary operations as above, but the coupling between sites is described by an exponentiated SWAP operator, which for $q=2$ is equivalent to a Heisenberg interaction. 
%We will refer to this as the random SWAP model, or RSM. 
 
The Floquet operator $W$ for this model can be expressed as $W=W_2 W_1$ where $W_1 = \bigotimes_{x \text{ odd}} V_{x,x+1}$ and $W_2 = \bigotimes_{x \text{ even}} V_{x,x+1}$; with periodic boundary conditions, which necessitates $L$ even, there are $L/2$ gates in each layer, and we identify $V_{L,L+1}=V_{L,1}$. With open boundary conditions and $L$ even there are $L/2$ gates in $W_1$ and $L/2-1$ in $W_2$, while with $L$ odd there are $(L-1)/2$ gates in each layer. The operators $V_{x,x+1}$ take the form
\begin{align}\label{SWAP}
    V_{x,x+1} = \exp\big(i \pi J\, \text{SWAP}\big) U_x \otimes U'_{x+1},
\end{align}
where $\text{SWAP}$ is the two-site swap operator, and each of $U_x$ and $U'_{x+1}$ are independent $2 \times 2$ Haar-random unitary matrices, which can be viewed as describing the precession of qubits in  random fields. The properties of this model are a periodic function of $J$, with $[0,1/2]$ as the fundamental interval. Behaviour at the two ends of this range is simple: at $J=0$ the system separates into uncoupled qubits, while at $J=1/2$ it can also be viewed as consisting of uncoupled qubits, half propagating with unit velocity in the positive direction and half in the negative direction. For numerical calculations we use the $q=2$ RSM with $J=0.2$. This value is chosen to be far from both the many-body localised regime at small $J$ \cite{Garratt_Local_2021b} and the straightforwardly integrable point at $J=1/2$. The boundary conditions are periodic except in Fig.~\ref{fig:obcscaling}, where they are open.

In order to investigate dependence of behaviour on the choice of model, we also study numerically a set of models in which $U_x$ and $U'_{x+1}$ are \emph{not} Haar distributed, comparing results for Floquet systems with those for systems in which the evolution operators are constructed in a similar way and each time step but are uncorrelated at different time-steps. Details are given in Sec.~\ref{modeldependence}.

%%%%%%%%%%%%%%%%%%%%%%%%%%%%%%%%%%%%%%%%%%%%%%%%%%%%%%%%%%%%%%%%%%%%%%%%%%%%%%%%%%%
\subsection{Autocorrelation functions}\label{sec:overview_corr}%%%%%%%%
%%%%%%%%%%%%%%%%%%%%%%%%%%%%%%%%%%%%%%%%%%%%%%%%%%%%%%%%%%%%%%%%%%%%%%%%%%%%%%%%%%%

Here we outline our results on the behaviour of autocorrelation functions. First we sketch the behaviour in systems whose Floquet operators are $D \times D$ Haar-random unitary matrices --- we will refer to these as `random matrix Floquet' (RMF) systems --- and then we offer a qualitative picture of the behaviour we should expect in a system with local interactions. Following this we describe our results in the RPM at large $q$.

\subsubsection{Qualitative approach}\label{sec:sec_qualitative}
In RUC systems with either global or local Haar-random unitary operations, and without time-translation invariance, ensemble averages of autocorrelation functions vanish. In a RMF system, which by construction has no spatial structure, for any operator $P_\alpha\in\mathcal{P}$ the autocorrelation function behaves as
 \begin{equation}\label{eq:auto_rmt}
      \overline{C_{\alpha\alpha}(t)}=\frac{\overline{K(t)}-1}{D^2-1}. %\label{eq:CRMT}
\end{equation}
This can be derived from Eq.~\eqref{eq:sff1} by using the fact that $\overline{C_{\alpha\alpha}(t)}$ is independent of $\alpha$ for the RMF, with the number of operators $\alpha$ being $D^2-1$.
In turn, $\overline{K(t)}$ is given by Eq.~\ref{eq:SFF_RMT}. As a consequence,
%\cite{Mehta_Random_2004}
%\begin{equation}\label{eq:SFF_RMT}
%    \overline{K(t)}=\begin{dcases*}
%    D^2 & $t=0$\\
%    t & $1\leq t< D$ \\
%    D & $D\leq t$.
%    \end{dcases*}
%\end{equation}
%That is, 
the average autocorrelation function drops from unity to zero over the first time step, and then increases linearly (at a rate that is exponentially small in the number of degrees of freedom) before an eventual plateau beyond $t=D$.

As a step towards understanding the behaviour in a system with spatial structure, first consider a system of uncoupled sites. That is, take the evolution operator for the $L$-spin system to be a tensor product of $q \times q$ Haar-random unitary operators (note that this corresponds to the $\varepsilon=0$ limit of the RPM). The evolution operator for each spin is fixed in time, and so the dynamics corresponds simply to the precession of spins in local fields. For single-site operators we can apply the above results for RMF systems, replacing the global Hilbert space dimension $D$ with the local one, giving
\begin{align}
    \overline{C_{\alpha \alpha}(t)} = 
    \begin{cases} 
    1 & t=0 \\
    (t-1)/(q^2-1) & 1 \leq t \leq q \\
    (q-1)/(q^2-1) & q \leq t.
    \end{cases} \label{eq:Cdecoupled}
\end{align}
For $q=2$ the result is particularly simple: $\overline{C_{\alpha \alpha}(0)}=1$, $\overline{C_{\alpha \alpha}(1)}=0$ and $\overline{C_{\alpha \alpha}(t\geq 2)}=1/3$.

Now we can ask how this behaviour is modified by intersite coupling that is weak but chosen sufficiently large that the coupled system is in the ergodic rather than the many-body localised phase (a condition that is easy to satisfy if $q \gg 1$ \cite{Chan_Spectral_2018}).
%weak local interactions. Assuming the system is not many-body localised, its 
The behaviour should be approximately described by Eq.~\eqref{eq:Cdecoupled} at early times and by the 
RMF result in Eq.~\eqref{eq:auto_rmt} at late times. Physically, we can understand this interpolation as corresponding to the dephasing of spins' precession by interactions with their neighbors. For this reason, we should expect that averaged autocorrelation functions first increase with time as in Eq.~\eqref{eq:Cdecoupled}, but that the late-time plateau which occurs there for $t \geq q$ is replaced by gradual decay of the autocorrelation function to a value that is exponentially small in $L$, as in Eq.~\eqref{eq:auto_rmt}.

As we will demonstrate, both in the RPM at large $q$ and numerically at $q=2$, the ideas above provide an qualitative characterisation of the dynamics of local observables. However, it is not immediately clear how they should be extended to describe the behaviour of observables with support on multiple sites, and how the autocorrelation function will depend on the spatial structure of the observable. This can be understood through a study of the RPM.

\subsubsection{RPM at large $q$}

In the RPM we can calculate ensemble-averaged autocorrelation functions analytically in the limit of large $q$. Our basis operators $P_{\alpha}$ are tensor products of identity and non-identity operators; it is useful to view each $P_{\alpha}$ as consisting of `clusters' of non-identity operators which are separated by identities. We denote the number of clusters in $P_{\alpha}$ by $n$, use $m=1,\ldots,n$ to label the clusters, and let $a_m$ be the length of the cluster $m$. The total number of sites on which the
operator $P_{\alpha}$ differs from the identity is then $a = \sum_{m=1}^n a_m$; we call this the weight of the operator. As we will show in Sec.~\ref{sect:rpm_derivation}, in the limit of large $q$ the ensemble-averaged autocorrelation functions depend only on $a$ and $n$ and are given for integer $t>0$ by
\begin{equation}\label{eq:corr_rpm_exact}
    \overline{C_{\alpha\alpha}(t)}=q^{-2a}e^{-2n\varepsilon t}(t-1)^n\left[1+(t-2)e^{-\varepsilon t}\right]^{a-n}.
\end{equation}
 Clearly $\overline{C_{\alpha\alpha}(t)}=0$ at the decay time $t_\mathrm{d}=1$, but from Eq.~\eqref{eq:corr_rpm_exact} we see that this is followed by a peak. In the simplest case of $n=1$, i.e. an operator $P_{\alpha}$ with support on $a$ contiguous sites, the late-time tail of this peak behaves as 
 \begin{align}
    \overline{C_{\alpha\alpha}(t)} = &\, q^{-2a}e^{-2\varepsilon t}(t-1) \\ &\times [1 + e^{-\varepsilon t}(a-1)(t-2)+\ldots]. \notag
 \end{align}
From this expression we can identify the time scale at which $\overline{C_{\alpha\alpha}(t)}$ falls to a value of order $q^{-2a}$ as $t_{\text{Th},\alpha}$, where
\begin{equation}
   t_{\text{Th},\alpha}=\frac{\ln a/\varepsilon}{\varepsilon},
\end{equation}
and we have assumed that $a\gg n$.

More generally, the late-time behaviour of $\overline{C_{\alpha \alpha}(t)}$ can be understood by viewing each cluster (of size $a_m$) as having random matrix dynamics.  Approximating the evolution of a cluster of size $a_m$ by its evolution in a $q^{a_m}$-dimensional RMF system, its contribution to autocorrelation functions is $q^{-2a_m}(t-1)e^{-2\varepsilon t}$ in the large-$q$ limit, where factors $e^{-\varepsilon t}$ at each boundary capture dephasing induced by coupling to the remainder of the system. Combining the contributions from each cluster multiplicatively, we recover the late time asymptotics of autocorrelation functions in the RPM.

Note that the value $t_\mathrm{d}=1$ for the relaxation time is a consequence of Haar-random local unitary rotations included in the Floquet operator, and is not generic. In Appendix~\ref{sec:spinful} we modify the RPM so that the single-site operations are not Haar random, giving a value for $t_\mathrm{d}$ that varies with model parameters. 

We can gain more intuition about the late-time behaviour of autocorrelation functions at $q\to\infty$ by interpreting them as sums over amplitudes of operator paths using Eq.~\eqref{eq:paths}. In Sec.~\ref{sec:paths} we will show that in the RPM at large $q$ the leading-order contribution to ensemble-averaged autocorrelation functions comes from operator paths that, despite the ergodicity of the model, are localised in the sense that all operators in the path have clusters of the same length and in the same spatial locations.

%%%%%%%%%%%%%%%%%%%%%%%%%%%%%%%%%%%%%%%%%%%%%%%%%%%%%%%%%%%%%%%%%%%%%%%%
\subsection{The PSFF in the RPM}\label{sect:overview_psff_rpm}%%%%%%%%%%
%%%%%%%%%%%%%%%%%%%%%%%%%%%%%%%%%%%%%%%%%%%%%%%%%%%%%%%%%%%%%%%%%%%%%%%%
Summing over autocorrelation functions of all operators supported within a region $A$, we arrive at the PSFF via Eq.~\eqref{eq:defPSFF}. We will restrict ourselves to a contiguous regions $A$ of length $L_A$, and we are primarily interested in the regime $L_A \ll L$. In the limit of large $q$, the average PSFF is [see Sec.~\ref{sec:psff_rpm}]
\begin{equation}\label{eq:psff_rpm_inf_q}
   \overline{K_A(t)}=\frac{1}{t}\left[\lambda_0^{L_A+1}+(t-1)\lambda^{L_A+1}\right]
\end{equation}
where $\lambda_0=1+(t-1)e^{-\varepsilon t}$ and $\lambda=1-e^{-\varepsilon t}$. Like autocorrelation functions, the PSFF exhibits an asymmetric peak after the decay time $t_{\rm d}$, and the height of the peak grows exponentially with $L_A$. For $L_A \ll L$ we then have $K_A(t) \simeq 1$ beyond a time scale which we denote by $t_{\text{Th},A}$. The Thouless time $t_{\text{Th},A}$ for the region $A$ is given by
\begin{equation}\label{eq:tloc_psff}
    t_{\text{Th},A} =\frac{\ln L_A/\varepsilon}{\varepsilon}.
\end{equation}
Although this time scale coincides with $t_{\text{Th},\alpha}$ for operators having $n=1$ and $a=L_A$, the late-time behaviour of the PSFF has contributions from all $n=1$ operators with $1 \leq a \leq L_A$. For example the contribution from single-site operators is $L_A e^{-2\varepsilon t}$, where the factor of $L_A$ comes from summing over all possible locations of the operator within $A$. Note that there is an additional factor $\varepsilon^{-1}$ within the logarithm in the Thouless time $t_{\text{Th},A}$ compared to that associated with the SFF $t_{\text{Th}}=\varepsilon^{-1}\ln L$ reported in Ref.~\cite{Chan_Spectral_2018}. This is because in the paper the coupling strength $\varepsilon$ was treated as an $O(1)$-valued number.
%In the case $L_A=L$, i.e. $K_A(t)=K(t)$, we recover $t_{\text{Th},A}=t_{\text{Th}} = \varepsilon^{-1}\ln L/\varepsilon$ from Ref.~\cite{Chan_Spectral_2018}.

The behaviour above should be contrasted with that in RMF systems without spatial structure. In that setting one finds a {\it shift-ramp-plateau} structure \cite{Joshi_Probing_2022}
\begin{equation}\label{eq:rmt_psff}
 \overline{K_A(t)}=\frac{D_A^2-1}{D^2-1}\overline{K(t)}+\frac{D^2-D_A^2}{D^2-1},
 \end{equation}
where we recall that the SFF behaves as in Eq.~\eqref{eq:SFF_RMT}. Note that Eq.~\eqref{eq:rmt_psff} can also be obtained by summing over averaged autocorrelation functions of the form in Eq.~\eqref{eq:auto_rmt}. So far we have provided only analytic results, but in Secs.~\ref{sec:Cnumerics} and \ref{sec:numerics_path}, we show numerically that the $q=2$ RSM has similar dynamics to the RPM at large $q$. 

%%%%%%%%%%%%%%%%%%%%%%%%%%%%%%%%%%%%%%%%%%%%%%%%%%%%%%%%% 
 \subsection{The OTOC in the RPM}\label{sect:RPM-OTOC}
%%%%%%%%%%%%%%%%%%%%%%%%%%%%%%%%%%%%%%%%%%%%%%%%%%%%%%%%%

The notion of operator spreading provides a different perspective on dynamics, and here the central quantity is the OTOC in Eq.~\eqref{eq:otoc}. The basic phenomenology, supported by calculations in RUCs \cite{Nahum_Operator_2018,Keyserlingk_Operator_2018}, is that operators grow ballistically, with the left- and right-hand operator `fronts' broadening diffusively in time, and so
\begin{equation}\label{otoc2}
    \overline{\mathcal{C}(x,t)}\simeq \Phi\left(\frac{v_Bt-|x|}{\sqrt{2\mathcal{D}t}}\right).
\end{equation}
Here $\Phi(u)$ is the error function, with $\Phi(u \to -\infty) \to 0$ and $\Phi(u \to \infty) \to 1$, and the quantity $v_B$ is known as the butterfly velocity. In RUCs with Haar-random two-site gates, $v_B$ and $\mathcal{D}$ are functions of $q$. A pathology of the large-$q$ limit in these models \cite{Nahum_Operator_2018,Keyserlingk_Operator_2018} (and in the RFC with Haar-distributed two-site gates \cite{Chan_Solution_2018}) is that $v_B$ approaches the lightcone velocity $v_L$ \cite{Lieb_Finite_1972} from below, and the constant $\mathcal{D}$ goes to zero. One also finds $v_B=v_L$ in dual unitary circuits \cite{Claeys_Maximum_2020}, although the generic behaviour $v_B < v_L$ has recently been recovered in a perturbative expansion around the dual unitary limit \cite{Rampp_From_2023}.

An attractive feature of the RPM is that diffusive broadening of the operator front survives even at $q\to\infty$, and $v_B < v_L$ (see also Ref.~\cite{McCulloch_Operator_2022}). To be more precise, again on a coarse-grained space-time scale, the OTOC in the RPM at large $q$ has exactly the same functional form as in Eq.~\eqref{otoc2} where the butterfly velocity $v_\mathrm{B}$ and the diffusion constant $\mathcal{D}$ are in this case
\begin{equation}\label{eq:rpm_transport}
    v_B=1-e^{-2\varepsilon},\quad \mathcal{D}= \frac{1}{2} e^{-2\varepsilon}(1-e^{-2\epsilon}).
\end{equation}
Note that the structure above shows that the operator paths contributing to OTOCs are qualitatively different to those contributing to autocorrelation functions. In particular, while operator paths contributing to autocorrelation functions do not change their spatial structure in time, those contributing to OTOCs grow ballistically; we elaborate on this difference in Sec.~\ref{sec:paths}.

%%%%%%%%%%%%%%%%%%%%%%%%%%%%%%%%%%%%%%%%%%%%%%%%%%%%%%%%%%%%%%%%%%%%%%%%%%%%%%%%
\section{Autocorrelation functions}\label{sect:rpm_derivation}%%%%%%%%%%%%%
%%%%%%%%%%%%%%%%%%%%%%%%%%%%%%%%%%%%%%%%%%%%%%%%%%%%%%%%%%%%%%%%%%%%%%%%%%%%%%%%

Our main source of insight into the behaviour of autocorrelation functions comes from the RPM. In this model the effects of locality on the SFF are evident even for $q \to \infty$ \cite{Chan_Spectral_2018}, and through Eq.~\eqref{eq:sff1} we can infer that the phenomena encountered in that setting should also manifest themselves in dynamics. In this section we calculate autocorrelation functions, finding features  that are a consequence of locality and are most dramatic for local operators. First, in Sec.~\ref{sect:diagrammatics}, we summarise the diagrammatic techniques we use for evaluating Haar averages in autocorrelation functions and the PSFF. We then calculate averaged autocorrelation functions in the RPM at large $q$ in Sec.~\ref{sect:correlation}, generalising these results to a model with a parameter-dependent relaxation time in Appendix~ \ref{sec:spinful}. As noted above, PSFFs can be expressed as partial sums over autocorrelation functions, and we calculate the averages of PSFFs in Sec.~\ref{sec:psff_rpm}. 
In addition, we connect our discussion of autocorrelation functions to the behaviour of the SFF, and in this way provide an intuitive perspective on the deviations of spectral statistics from RMT which are known to arise in spatially extended systems with only local couplings. 
Guided by these calculations, in Sec.~\ref{sec:Cnumerics} we turn to numerics on qubits chains ($q=2$), and show that the behaviour of averaged autocorrelation functions in these chains is very similar to that at large $q$ in the RPM. In Sec.~\ref{sec:finiteq}, we discuss an approximate description of the behaviour of the PSFF of the RPM at finite $q$, including a treatment of systems of finite size. 
%Finally, in Sec.~\ref{sec:sff}, we connect our discussion of autocorrelation functions to the behaviour of the SFF, and in this way provide an intuitive perspective on the deviations of spectral statistics from RMT which are known to arise in spatially extended systems with only local couplings.

\subsection{Diagrammatics and Haar averages}\label{sect:diagrammatics}

In the RPM, the single-site unitary operators $U_x$ are independent for different sites, so Haar averages can be performed independently for each site. To evaluate single-site Haar averages, it is convenient to express the averages diagrammatically. In this section, we outline the diagramatic scheme of Brouwer and Beenakker \cite{Brouwer_Diagrammatic_1996}. In brief, an average over a product of matrix elements of a unitary and its conjugate can be expressed as a sum over `pairings' of their indices, and a subset of pairings contributes to physical quantities at large $q$.

To illustrate the idea, consider first a simple example $q^{-1}\overline{\text{Tr}[ 
 UUU^\dagger U^\dagger]}$, where $U$ is a $q\times q$ Haar-distributed unitary matrix. This quantity is trivial because, without taking a Haar average, the unitarity of $U$ implies that it is simply unity. It is however instructive to reproduce this result by explicitly Haar averaging the unitaries rather than resorting to the unitarity condition $UU^\dagger=1$. The calculation highlights the importance of pairings that are subleading and hence are often neglected at large $q$.

\begin{figure}[h!]
\centering
\includegraphics[width=8cm]{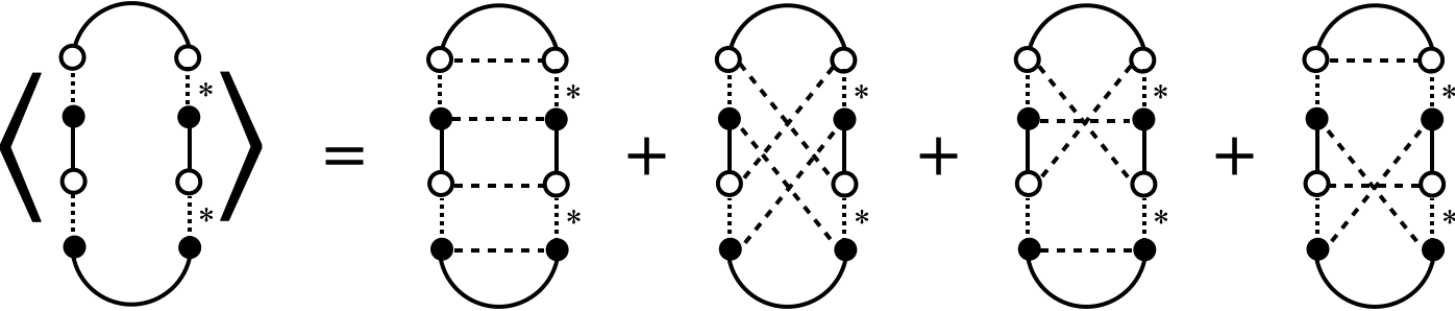}
\caption{The average of $q^{-1}\overline{\text{Tr}[ 
 UUU^\dagger U^\dagger]}$ is a sum of two Gaussian (first and second) and two non-Gaussian (third and fourth) pairings.} 
\label{fig:t2diag}
\end{figure}

We represent matrix elements $U_{aa'}$ of a unitary $U$ by a pair of black and white dots, with the first index ($a$) black and the second ($a'$) white, connected by a dotted line. $U^*_{bb'}$ is also represented in the same way, but to distinguish it from $U_{aa'}$, an asterisk is shown alongside the dotted line. With these rules, Haar averaging induces pairings of indices used for unitaries $U$ and their conjugates $U^*$, and there are four possible pairings as depicted in Fig.~\ref{fig:t2diag}. These pairings can be classified into two categories: Gaussian pairings and non-Gaussian pairings. In Fig.~\ref{fig:t2diag}, the first two pairings are referred to as Gaussian: the first and second indices of a unitary $U$ are paired with those of a unique $U^*$. This is not the case in the last two pairings in Fig.~\ref{fig:t2diag}, where at least one unitary $U$ is paired with two different conjugates $U^*$. These pairings are thus called non-Gaussian pairings.
It is known that averages of unitaries, such as those in Fig.~\ref{fig:t2diag}, can be expressed in terms of the Weingarten function \cite{Brouwer_Diagrammatic_1996}. The pairings on the right side of the equality in Fig.~\ref{fig:t2diag} have respective values
\begin{equation}
    \frac{q^2}{q^2-1},\quad \frac{1}{q^2-1},\quad -\frac{1}{q^2-1},\quad -\frac{1}{q^2-1}.
\end{equation}
The four pairings sum to one; although the first Gaussian pairing is the leading contribution at large $q$, taking into account the others is necessary for the result to be compatible with unitarity at finite $q$. In Sec.~\ref{sec:otoc_derivation_nospin} we will see that non-Gaussian pairings are important for the calculation of the OTOC even in the large-$q$ limit. In the current section, however, we will only have to consider Gaussian pairings. For this reason, it will be convenient to simplify our notation further, and below we will represent both the first and second indices of a unitary together as a single black dot. 

\subsection{Autocorrelation functions in the RPM}\label{sect:correlation}
Having the diagrammatics involved in Haar averaging in mind, we move to evaluate autocorrelation functions $\overline{\langle  P_\alpha(t) P_\alpha\rangle}$ at large $q$. Since unitaries are acting on single sites in the RPM, we can take an independent Haar average at each site and study which pairings are relevant in the large-$q$ limit. As we alluded to earlier, contributions from non-Gaussian diagrams are suppressed at large $q$. Moreover, of the $t!$ Gaussian pairings, only the $t$ {\it cyclic pairings} will contribute.

\begin{figure}[h!]
\subfloat[\label{fig:corr_gauss_diag1}]{\includegraphics[width=1.cm]{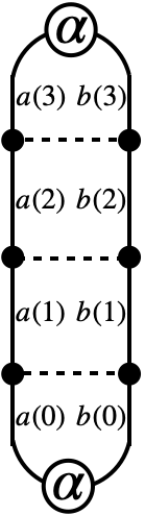}}
\hfill
\subfloat[\label{fig:corr_gauss_diag2}]{\includegraphics[width=1.cm]{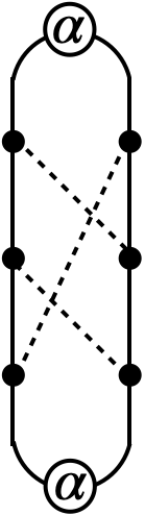}}
\hfill
\subfloat[\label{fig:corr_gauss_diag3}]{\includegraphics[width=1.cm]{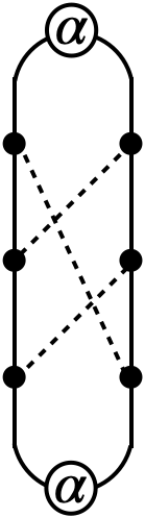}}
\hfill
\subfloat[\label{fig:corr_gauss_diag4}]{\includegraphics[width=1.cm]{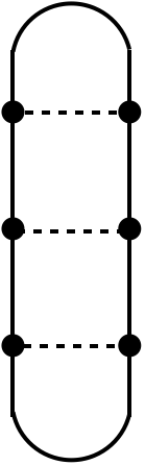}}
\caption{Leading local pairings in autocorrelation functions at $t=3$. When $P_\alpha$ acts on the site nontrivially, we have three pairings: (a) $s=0$, (b) $s=1$, and (c) $s=2$. Note that the first pairing has a vanishing contribution as mentioned in the main text. (d) The only dominant local pairing on the site where the operator acts on trivially, $s=0$. Black dots represent on-site unitaries $U_x$ and $U^*_x$.}
\end{figure}

To start, suppose the operator $P_\alpha$ acts on the site $x$ nontrivially, in the sense that $P_{\alpha_x}\not= I_q$ (the identity). Figs.~\ref{fig:corr_gauss_diag1}-\ref{fig:corr_gauss_diag3} then show the leading pairings when $t=3$. Note that every unitary $U_x$ and $U^*_x$ is now represented by a single black dot, and for brevity, we have also removed the asterisk that was previously used for indicating conjugation. If we label the unitaries $U_x$ and $U^*_x$ on the left and right edge of the diagrams as $[U_x]_{a(0) a(1)},[U_x]_{a(1) a(2)},\cdots,[U_x]_{a(t-1)a(t)}$ and $[U^*_x]_{b(0) b(1)},[U^*_x]_{b(1) b(2)},\cdots,[U^*_x]_{b(t-1)b(t)}$ from bottom to top, these pairings can be conveniently parameterised by $s_x=0,\cdots,t-1$, corresponding to indices paired cyclically as $a(r)=b(r+s_x)$. 

 Importantly, when $P_{\alpha_x}\not= I_q$, the contribution from the $s_x=0$ pairing vanishes because our basis operators are traceless. Therefore $s_x\neq0$ pairings become the leading pairings at site $x$. With such pairings, the contribution from the choice of operator is fixed by our normalisation condition $q^{-1} {\rm Tr} [P_{\alpha_x}^2] = 1$. On the other hand, if $P_{\alpha_x}=I_q$, the opposite happens: $s_x=0$ is the only pairing (with the weight $q$) compatible with the boundary condition, and as such $s_x\neq0$ pairings are all subleading at large $q$. As a result, averaged autocorrelation functions of basis operators can be expressed as sums over `pairing configurations' $s_1,\ldots,s_L$, with $s_x=0$ at any site $x$ where the operator of interest acts as the identity, and $s_x=1,\ldots,(t-1)$ otherwise.

In other words, the averaged autocorrelation function then has the form of a partition function for a one-dimensional system of $L$ sites, whose on-site degrees of freedom are pairings $s_x$. Averaging over the phases $\varphi$ on each bond [see Eq.~\eqref{eq:RPM_W2}] for large $q$ we find a factor of $e^{-\varepsilon t}$ wherever $s_x \neq s_{x+1}$, and a factor of unity otherwise. To connect with the calculation of the SFF in Ref.~\cite{Chan_Spectral_2018} we define a $t\times t$ transfer matrix which acts on a space in which the $t$ cyclic pairings correspond to orthonormal basis vectors. The matrix elements of $T$ in this basis are
\begin{align}\label{T}
    T_{ss'}=\delta_{ss'}+(1-\delta_{s s'})e^{-\varepsilon t},
\end{align}
where for brevity we have dropped the site label $x$ from the pairings $s,s'$. This object is related to the SFF via $\overline{K(t)}=\text{Tr}T^L$ in a system with periodic boundary conditions \cite{Chan_Spectral_2018}. 
A key concept which emerges from the structure of the transfer matrix is the notion of a domain wall, which is simply a configuration of unequal pairings on two adjacent sites. Domain walls do not exist in RMF systems, and they provide a way to understand the effects of locality in dynamics and on many-body spectra \cite{Chan_Spectral_2018,Garratt_Local_2021}.

To calculate autocorrelation functions, in addition to the transfer matrix we require the on-site matrices 
\begin{align}
   D_0=|0\rangle\langle 0|, \quad D_\mathrm{c}=I_t-|0\rangle\langle 0|=\sum_{s=1}^{t-1}|s\rangle\langle s|,
\end{align}
where $|s\rangle\langle s|$ is the projector onto the pairing $s$, and $I_t$ is the identity matrix of size $t$. If the operator acts nontrivially on the site we will insert a matrix $D_{\mathrm{c}}$ into the product of transfer matrices, whereas if it acts trivially we insert $D_0$. Suppose now that the operator $P_\alpha$ is made from $n$ clusters of non-identity operators and each cluster contains $a_m$ on-site operators ($m=1,\cdots,n$). The autocorrelation function of the operator $P_\alpha$ at large $q$ is then given by
\begin{align}
\overline{C_{\alpha\alpha}(t)}&=q^{-2a}\prod_{m=1}^n\left[(TD_\mathrm{c})^{a_m}T\right]_{00} \label{eq:corr_rpm_exact2}\\
     &=q^{-2a}e^{-2n\varepsilon t}(t-1)^n\left[1+(t-2)e^{-\varepsilon t}\right]^{a-n},\notag
\end{align}
provided we do not simultaneously have $n=1$ and $a=L$. In that case $\overline{C_{\alpha\alpha}(t)}=q^{-2L}(\tilde{\lambda}_0^L+(t-2)\lambda^L)$, where $\tilde{\lambda}_0=1+(t-2)e^{-\varepsilon t}$, at leading order in $q$. 

Eq.~\eqref{eq:corr_rpm_exact2} is one of the main technical results of this paper. Correlation functions as a function of $t$ have a minimum at $t$ (here $t_\mathrm{d}=1$) and for $t > t_\mathrm{d}$ initially increase. At later times this behaviour crosses over into a decrease controlled by the factors $e^{-\varepsilon t}$ which come from domain walls. The timescale that separates these two behaviours is $2+\varepsilon^{-1}$, at which $(t-2)e^{-\varepsilon t}$ is maximized. Before this time, Eq.~\eqref{eq:corr_rpm_exact2} can be approximated by
\begin{equation}\label{eq:corr_early_time}
    \overline{C_{\alpha \alpha}(t)} \simeq q^{-2a}(t-1)^n(t-2)^{a-n} e^{-(a+n)\varepsilon t}.
\end{equation}
At late times, on the other hand, we have
\begin{align}\label{eq:corr_late_time}
     \overline{C_{\alpha\alpha}(t)}\simeq q^{-2a}&e^{-2n\varepsilon t}(t-1)^n \notag\\ &\times\left[1
     %+(a-n)(t-2)e^{-\varepsilon t} + \ldots
     + {O}(te^{-\varepsilon t})\right].
\end{align}
%where the ellipsis denotes terms decaying faster than $te^{-\varepsilon t}$. 
While the early-time behaviour of autocorrelation functions depends on the internal structure of the operator (in the sense that both $a$ and $n$ enter), the late-time behaviour is simpler. Equation~\eqref{eq:corr_late_time} shows that autocorrelation functions approach $q^{-2a}e^{-2n\varepsilon t}(t-1)^n$; up to an overall $a$-dependent prefactor, in this regime the autocorrelation function becomes insensitive to the sizes of individual clusters, since it depends on $n$ but not $a$.

The approach to this asymptotic value is controlled by the timescale $t_\mathrm{Th,\alpha}$, which can be inferred from the late-time expansion in Eq.~\eqref{eq:corr_late_time}. Over the timescale $t_\mathrm{Th,\alpha}$ the average autocorrelation function drops to a value of order $q^{-2a}$, which implies $t_\mathrm{Th,\alpha}=\varepsilon^{-1}\ln(a-n)/\varepsilon$. As mentioned in Sec.~\ref{sec:overview_corr}, $t_\mathrm{Th,\alpha}$ bears a concrete physical meaning: beyond this timescale the structure within each cluster becomes unimportant, and a cluster of size $a_m$ behaves as though it evolves as a RMF system of dimension $q^{a_m}$ that is subjected to dephasing induced by coupling at the boundary to the rest of the system.

\subsection{PSFF in the RPM}\label{sec:psff_rpm}

\begin{figure}[h!]
\subfloat[\label{fig:psff_gauss_diag1}]{\includegraphics[width=1.2cm]{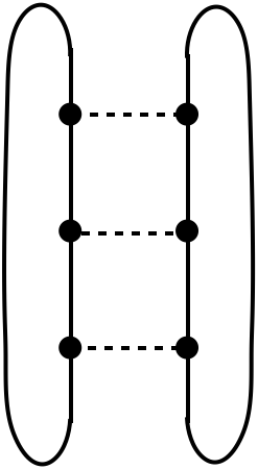}}
\hfill
\subfloat[\label{fig:psff_gauss_diag2}]{\includegraphics[width=1.2cm]{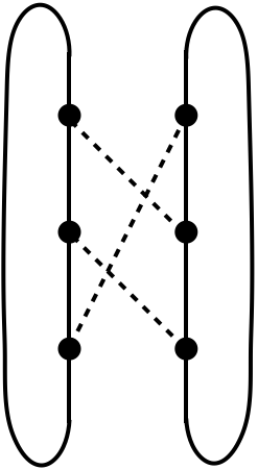}}
\hfill
\subfloat[\label{fig:psff_gauss_diag3}]{\includegraphics[width=1.2cm]{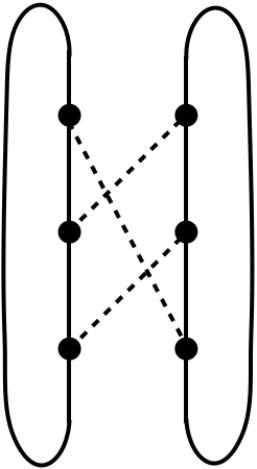}}
\hfill
\subfloat[\label{fig:psff_gauss_diag4}]{\includegraphics[width=0.67cm]{coor_diag4.pdf}}
\caption{Leading local pairings in the PSFF at $t=3$. On $A$ we have again three cyclic pairings: (a) $s=0$, (b) $s=1$, and (c) $s=2$. (d) The only leading local pairing $s=0$ on $\bar{A}$.}
\end{figure}

In this section, we provide a derivation of the exact PSFF in the RPM at $q\to\infty$ reported in Sec.~\ref{sect:overview_psff_rpm}. For the exact computation, it is convenient to express the PSFF as in Eq.~\eqref{eq:defPSFF2}.
%\begin{equation}\label{eq:psff_rpm}
 %  K_A(t) = q^{-(L-L_A)}\mathrm{Tr}_{\bar{A}}\left[(\mathrm{Tr}_A W(t)^\dagger)(\mathrm{Tr}_A W(t))\right].
%\end{equation}
Averaging this object, we find that in the region $A$ the leading pairings are cyclic with $s=0,\ldots,(t-1)$; see Figs.~\ref{fig:psff_gauss_diag1}-\ref{fig:psff_gauss_diag4}. In region $\bar{A}$, on the other hand, the leading pairing is $s=0$. The average PSFF is a sum over configurations of such pairings, and hence can be expressed in terms of the transfer matrix $T$ introduced above,
\begin{equation}\label{eq:psff_vanilla_rpm}
    \overline{K_A(t)}=[T^{L_A+1}]_{00}=\frac{1}{t}\left[\lambda_0^{L_A+1}+(t-1)\lambda^{L_A+1}\right]
\end{equation}
where $\lambda_0=1+(t-1)e^{-\varepsilon t}$ and $\lambda=1-e^{-\varepsilon t}$ are eigenvalues of $T$,  with degeneracies $1$ and $t-1$ respectively, and $L_A < L$. Note that, unlike in autocorrelation functions, here we do not have a $q$-dependent prefactor. The PSFF approaches unity at late times, with the asymptotic behaviour
\begin{equation}\label{eq:psff_decay}
   \overline{K_A(t)}= 1+\frac{t-1}{2}e^{-2\varepsilon(t-t_{\mathrm{Th},A})}+\cdots.
\end{equation}
The Thouless time $t_{\mathrm{Th},A}$ for the region $A$ is 
%then easily determined to be 
$t_{\mathrm{Th},A}=\varepsilon^{-1}\ln L_A$. The infinite-time value is controlled by the identity operator which does not evolve in time, hence $\lim_{t\to\infty}\overline{K_A(t)}=q^{-L}\Tr I_{q^L}=1$. 

In fact, the decomposition of the PSFF as a sum over autocorrelation functions in Eq.~\eqref{eq:defPSFF}, along with our result in Eq.\eqref{eq:corr_rpm_exact2},  provides us a more detailed understanding of its behaviour. At first $\overline{K_A(t)}$ increases due to the factor $t-1$, but this is eventually cut off by the domain wall cost $e^{-\varepsilon t}$. The timescale associated with this corresponds to the maximum of $(t-1)e^{-\varepsilon t}$, and we can infer which operators contribute to the build-up of this peak from Eq.~\eqref{eq:corr_rpm_exact2}.  First, recall that the relevant operator strings are supported only in region $A$. From Eq.~\eqref{eq:corr_early_time}, it is clear that the contributions from operator strings are larger when $a$ is large and $n$ is small. Note, however, that we also need to take into account the combinatorial factor associated with the number of ways of distributing clusters when $n>1$. Since for a given $n=k$ (with $k=1,\cdots, \lfloor n/2\rfloor-1$) the largest possible $a$ is $a=L_A+1-k$, the total contribution from the operator strings with $a=L_A+1-k$ and $n=k$ is
 \begin{align}
      & \begin{pmatrix}
        L_A-2\\k-1
    \end{pmatrix}(t-2)^{L_A+1-k}e^{-(L_A+1)\varepsilon t}\n
    &\quad\approx \frac{1}{k!}\times\left(\frac{L_A}{t-2}\right)^{k-1}\left[(t-2)e^{-\varepsilon t}\right]^{L_A+1},
 \end{align}
 where the factor of $q^{-2a}$ in individual autocorrelation functions has, at leading order in $q$, cancelled with the number $q^{2a}-1$ of contributing operators. Assuming that $t\ll L_A$, it is readily seen that the expression is maximized when $k=L_A/(t-2)$. Thus the operator strings that contribute to the onset of the second peak in the PSFF vary as the dynamics proceeds, and at a given time $t$ they are specified by
\begin{equation}
    a=1+\left\lfloor L_A\frac{t-3}{t-2}\right\rfloor,\quad n=\left\lfloor\frac{L_A}{t-2}\right\rfloor.
\end{equation}

Similarly, the late-time behaviour Eq.~\eqref{eq:psff_decay} can be recovered by recalling that the operator strings consisting of a single cluster decay in the slowest way. These operators have $a$ from $1$ to $L_A$, and summing over all such operators we find factors of $L_A-a+1$ corresponding to their possible locations within $A$. Combining these with the factor $q^{2a}$, the net contribution of these operator strings to $\overline{K_A(t)}$ is
\begin{equation}
    \sum_{a=1}^{L_A}(L_A-a+1)e^{-2\varepsilon t}(t-1)=\begin{pmatrix}
        L_A+1\\2
    \end{pmatrix}e^{-2\varepsilon t}(t-1),
\end{equation}
which matches Eq.~\eqref{eq:psff_decay}. 

\subsection{General scenario for PSFF at finite $q$}\label{sec:scenario}

A similar transfer matrix structure to the one that appears in the exact calculations we have presented of autocorrelation functions, the PSFF and the SFF for the RPM at large $q$ [see Eq.~\eqref{T}] also arises in calculations of the same quantities for any model and at finite $q$. Here we outline some key features of this finite-$q$ tranfer matrix that are useful in understanding the numerical results of Sec.~\ref{sec:Cnumerics}; a full account is given in Ref.~\cite{Garratt_Local_2021}. 

The starting point for the approach is to recognise that the tensor network that gives the value of an autocorrelation function, the PSFF or the SFF may be contracted in different ways but with the same result \cite{Garratt_Local_2021,Lerose_Influence_2021}. Performing this contraction by working in the time direction leads to the Floquet operator $W$. Alternatively, the contraction can be made working in the space direction, leading to a transfer matrix. The ensemble-averaged version of this transfer matrix, which we denote by $\overline{\mathcal T}$, is translation invariant and the PSFF can be expressed in terms of a matrix element of ${\overline{\mathcal T}}^{L_A+1}$ in analogy with Eq.~\eqref{eq:psff_vanilla_rpm}. 

The most important difference between the transfer matrix $\mathcal T$ for finite $q$ and the transfer matrix $T$ defined in Eq.~\eqref{T} is that the former has a much larger dimension than the latter. Specifically, since $\overline{\mathcal T}$ acts on spin configurations at a site, which are defined for every time-step and enter both $W(t)$ and $W^\dagger(t)$, it is of dimension $q^{2t}$ in the RPM (and of dimension $q^{4t}$ for a brickwork circuit, where the spin configuration is defined every half time-step). The reduction in the large $q$ limit to the $t$-dimensional matrix $T$ arises because in this limit only $t$ eigenvalues of $\overline{\mathcal T}$ are non-zero. 

Despite the difference in dimension, phenomenological arguments and numerical evidence \cite{Garratt_Local_2021,Garratt_Local_2021b} suggests that behaviour of the PSFF at finite $q$, calculated using $\cal T$, mirrors closely that for $q\to \infty$, calculated using $T$, if $L_A$ is large. The reason for this is that $\overline{\mathcal T}$ has $t$ leading eigenvalues, all with a magnitude that approaches unity for large $t$, and contributions from these eigenvalues alone dominate ${\overline{\mathcal T}}^{L_A+1}$ for large $L_A$. Moreover, at large but finite $t$, since the leading eigenvalues are in this regime non-degenerate, only the leading eigenvalue contributes to ${\overline{\mathcal T}}^{L_A+1}$ if $L_A$ is sufficiently large, just as the contribution from $\lambda_0$ dominates Eq.~\eqref{eq:psff_rpm_inf_q} in this regime.

\subsection{Numerics}\label{sec:Cnumerics}

\begin{figure}[h]
\includegraphics[width=0.47\textwidth]{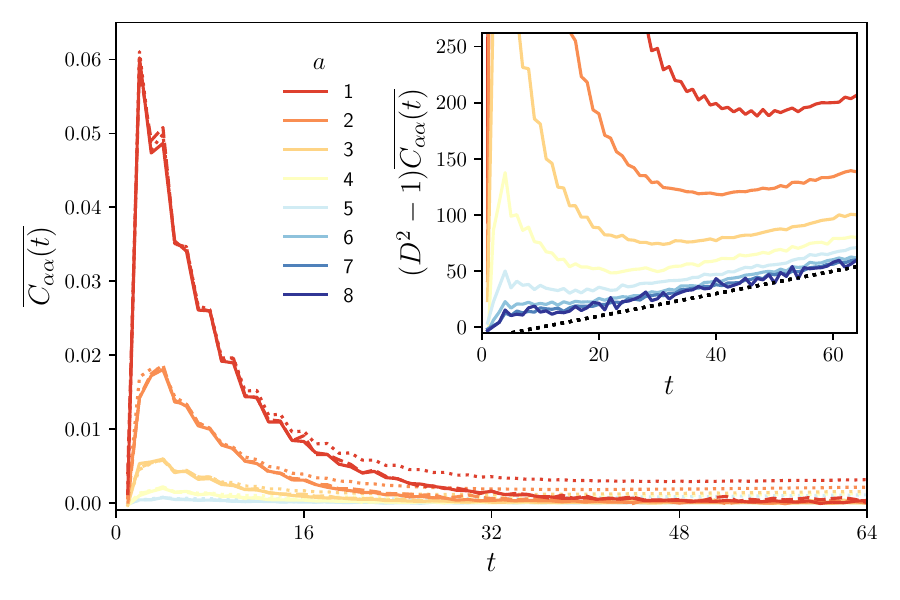}
\caption{Averaged autocorrelation functions $\overline{C_{\alpha\alpha}(t)}$ for operators consisting of a single cluster ($n=1$). The various supports $a$ are shown on the legend (with the main panel showing $a \leq 5$ only), and we use system sizes $L=8$ (dotted), $L=10$ (dashed) and $L=12$ (solid). The inset shows the $L=8$ data with the vertical axis scaled by $D^2-1=2^{2L}-1$ in order to highlight the ramp. With this scaling, we would find a ramp $t$ within RMT; for clarity we show this behavior shifted relative to the data as $t-10$ (dotted black line).}
\label{fig:C_swap}
\end{figure}

As shown earlier in this section, in the limit of large $q$ autocorrelation functions develop considerable structure as a consequence of time-translation symmetry. It is natural to ask which of these features carry over to finite $q$, and in particular to the dynamics of chains of qubits, and so here we discuss the behavior of autocorrelation functions in the kicked random-field Heisenberg model with $q=2$. Our numerical methods are described in Appendix~\ref{sec:numerics}.

First, in Fig.~\ref{fig:C_swap}, we show $\overline{C_{\alpha \alpha}(t)}$ for operator strings consisting of single clusters ($n=1$) of various lengths $a$. As predicted by our large-$q$ calculations in the RPM, there is a significant peak in $\overline{C_{\alpha \alpha}(t)}$ at early times, and the height of this peak is independent of $L$. At finite $q$ the autocorrelation function additionally has a ramp $\sim 2^{-2L}(t-1)$ at late times, and we emphasize this feature in the inset for small $L$. For small $a$ we see a significant vertical shift of the ramp relative to the RMT prediction, although even for moderately large $L$ this feature is negligible compared with the early-time peak. 

\begin{figure}[t]
\includegraphics[width=0.47\textwidth]{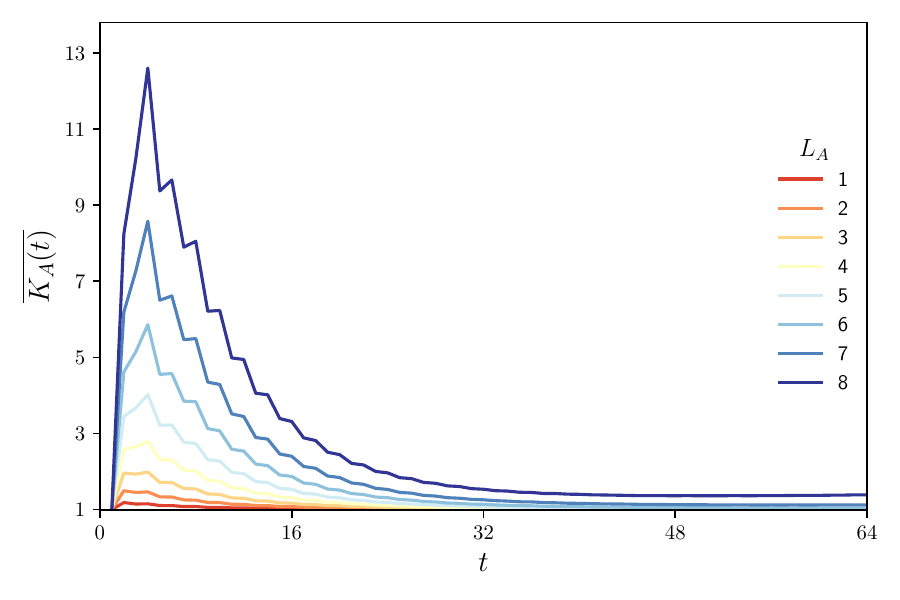}
\caption{Averaged PSFFs $\overline{K_A(t)}$ for various $L_A$ (legend). Here $L=12$.}
\label{fig:PSFF_swap}
\end{figure}

In Fig.~\ref{fig:PSFF_swap}, we show the average PSFF. Perhaps the most notable feature, which corroborates the large-$q$ result in Eq.~\eqref{eq:psff_vanilla_rpm}, is the growth of the PSFF with $L_A$. In the RPM, at any finite $t$ and at sufficiently large $L_A$, the fact that $\lambda_0(t) \geq \lambda(t)$ implies that the growth with $L_A$ is exponential. This behaviour can be seen to persist at finite $q$ by 
the arguments outlined in Sec.~\ref{sec:scenario}. 
%expressing the averaged PSFF in terms of the transfer matrix generating the averaged SFF \cite{Garratt_Local_2021,Garratt_ManyBody_2021} (see also Refs.~\cite{Akila_Particle_2016,Bertini_Exact_2018,Braun_Transition_2020} for studies of this transfer matrix in related models) and the influence matrix introduced in Ref.~\cite{Lerose_Influence_2021}. 
At finite $t$ the leading eigenvalue $\lambda_0(t)$ of the  transfer matrix is unique so, at large $L_A$, $\overline{K_A(t)} \sim \lambda^{L_A}_0(t)$ in a system with periodic boundary conditions. 

\begin{figure}
\includegraphics[width=0.47\textwidth]{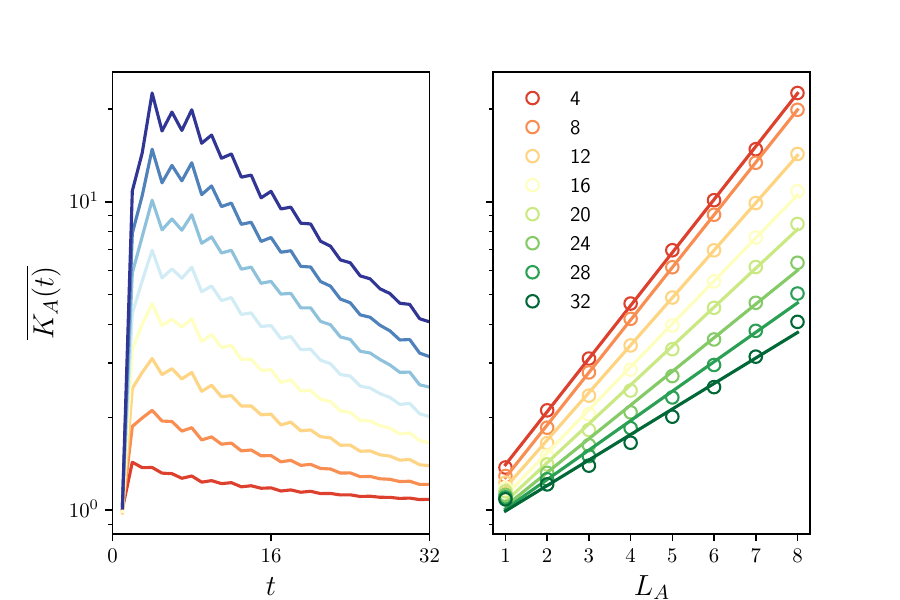}
\put(-225,132){(a)}
\put(-123,132){(b)}
\caption{Scaling of the PSFF with open boundary conditions, and with subregions $A$ and $\bar{A}$ each involving a system boundary. (a) $\overline{K_A(t)}$ versus $t$ for various $L_A$ (with the same color indexing as in Fig.~\ref{fig:PSFF_swap}). (b) Exponential growth of $\overline{K_A(t)}$ with increasing $L_A$ for various $t$ (see legend); the rate of exponential growth is set by the leading eigenvalue of the transfer matrix which also generates the average SFF (see Ref.~\cite{Garratt_ManyBody_2021} for the analysis of the transfer matrix for this model). Here $L=14$ and the vertical scale of the figure is logarithmic.}
\label{fig:obcscaling}
\end{figure}

While the data shown in Figs.~\ref{fig:C_swap} and \ref{fig:PSFF_swap} are for systems with periodic boundary conditions, 
verifying the exponential growth of $\overline{K_A(t)}$ with $L_A$ numerically is simplest in a system with open boundary conditions. This is because, with periodic boundary conditions, $\overline{K_A(t)}$ has contributions from all $t$ of the leading eigenvalues of the transfer matrix, and large values of $L_A$ are necessary to resolve the gaps between them even at moderate $t$. With open boundary conditions, on the other hand, taking $A$ to be e.g. the left-hand $L_A$ sites of the system, we only have a contribution from the leading eigenvalue (see the discussion of the effects of boundary conditions on the SFF in Ref.~\cite{Garratt_Local_2021}). For example, in the RPM at large $q$ we have the equality $\overline{K_A(t)}=\lambda^{L_A}_0(t)$. At finite $q$, Refs.~\cite{Garratt_Local_2021,Garratt_ManyBody_2021} and Ref.~\cite{Lerose_Influence_2021} imply that $\overline{K_A(t)} \simeq\mathcal{B}(t)\lambda^{L_A}_0(t)$ for large $L$ and $L_A$. Here $\mathcal{B}(t)$ is controlled by the overlap of leading eigenvectors of transfer and influence matrices, and so is independent of $L$ and $L_A$, while the ellipses denotes contributions from subleading eigenvalues. This relation implies that for large $L_A$ we should observe the scaling $\log \overline{K_A(t)} \propto L_A \log\lambda_0(t)$, and we demonstrate this numerically in Fig.~\ref{fig:obcscaling}. At large $q$ we have $\mathcal{B}(t)=1$, and although at $q=2$ there are small variations of $\mathcal{B}(t)$ with $t$, the changes they induce in $\overline{K_A(t)}$ are much less dramatic than those induced by the evolution of $\lambda_0(t)$, which was studied in this model in Ref.~\cite{Garratt_ManyBody_2021}: at $t=1$ we have $\lambda_0(1)=1$, with the eigenvalue then passing through a broad maximum extending from $t=2$ to $t=10$, and then gradually decreasing to unity at late times.

\subsubsection{Model dependence}\label{modeldependence}

The kicked random field Heisenberg model has the property that its ensemble of Floquet operators is invariant under arbitrary on-site unitary rotations. This ensures that $\overline{C_{\alpha \alpha}(t)} = 0$ at $t=1$. For circuits that are not Floquet, but instead have gates chosen independently at each time step, a similar invariance implies $\overline{C_{\alpha \alpha}(t)} = 0$ for all time-steps $t>0$. In this subsection we examine behaviour in models with either Floquet or random time evolution operators when the matrices $U_x$ and $U^\prime_{x+1}$ that appear in Eq.~\eqref{SWAP} are chosen from distributions other than the Haar distribution. For each choice of distribution, we compare behaviour in a Floquet model with behaviour in a model that is similar but has the evolution operator chosen independently at each time-step. In this way we aim to expose the consequences of time-translation symmetry.

\begin{figure}
\includegraphics[width=0.47\textwidth]{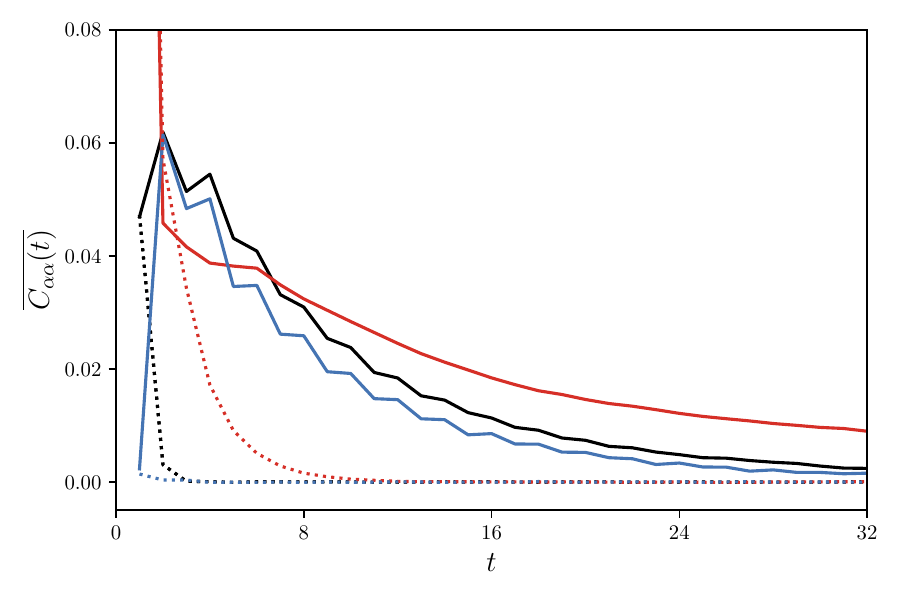}
\caption{Averaged autocorrelation functions of single-site Pauli operators for different ensembles of $U_x$ [Eq.~\eqref{SWAP}]. Solid and dashed lines show results for systems with and without time-translation symmetry, respectively. In the first case the evolution operator for time $t$ is $W(t)=W^t$ and we average over choices of $W$. In the second case we instead have $W(t)=\prod_{r=0}^{t-1} W_r$, where each `layer' $W_r$ is drawn independently from the same ensemble as $W$. 
For the black data $U_x$ is given by the square of a Haar-random unitary, while red ($g=1$) and blue ($g=4$) data show results for $U_x = e^{-igh_x}$, with $h_x$ drawn from the GUE with normalization $\overline{\text{Tr}h_x^2}=1$. For $g=1$ we have $\overline{C_{\alpha\alpha}(1)} = 0.2575(5)$ (not shown). We fix $J=0.2$ and $L=12$ for all circuits shown.}
\label{fig:Cvariations}
\end{figure}

Results for averaged autocorrelations functions of single-site operators are shown in Fig.~\ref{fig:Cvariations} for three choices of distribution, as detailed in the figure caption. In one case the early-time minimum in $\overline{C_{\alpha \alpha}(t)}$ is eliminated, in a second case the minimum is much shallower than with Haar-distributed gates, and in the third case the minimum remains deep. Despite these variations, all three choices of distribution lead to a tail in the autocorrelation function that is much more prominent in the Floquet model than in the corrseponding model without time-translation symmetry.

\begin{figure}
\includegraphics[width=0.47\textwidth]{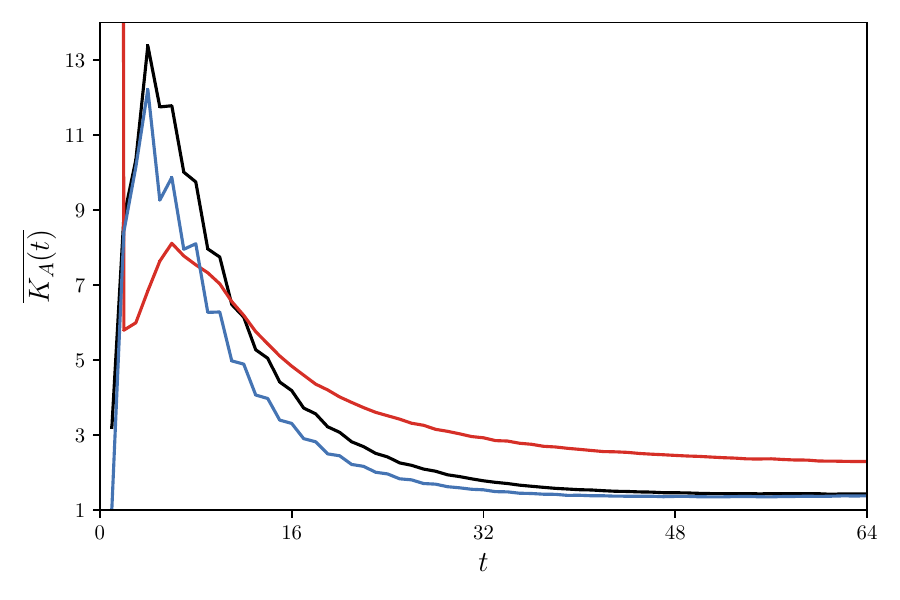}
\caption{PSFFs for the Floquet models studied in Fig.~\ref{fig:Cvariations}, with a subregion size of $L_A=8$ and total system size $L=12$. For $g=1$ we have $\overline{K_A(1)}=297.84(2)$ (not shown).}
\label{fig:Kvariations}
\end{figure}

In order to investigate the behaviour for operators with larger supports, we consider the PSFF $K_A(t)$, which can be expressed as a sum over autocorrelation functions for a complete set of operators with support on subsystem $A$ [see Eq.~\eqref{eq:defPSFF}]. Results for $L_A=8$ are shown in Fig.~\ref{fig:Kvariations}. For this case of operators having support on multiple sites, all three choices of distribution for single-site gates generate a short-time minimum and an asymmetric intermediate-time peak. Model-dependent differences from the case of Haar-distributed single-site gates consist of a varying depth to the minimum and a varying timescale for the late-time decay of the peak (in all cases many times the Floquet period).

Finally, in Appendix.~\ref{sec:rpm_more_layers}, we also derive the ensemble-averaged autocorrelation functions in the large-$q$ RPM with $n$ layers of single-site Haar unitaries at each time step. The consequence of such an alteration turns out to be rather straightforward: the value of $\overline{C_{\alpha\alpha}(t)}$ in the multi-layer RPM is simply given by the autocorrelation function in the standard RPM at $nt$. This implies that the early-time dip will disappear when the coupling strength $\varepsilon$ is large enough, 
%in which case the intermediate-time peak reaches its height in a few Floquet periods, but otherwise the only effect is the prolonged tail of the peak.
but the tail in the autocorrelation function remains.

In summary, we believe that the data shown in Figs.~\ref{fig:Cvariations} and \ref{fig:Kvariations}, and also the analytic results on the multi-layer RPM demonstrate that the intermediate-time peak or tail in autocorrelation functions and the PSFF of Floquet circuits, which we have emphasised as our main finding, is a robust feature and is not limited to simple models for which $\overline{C_{\alpha \alpha}(t)} = 0$ at $t=1$. In particular, this feature is a consequence of time-translation symmetry and is absent for circuits that are random in time.

\subsection{PSFF at finite $q$}\label{sec:finiteq}
So far we have focused exclusively on the large $q$ limit in the RPM. In the present section we attempt to go beyond this limit and capture in a quantitative way effects for $q$ large but finite. We find that finite $q$ has important consequences in finite-size systems for both autocorrelation functions and the PSFF as it restores their late-time ramps.

Accounting for the full ramifications of finite $q$ in a systematic way is a rather demanding task and we leave it for future studies. Rather, here we propose a possible mechanism by which the ramp is induced in the PSFF. To this end, it is illuminating to recall how the {\it diagonal approximation} \cite{Garratt_Local_2021}, which we define shortly, gives rise to the ramp at large but finite $D$ in RMF systems.

\subsubsection{Diagonal approximation in RMF systems}\label{sec:diagonal_approximation}
We start from the expression Eq.~\eqref{eq:defPSFF2} with $W$ is a $D \times D$ Haar random matrix. To perform the average it will be convenient to make the various index summations explicit; here indices $a(r)=0,\ldots,(D_A-1)$ with $r=0,\ldots,(t-1)$ label basis states in region $A$, while $\bar{a}_r=0,\ldots,(D/D_A-1)$ label basis states its complement; we use $b$ rather than $a$ for indices in the product of $W^*$. Using this notation we have
\begin{align}\label{eq:psff_fock}
    &K_A(t)=\frac{D_A}{D}\sum_{\{a\},\{b\}\in A}\sum_{\{\bar{a}\},\{\bar{b}\}\in \bar{A}}\delta_{\bar{a}(0)\bar{b}(0)}\delta_{\bar{a}(t)\bar{b}(t)} \n
    &\quad\times W_{a(0)\bar{a}(0)}^{a(1)\bar{a}(1)}W_{a(1)\bar{a}(1)}^{a(2)\bar{a}(2)}\cdots W_{a(t-1)\bar{a}(t-1)}^{a(0)\bar{a}(t)} \n
&\quad\times [W^*]_{b(0)\bar{b}(0)}^{b(0)\bar{b}(1)}[W^*]_{b(1)\bar{b}(1)}^{b(2)\bar{b}(2)}\cdots [W^*]_{b(t-1)\bar{b}(t-1)}^{b(0)\bar{b}(t)}
\end{align}
where $W_{ab}^{cd}=\langle a,b|W|c,d\rangle$. The factor $\delta_{\bar{a}(0)\bar{b}(0)}\delta_{\bar{a}(t)\bar{b}(t)}$ implements the contraction of $\bar{A}$ indices between $W(t)$ and $W^*(t)$ at the initial and final times. Haar averaging then induces pairings among indices, and as above we are interested in the {\it cyclic} pairings, i.e. those parameterized by $(a(r),\bar{a}(r))=(b(r+s),\bar{b}(r+s))$ for all $r$ and fixed $s=0,\cdots,t-1$; it can be shown that in this setting the cyclic pairings are dominant for $D_A,D/D_A\gg 1$. Note that, since RMF systems have no spatial structure the pairing $s$ is here uniform throughout the system. By contrast, in the RPM at large $q$ we found $s=0$ in region $\bar{A}$ while $s=0,\ldots,(t-1)$ in region $A$.

In the computation of the SFF, the restriction to cyclic pairings without spatial structure is known as the diagonal approximation. The sum over the $t$ cyclic pairing generates the characteristic ramp $\overline{K(t)}=t$ in RMF systems, with each pairing contributing unity. In the computation of the PSFF, on the other hand, the $s=0$ and $s \neq 0$ pairings are inequivalent. First, the contribution from $s=0$ pairing in the summation in Eq.~\eqref{eq:psff_fock} simply gives
\begin{equation}
  \frac{1}{D^t}\delta_{\bar{a}(t)\bar{b}(t)}\prod_{r=0}^{t-1}\delta_{a(r)b(r)}\delta_{\bar{a}(r)\bar{b}(r)}.
\end{equation}
Summing over $\{b\}$ and $\{\bar{b}\}$, we are left with $t$ and $t+1$ free indices on $A$ and $\bar{A}$. Hence after the sum over the remaining indices, we get $D/D_A$.
When pairings are $s\neq0$, the indices used in imposing the boundary condition on $\bar{A}$, i.e. $\bar{a}(0), \bar{b}(0), \bar{a}(t), \bar{b}(t)$ also appear in the pairing conditions $\delta_{\bar{a}(r)\bar{a}(r+s)}$, leaving only $t-1$ free indices on $\bar{A}$. This means that the net contribution upon taking the sum is $(t-1)D_A/D$. Combining both cases we obtain the diagonal approximation to the PSFF
\begin{equation}\label{eq:rmt_psff_diag}
    \overline{K_A(t)}\simeq \frac{D^2_A}{D^2}(t-1)+1.
\end{equation}
Comparing this with Eq.~\eqref{eq:rmt_psff} it can be verified that the diagonal approximation is appropriate in this setting for $1 \ll t \ll D$ and $1\ll D_A \ll D$.

%which is consistent with \eqref{eq:rmt_psff} when $t>0$ at large $D$. \color{red} S: Is this true? \color{black} Note that, in contrast to the SFF, the diagonal approximation is not exact for the PSFF even in RMF systems.

\subsubsection{Local diagonal approximation in the RPM}
Motivated by the above result, we make a similar approximation in computing the finite $q$ PSFF in the RPM, which we call the {\it local diagonal approximation}. Namely, we include the contribution from cyclic Gaussian pairings on $\bar{A}$ with $s \neq 0$, as in the diagonal approximation for RMF systems. Of course, at finite $q$ there are contributions from many other possible pairings both on $A$ and $\bar{A}$ which we simply neglect. The approximation generates a ramp at late times.

We first notice that on-site diagrams on $\Bar{A}$ with $s\neq0$ local pairing (imagine replacing the $s=0$ pairing in Fig.~\ref{fig:psff_gauss_diag4} with that in Fig.~\ref{fig:psff_gauss_diag2} and Fig.~\ref{fig:psff_gauss_diag3}) carry the factor $q^{-1}$. Further, accounting for the extra $q^{-1}$ from the normalisation of trace, each $s\neq0$ pairing carries a factor of $q^{-2}$ whereas the $s=0$ pairing carries a factor of unity. Writing the average PSFF as a sum over all cyclic pairings, including all values of $s$ at each site, we find
\begin{equation}\label{eq:psff_finite_q}
    \overline{K_A(t)}=\mathrm{Tr}(T^{L_A} \hat{T}^{L-L_A}),
\end{equation}
where $\hat{T}=Td$ and the on-site diagonal matrix $d$ has entries $d_{ss'}=\delta_{ss'}(\delta_{s0}+(1-\delta_{s0})q^{-2})$ where $s,s'=0,\cdots,t-1$. It is immediate to see that this gives the desired RMT behaviour at $t\to\infty$. At infinite time, the transfer matrix becomes the identity matrix, hence
\begin{equation}\label{eq:finite_q_infty}
    \lim_{t\to\infty}\overline{K_A(t)}=\mathrm{Tr}(d^{L-L_A})=1+\frac{t-1}{q^{2(L-L_A)}}.
\end{equation}
The result in Eq.~\eqref{eq:finite_q_infty} corresponds to uniform pairing throughout the system.

\begin{figure}[h!]
\subfloat[\label{fig:domainwall1}]{\includegraphics[width=3.cm]{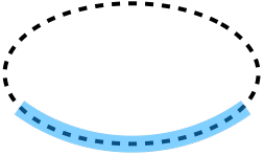}}
\hfill
\subfloat[\label{fig:domainwall2}]{\includegraphics[width=3.cm]{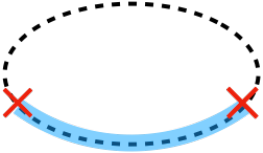}}
\hfill
\subfloat[\label{fig:domainwall3}]{\includegraphics[width=3.cm]{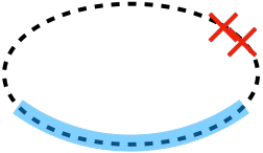}
}
\hfill
\subfloat[\label{fig:domainwall4}]{\includegraphics[width=3.cm]{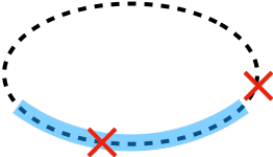}
}
\caption{Leading domain wall configurations contributing to the late-time behaviour of the PSFF at large but finite $q$ with periodic boundary conditions. A red cross indicates the location of a domain wall and the shaded blue region corresponds to region $A$. (a), (b), and (c) and (d) give the second, third, and fourth terms in Eq.~\eqref{eq:finite_q_domainwall}, respectively.}
\end{figure}

We must now ask when the behaviour in Eq.~\eqref{eq:finite_q_infty} sets in; this defines the time scale $\trmt$. In fact, that can be inferred readily from an argument based on domain walls. As we already know, at large $q$, the decay of the PSFF is governed by two domain walls supported on $A$ where each domain wall comes with a statistical cost $e^{-\varepsilon t}$ (see Fig.~\ref{fig:domainwall2}, where each domain wall is indicated by a red cross). This is still the case at finite $q$ within the local diagonal approximation, but now domain walls can be supported anywhere in the system.

The next-to-leading order corrections within the local diagonal approximation correspond to two different kinds of domain wall configurations. The first corresponds to a single-site $s\neq0$ domain within $\Bar{A}$, with a $s=0$ domain covering the rest of the system (see Fig.~\ref{fig:domainwall3}). The number of such pairing configurations is $(L-L_A)(t-1)$. Thus, their total contribution to the average PSFF is
\begin{equation}
    (L-L_A)(t-1)\frac{e^{-2\varepsilon t}}{q^2}.
\end{equation}
In the second kind of configuration, there is a $s \neq 0$ domain having one of its walls in $A$ and one in $\bar{A}$. Since each site having $s\neq 0$ in $\bar{A}$ costs a factor $q^{-2}$, the leading contribution of this kind corresponds to a domain that has only a single site in $\bar{A}$ [see Fig.~\ref{fig:domainwall4}]. Summing over the $L_A$ possible domain wall locations we find a contribution
\begin{equation}
    2L_A(t-1)\frac{e^{-2\varepsilon t}}{q^2},
\end{equation}
and for the PSFF we then have
\begin{align}
   & \overline{K_A(t)}\simeq1+\frac{t-1}{q^{2(L-L_A)}} \label{eq:finite_q_domainwall}  \\
&\quad+\begin{pmatrix}
    L_A+1\\ 2
\end{pmatrix}(t-1)e^{-2\varepsilon t}+(L+L_A)(t-1)\frac{e^{-2\varepsilon t}}{q^2}. \notag
\end{align}
at late times within the local diagonal approximation. The first and second terms come from configurations without domain walls, and correspond to $s=0$ and $s \neq 0$, respectively. The third term, which is leading order in $q$ comes from domain walls in or on the boundary of $A$. The fourth term, which we now see is always smaller than the third, corresponds to domain wall configurations of the kind described above.

From Eq.~\eqref{eq:finite_q_domainwall} we can identify $\trmt$. In brief, the domain wall contribution to the PSFF, which arises even at leading order in $q$, decays as $e^{-2\varepsilon t}$, and eventually this becomes smaller than the ramp expected from RMT, which here has prefactor $q^{2(L_A-L)}$. Therefore 
\begin{equation}
    \trmt\sim \frac{\ln q}{\varepsilon}(L-L_A).
\end{equation}
Although this timescale is proportional to the distance $L-L_A$, it should not be interpreted in terms of a velocity since it is simply the time at which two distinct contributions to the PSFF exchange dominance.

The situation above should be compared with that in Floquet systems having scalar charge conservation. Working within the diagonal approximation, Refs.~\cite{Friedman_Spectral_2019,Roy_2020_Random,Moudgalya_2021_Spectral} found $t_{\text{RMT}}\sim L^2$, and so there is a very clear connection between time scales appearing the spectrum and in the (diffusive) dynamics of a local charge densities. 

\section{Out-of-time-order correlators}\label{sec:otoc_derivation_nospin}

In this section we calculate the average OTOC for the RPM. In RUCs the standard calculation of the OTOC \cite{Nahum_Operator_2018,Keyserlingk_Operator_2018} involves mapping the dynamics of operators to a kind of classical stochastic dynamics. There one finds that operators grows ballistically, and that the operator fronts broaden diffusively as $\sim (\mathcal{D}t)^{1/2}$. Here we recover this behaviour, including the broadening of the front, in the large-$q$ limit of a Floquet model [c.f. Ref.~\cite{Chan_Solution_2018}]. 

In generic Floquet systems it is not possible to directly map the calculation of the OTOC to a property of a stochastic process, simply because unitary operations cannot be averaged independently at different time steps, and so here we will instead evaluate the OTOC using a spatial transfer matrix. 

\subsection{Diagrammatics and Haar averages}

Compared to the computation of averaged autocorrelation functions and the PSFF at large $q$, that of the OTOC is more complicated. There are two reasons for this. First, the fact that the OTOC involves two copies of both $W(t)$ and its conjugate means that the transfer matrix has a dimension that is twice as large as that for the PSFF. Second, the pairings of unitary operators which contribute at leading order in $q$ can be non-Gaussian \cite{Brouwer_Diagrammatic_1996}, whereas for the PSFF are all contributing pairings are Gaussian at this order. 

\begin{figure}[h!]
\subfloat[\label{fig:diag1}]{\includegraphics[width=5cm]{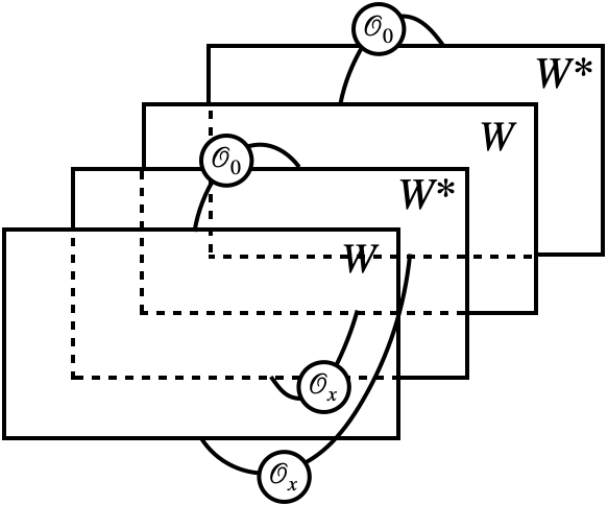}}
\hfill
\subfloat[\label{fig:diag2}]{\includegraphics[width=2.5cm]{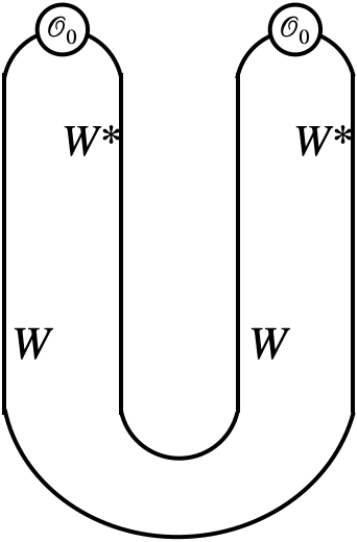}}
\hfill
\subfloat[\label{fig:diag3}]{\includegraphics[width=2.5cm]{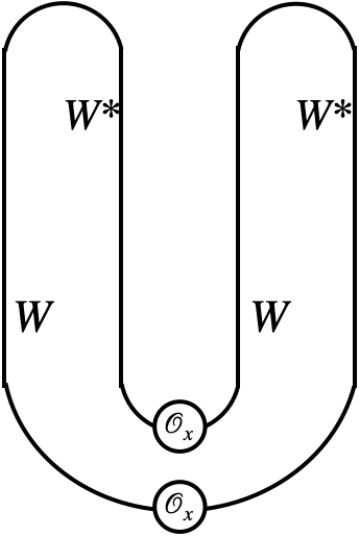}}
\hfill
\subfloat[\label{fig:diag4}]{\includegraphics[width=2.5cm]{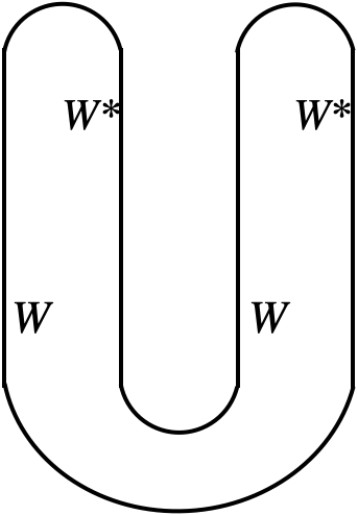} 
}
\caption{(a): Diagrammatic representation of $\mathcal{C}(x,t)$. (b, c, d): Single-site diagram on $0$, $x$, and other sites.}
\end{figure}

In Fig.~\ref{fig:diag1} we illustrate the basic structure of the OTOC, as defined in Eq.~\eqref{eq:otoc}, and in Fig.~\ref{fig:diag2}-\ref{fig:diag4} we show the contractions of indices relevant to the different sites. 
The strategy we follow is the same as before. We first evaluate averages over one-site unitary operators $U_x$ and enumerate all relevant leading-order pairings. This will allow us to express the ensemble-averaged OTOC as a product of transfer matrices, where these objects can be viewed as acting on the space of pairings. However, since non-Gaussian pairings now contribute at leading order, it will be useful to represent diagrams in a slightly different way.

\begin{figure}[h!]
\subfloat[\label{fig:square_nospin1}]{\includegraphics[width=4.cm]{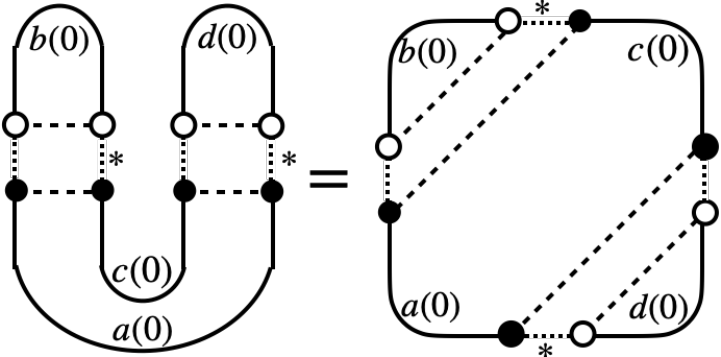}}
\hfill
\subfloat[\label{fig:square_nospin2}]{\includegraphics[width=4cm]{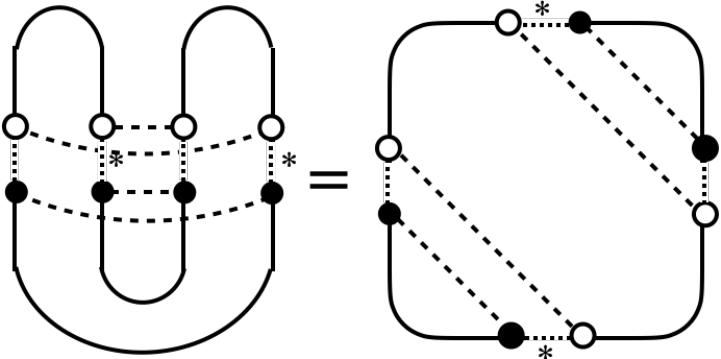}}
\hfill
\subfloat[\label{fig:square_nospin3}]{\includegraphics[width=4cm]{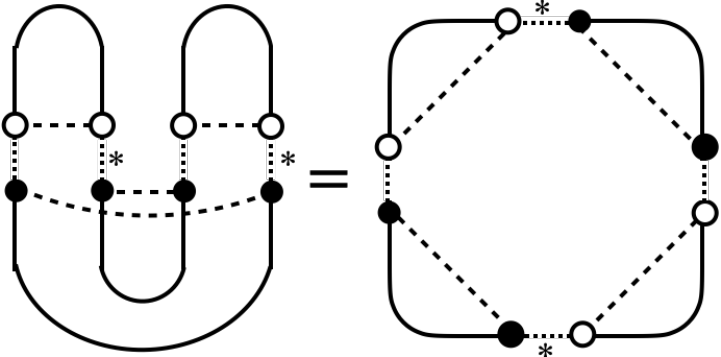}
}
\hfill
\subfloat[\label{fig:square_nospin4}]{\includegraphics[width=4cm]{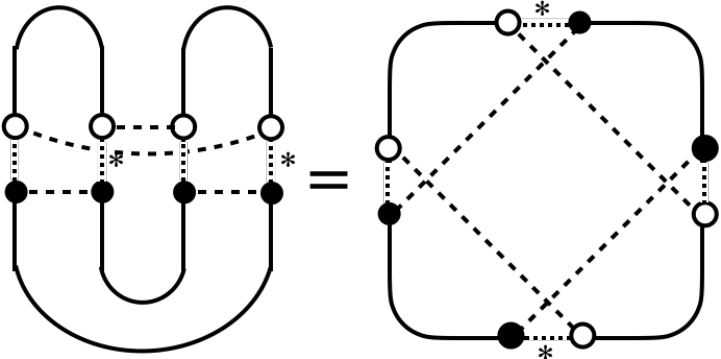}}
\caption{(a,b): Gaussian pairings carrying the value $q$ and (c, d): non-Gaussian pairings carrying the values $-q$ and $-q^{-2}$, respectively, for $t=1$.}
\end{figure}

Let us start with a site $x' \neq 0,x$, which contains no local operators [see Fig.~\ref{fig:diag4}]. The on-site diagram consists of two Floquet operators $W$ and their conjugates $W^*$, with indices contracted as in the figure. An equivalent way of representing this is to draw a square and assign a Floquet operator (or its conjugate $W^*$) to each edge. To illustrate this, consider first $t=1$: in Fig.~\ref{fig:square_nospin1} we show the contractions of unitary operators using both ``U-shape'' and ``square-shape'' diagrams, where indices $a(0),b(0),c(0),d(0)$ label the boundaries of the edges in both diagrams. Comparing the location of the labels in both diagrams, one can also infer the correspondence between edges. 

At $t=1$, there are four pairings: two are Gaussian (see Figs.~\ref{fig:square_nospin1} and \ref{fig:square_nospin2}), and two are non-Gaussian (see Figs.~\ref{fig:square_nospin3} and \ref{fig:square_nospin4}). From Ref.~\cite{Brouwer_Diagrammatic_1996} the values of these pairings can be readily obtained for Figs.~\ref{fig:square_nospin1} and \ref{fig:square_nospin2} as $q^3/(q^2-1)$, for \ref{fig:square_nospin3} as $-q^3/(q^2-1)$, and for \ref{fig:square_nospin4} as$-q/(q^2-1)$. At large $q$ these become $q,-q$, and $-q^{-2}$, respectively. This indicates that, at large $q$, we have contributions from the Gaussian diagrams \ref{fig:square_nospin1} and \ref{fig:square_nospin2}, as well as non-Gaussian diagrams \ref{fig:square_nospin3} which have no crossings in the square representation. Non-Gaussian diagrams with crossings, such as \ref{fig:square_nospin4}, are subleading. Note that, if we keep track of only the Gaussian diagrams and non-crossing non-Gaussian diagrams, we can again avoid distinguishing first and second indices of unitaries $U_x$ provided we use square diagrams. For this reason, as in the previous section, we will just use black dots to indicate the unitaries $U_x$. As we will see shortly, the inclusion of non-Gaussian diagrams turn out to be crucial for the OTOC to behave in the way we expect.

\begin{figure}[h!]
\subfloat[\label{fig:t3square1}]{\includegraphics[width=1.8cm]{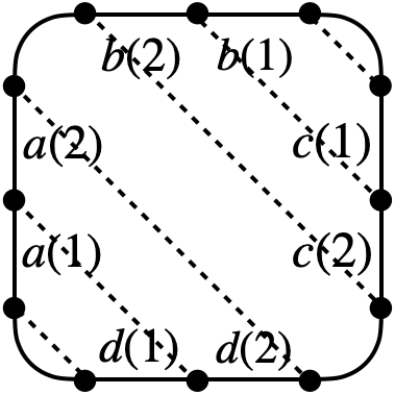}
}
\hfill
\subfloat[\label{fig:t3square2}]{\includegraphics[width=1.8cm]{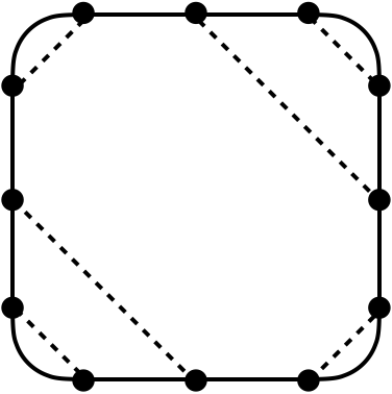}
}
\hfill
\subfloat[\label{fig:t3square3}]{\includegraphics[width=1.8cm]{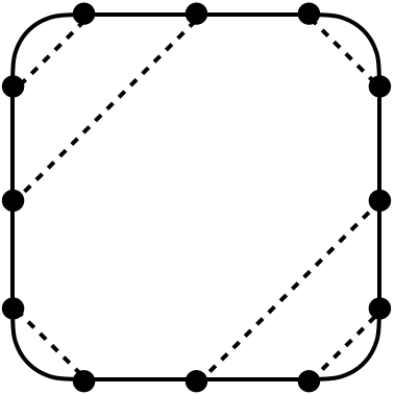}
}
\hfill
\subfloat[\label{fig:t3square4}]{\includegraphics[width=1.8cm]{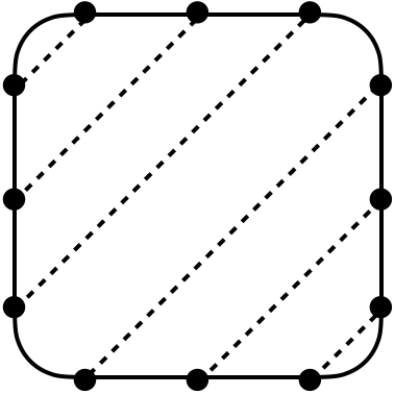}
}
\hfill
\subfloat[\label{fig:t3square5}]{\includegraphics[width=1.8cm]{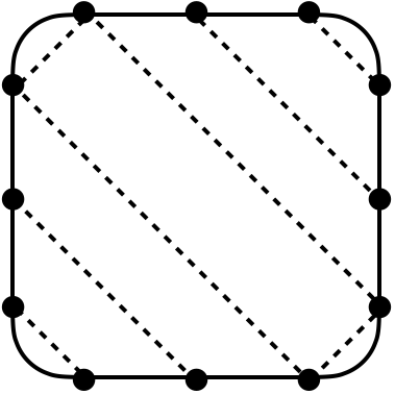}
}
\hfill
\subfloat[\label{fig:t3square6}]{\includegraphics[width=1.8cm]{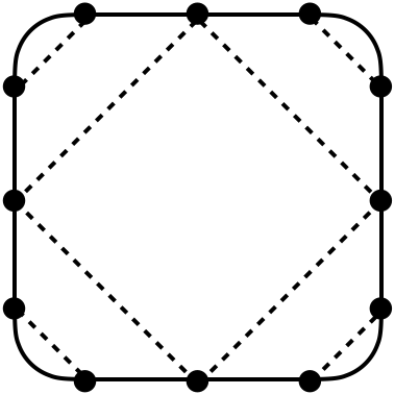}
}
\hfill
\subfloat[\label{fig:t3square7}]{\includegraphics[width=1.8cm]{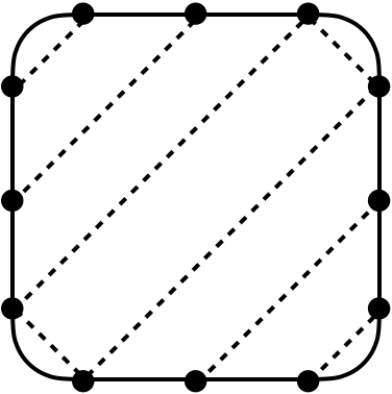}
}
\caption{(a)-(d): Gaussian pairings $g_0,\cdots,g_3$ and (e)-(g): the non-Gaussian pairings $n_0,n_1$, and $n_2$ obtained as superpositions of the Gaussian diagrams for $t=3$.}
%\label{fig:t1diag}
\end{figure}

The analysis above can be readily carried over to generic $t>1$, in which case we have $t$ unitaries on each edge. This means that on each edge there are $t-1$ bonds that connect unitaries. On the left and right edges of the square diagrams (which each correspond to $W(t)$) we assign indices $a(1),a(2),\cdots,a(t-1)$ and $c(1),c(2),\cdots,c(t-1)$ to these bonds, whereas on the upper and lower edges (which each correspond to $W^*(t)$) we assign indices
$b(t-1),b(t-2),\cdots,a(1)$ and $d(t-1),d(t-2),\cdots,d(1)$; our ordering convention is shown in Fig.~\ref{fig:t3square1}. The two ends of each edge also carry indices, which we denote e.g. $a(0)$ and $a(t)$, but the trace structure imposes conditions $a(0)=d(0), a(t)=b(t), b(0)=c(0), c(t)=d(t)$, leaving four free indices out of eight edge-attached indices. 

With the above convention the relevant Gaussian diagrams, which we denote by $g_p$ with $p=1,\cdots,t-1$, correspond to indices paired as
\begin{align}\label{gaussian_param}
        a(\tau)=b(\tau),\quad c(\tau)=d(\tau), & \quad p\leq \tau\leq t-1\\
        a(\tau)=d(\tau),\quad b(\tau)=c(\tau),  & \quad1\leq \tau\leq p.
\end{align}
This parametrisation, however, does not cover two diagrams that are given by $a(\tau)=b(\tau),\, c(\tau)=d(\tau)$ and $a(\tau)=d(\tau),\, b(\tau)=c(\tau)$ for $1\leq\tau\leq t-1$, which we denote by $g_t$ and $g_0$, respectively. We therefore have in total $t+1$ Gaussian diagrams $\{g_p\}_{p=0}^t$, and in Figs.~\ref{fig:t3square1}-\ref{fig:t3square4} we depict $g_0,\cdots,g_3$ at $t=3$. Evaluating the various diagrams, we find that each carries an overall weight $q$.

The above labelling convention is also useful for enumerating the contributing non-Gaussian diagrams. Those which generate contributions that are leading order in $q$ are obtained by stacking two consecutive Gaussian diagrams (e.g. $p=p_1, p_1+1$) on top of each other; see for example the stacking of Figs.~\ref{fig:t3square1} and \ref{fig:t3square2} to generate the non-Gaussian diagram Fig.~\ref{fig:t3square5}. The non-Gaussian diagrams, which we denote by $n_p$ for $0\leq p\leq t-1$, correspond to the pairings
\begin{align}\label{nongaussian_param}
      a(\tau)=b(\tau),\quad c(\tau)=d(\tau),\quad  & p< \tau\leq t\\
       a(\tau)=d(\tau),\quad b(\tau)=c(\tau),\quad    & 0\leq \tau\leq p.
\end{align}
The non-Gaussian pairings $n_1,n_2$, and $n_3$ are depicted in Figs.~\ref{fig:t3square5}, \ref{fig:t3square5}, and \ref{fig:t3square7}, and we find that each $n_p$ carries a weight $-q$.

Haar averaging for each site on $y\neq0,x$ therefore yields $2t+1$ leading pairings where $t+1$ of them are Gaussian carrying the factor $q$ and $t$ of them are non-Gaussian with the factor $-q$. We organise these pairings by introducing a new set of labels $\{h_p\}_{p=0}^{2t}$ defined by
\begin{equation}
    h_p:=\begin{cases}
        g_p & p= 0,1,\cdots,t \\
        n_p & p=t+1,t+2,\cdots,2t
    \end{cases}
\end{equation}
so that the first $t+1$ pairings are Gaussian whereas the latter $t$ pairings are non-Gaussian. For our transfer matrix calculation it will be convenient to associate the $2t+1$ pairings with an orthonormal basis $|p\ket$ of vectors labelled by the index $p=0,\ldots,2t$.

We now move on to evaluate the contributions from nearest-neighbour interactions in the model. It is worth stressing that, unlike in the case of autocorrelation functions, the phases could partially cancel out even if the pairings of two adjacent sites differ. For example, if the sites $y$ and $y+1$ have local pairings corresponding to the diagram Fig.~\ref{fig:t3square1} and \ref{fig:t3square7} at $t=3$, the non-vanishing contribution to the phase is
\begin{align}
      \Phi&=\varphi_{y;a_y(1),a_{y+1}(1)}+\varphi_{y;a_y(2),a_{y+1}(2)}-\varphi_{y;b_y(1),b_{y+1}(1)}\n
      &\quad-\varphi_{y;b_y(2),b_{y+1}(2)}+\varphi_{y;c_y(1),c_{y+1}(1)}+\varphi_{y;c_y(2),c_{y+1}(2)}\n
      &\quad-\varphi_{y;d_y(1),d_{y+1}(1)}-\varphi_{y;d_y(2),d_{y+1}(2)}.
\end{align}
Upon averaging over the phases, we find $\langle e^{\ii\Phi}\rangle=\rho^2$ where $\rho=e^{-2\varepsilon}$. The pattern can be readily inferred for arbitrary $t$, from which we obtain the $(2t+1)\times(2t+1)$ transfer matrix
\begin{equation}
    S=\begin{pmatrix}
        S_1 & S_2 \\
        S_2^\mathrm{T} & S_3
    \end{pmatrix},
\end{equation}
where $S_1$, $S_2$, and $S_3$ are $(t+1)\times(t+1)$, $(t+1)\times t$, and $t\times t$ matrices, respectively. Their matrix elements are given by $[S_1]_{pp'}=\delta_{pp'}+\rho^{|p-p'|-1}(1-\delta_{pp'})$, $[S_2]_{pp'}=\delta_{pp'}+\rho^{|p-p'|-1}(1-\delta_{pp'}+\Theta(p'-p)(\rho-1))$, and $[S_3]_{pp'}=\delta_{pp'}+\rho^{|p-p'|}(1-\delta_{pp'})$, where indices $p,p'$ run from $0$ to $t$ and $\Theta(p)$ is the step function with $\Theta(0)=0$. For instance, the transfer matrix for $t=3$ reads
\begin{equation}
    S=\begin{pmatrix}
        1 & 1& \rho & \rho^2 &1& \rho & \rho^2 \\
        1 & 1& 1& \rho & 1 & 1 & \rho \\
        \rho & 1 & 1 & 1 & \rho & 1 & 1\\
        \rho^2 & \rho & 1 & 1 & \rho^2 & \rho & 1 \\
         1 & 1& \rho & \rho^2 &1& \rho & \rho^2 \\
          \rho & 1& 1& \rho & \rho & 1 & \rho \\
          \rho^2 & \rho & 1 & 1 & \rho^2 & \rho & 1 
    \end{pmatrix}.
\end{equation}
The transfer matrix as defined above does not account for the different signs of pairings (non-Gaussian pairings can make negative contributions), so we will also introduce an on-site $(2t+1)\times(2t+1)$ matrix $qE$ that encodes this information
\begin{equation}
    E=\begin{pmatrix}
        I_{t+1} & 0 \\
        0 & -I_{t}
    \end{pmatrix}.
\end{equation}

\begin{figure}[h!]
\subfloat[\label{fig:diagram_0}]{\includegraphics[width=3cm]{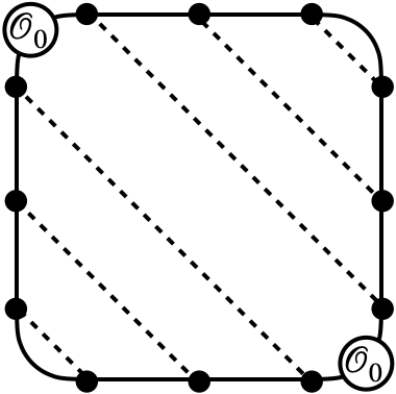} 
}
\hfill
\subfloat[\label{fig:diagram_x}]{\includegraphics[width=3cm]{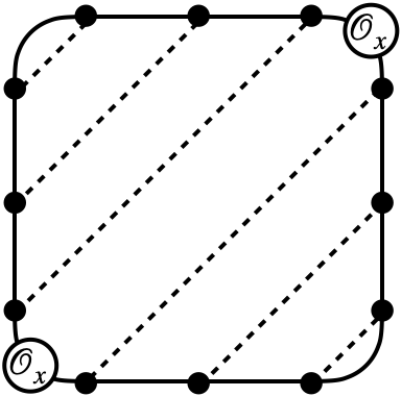} 
}
\caption{The only leading pairing on (a): site $0$ and (b): site $x$ for $t=3$.}
\label{fig:t1diag}
\end{figure}

Finally, we turn to Haar averages over unitary operators on sites $0$ and $x$. In this case the trace structures, shown in Fig.~\ref{fig:diag1} and \ref{fig:diag3}, respectively, admit only one leading pairing because $\mathcal{O}_0$ and $\mathcal{O}_x=0$ are traceless. More precisely, on site $0$ only the diagonal pairing $a(\tau)=d(\tau),\,b(\tau)=c(\tau)$ contributes, whereas on site $x$ the anti-diagonal pairing $a(\tau)=b(\tau),\,c(\tau)=d(\tau)$ contributes (see Fig.~\ref{fig:diagram_0} and Fig.~\ref{fig:diagram_x}, respectively); each of these pairings carries a weight $q$. Due to the different index contractions at sites $0$ and $x$ it will be convenient to introduce new on-site matrices for the transfer matrix calculation. These are $E_0=q|0\rangle\langle 0|$ and $E_t=q|t\rangle\langle t|$, respectively.
\subsection{Computation of the OTOC at large $q$}
Now we are ready to compute the OTOC at large $q$. Note first that a factor $q^{-L}$ in the definition of the OTOC cancels the $L$ factors of $q$ corresponding to on-site pairings. This leads us to the expressions
\begin{equation}
    \overline{\mathcal{C}(x,t)}=\begin{dcases*}
        1-\hat{S}^{L-x}_{t0}\hat{S}^x_{0t} & $x>0$\\
        1-\hat{S}^{L-|x|}_{0t}\hat{S}^{|x|}_{t0} & $x<0$
    \end{dcases*},
\end{equation}
where $\hat{S}=ES$, $\hat{S}_{pp'}=\langle p|\hat S |p'\rangle$, $E|0\rangle=|0\rangle$, and $E|t\rangle=|t\rangle$.

The transfer matrix $\hat{S}$ has a number of interesting properties. First, it is readily seen that the matrix always has an eigenvalue 1, as well as $2t$ degenerate eigenvalues equal to $0$. Since there are only two linearly independent eigenvectors, $v_1=|t\rangle+|2t\rangle$ and $v_2=|0\rangle+|t+1\rangle$, associated with the eigenvalue 0 (i.e. the geometric multiplicity of the eigenvalue $0$ is 2), the matrix $\hat{S}$ is not diagonalisable. However, it becomes diagonalisable if we take high enough powers:
\begin{equation}\label{power}
    \hat{S}^{k+t}=\hat{S}^t=P,
\end{equation}
where $k\geq0$ and the matrix elements of $P$ are given by $P_{pp'}=1-2\Theta(p'-t)$. Eq.~\eqref{power} implies that when the power $p$ reaches $t$, $S^p$ becomes a rank-one projection matrix $P$, which is diagonalisable. This is in accordance with what we expect: up to $t=|x|$ the operator front of $\mathcal{O}(0,t)$ will not have reached the site $x$, hence, for large enough $L$, the OTOC is zero. Note also that, in the thermodynamic limit, $\lim_{L\to\infty}\hat{S}^{L-|x|}_{t0}=P_{t0}=1$. The question now is how to access the matrix elements of $\hat{S}^\ell$ when $\ell<t$. 

While obtaining matrix elements is in general a formidable task, some patterns in their structure are apparent already for small $p$ and $t$. This motivates an ansatz $\hat{S}^\ell_{pt}=F(\ell,t-p)$, 
where 
\begin{equation}\label{eq:F_ansatz1}
    F(\ell,k)=1-(1-\rho)^\ell\sum_{i=0}^{k-\ell-1}\begin{pmatrix}
        \ell+i-1\\
        i
    \end{pmatrix}\rho^i,
\end{equation}
or equivalently
\begin{equation}\label{eq:F_ansatz2}
   F(\ell,k)= 1-\sum_{i=0}^{k-\ell-1}\begin{pmatrix}
        k-1\\
        i
    \end{pmatrix}\rho^i(1-\rho)^{k-i-1},
\end{equation}
which we now show is correct by induction. First consider $p=0$: using the above matrix elements of $\hat{S}$, the matrix element $\hat{S}_{0t}^{\ell+1}$ is 
\begin{equation}
    \hat{S}_{0t}^{\ell+1}=F(\ell,t)+\sum_{i=0}^{t-1}\left[F(\ell,t-i-1)-F(\ell,t-i)\right]\rho^i.
\end{equation}
Plugging Eq.~\eqref{eq:F_ansatz1} into the square bracket, it is readily seen that
\begin{align}
     &F(\ell,t)+\sum_{i=0}^{t-1}\left[F(\ell,t-i-1)-F(\ell,t-i)\right]\rho^i\n
&=F(\ell,t)+\begin{pmatrix}
        t-1\\
        t-\ell-1
    \end{pmatrix}\rho^{t-\ell-1}(1-\rho)^\ell.
\end{align}
Using Eq.~\eqref{eq:F_ansatz2}, the right hand side precisely coincides with $F(\ell+1,t)$, establishing that $\hat{S}_{0t}^{\ell+1}=F(\ell+1,t)$. The induction for $p>0$ can be carried out in the same way, and we can similarly show that $\hat{S}^\ell_{t0}=\hat{S}^\ell_{t0}$. From these results we
obtain the exact OTOC at $q\to\infty$. Outside of the light cone (i.e. for $|x|>t$) we have $\overline{\mathcal{C}(x,t)}=0$, whereas for $|x|\leq t$ we have
\begin{equation}\label{eq:otoc_nospin}
    \overline{\mathcal{C}(x,t)}=
        \sum_{i=0}^{t-|x|-1}
        \begin{pmatrix}
        t-1\\
        i
        \end{pmatrix}
        \rho^i(1-\rho)^{t-i-1}.
\end{equation}
Note that the same function appeared in the computation of the OTOC in RUCs \cite{Nahum_Operator_2018}. For large $x$ and $t$ with $|x|<t$ we have
\begin{align}
    \overline{\mathcal{C}(x,t)} &\simeq \frac{1}{\sqrt{4\pi \mathcal{D}t}}\int_0^{t-|x|}\dd y\,\exp[-\frac{(y-t\rho)^2}{4\mathcal{D}t}]\n
        &\simeq \Phi\left(\frac{v_Bt-|x|}{\sqrt{2\mathcal{D}t}}\right)
\end{align}
where $\Phi(x):=\int_{-\infty}^x\dd y\,e^{-y^2/2}/(\sqrt{2\pi})$ with the butterfly velocity $v_B=1-\rho$ and the diffusion constant $\mathcal{D}=\frac{1}{2}\rho(1-\rho)$ as highlighted above in Eq.~\eqref{eq:rpm_transport}. As mentioned above, even though the form of the OTOC is quite similar to that in finite-$q$ RUCs with Haar-random two-site gates, sending $q$ to infinity in that setting causes $\mathcal{D} \to 0$ and the butterfly velocity $v_B \to 1$, the geometrical lightcone velocity. This is markedly different from the case in the RPM where, even in this limit, the OTOC retains the characteristics expected for chaotic quantum many-body systems in one dimension.

\section{Operator paths}\label{sec:paths}

A useful perspective on our results in the previous sections comes from expressing quantities in terms of sums over paths in the space of operators. In RUCs with Haar-random gates and without time-translation symmetry, the structure of the paths contributing to correlation functions was studied in Ref.~\cite{Nahum_Real_2022}. In that setting ensemble-averaged correlation functions vanish, but rich structures arise in their magnitudes; for squared autocorrelation functions it was shown that $\overline{C^2_{\alpha \alpha}(t)}$ is dominated by operator paths whose widths grow as $\sim t^{1/2}$. These paths should be contrasted with those contributing to the OTOC, whose widths grow ballistically \cite{Nahum_Operator_2018,Keyserlingk_Operator_2018}. 
Separately, we note that Refs.~\cite{Rakovszky_Dissipation_2022,Keyserlingk_Operator_2022} have studied the operator paths relevant for the description of hydrodynamics in RUCs with conserved quantities; see also Refs.~\cite{Rakovszky_Diffusive_2018,Khemani_Operator_2018} for a discussion of their OTOCs. A related idea, which involves analysing operator paths in a Krylov space rather than in real space, has been introduced to characterise Hamiltonian systems \cite{Parker_Universal_2019}.

In Sec.~\ref{sec:op_path} we study the operator paths contributing to autocorrelation functions in random Floquet circuits. Our main result is to show that time-translation symmetry in the dynamics causes a specific subset of these paths to have real positive amplitudes, and therefore to survive an ensemble average. The paths we identify have a spatially localised support that does not change with time. Strikingly, contributions from these localised paths alone generate the full ensemble-averaged autocorrelation function $\overline{C_{\alpha \alpha}(t)}$ of the RPM at large $q$. Following this, in Sec.~\ref{sec:modsquareamplitudes} we show that, in the RPM at large-$q$, the operator paths contributing to the ensemble-averaged OTOC are independent of time-translation symmetry, in the sense that they are the same as in RUCs. In Sec.~\ref{sec:numerics_path} we show numerically that these pictures of the paths contributing to the autocorrelation function and to the OTOC hold also for qubit chains.

To streamline our discussion we will focus on systems of qubits, although the following discussion can be generalised. For $q=2$ we can use a tensor product basis of Pauli strings $P_{\alpha}=P_{\alpha_1}\cdots P_{\alpha_L}$, where each of $P_{\alpha_x}$ is a Pauli operator or the identity. The operators $P_{\alpha}$ are unitary, Hermitian, and have simple commutation and anticommutation relations $P_{\alpha}P_{\beta}=\pm P_{\beta}P_{\alpha}$. Where we discuss the limit of large $q$, we can for simplicity restrict ourselves to the case where $q$ is a power of $2$, since then each site $x=1,\ldots,L$ can viewed as a set of $\log_2 q$ qubits, and we can continue to use an operator basis of Pauli strings. 

\subsection{Amplitudes}\label{sec:op_path}

Here we ask which kinds of operator paths give rise to structure in ensemble-averaged autocorrelation functions. First we discuss the case where the Floquet operator is a single $D \times D$ Haar-random unitary matrix, and hence the only nontrivial feature is the ramp. Since the SFF and PSFF are sums over autocorrelation functions, this discussion will also provide an explanation for the ramps appearing in those quantities. Following this we study the role of locality via the RPM at large-$q$. In this setting the operator paths contributing to autocorrelation functions of local operators will turn out to be themselves localised. 

Our discussion is based on Eq.~\eqref{eq:paths}.
It will be useful to expand the correlators $C_{\alpha_r\alpha_{r+1}}$ for individual time steps as
\begin{align}
C_{\alpha(r)\alpha(r+1)}&=D^{-1} \sum_{\substack{a(r),a'(r)\\b(r),b'(r)}}W^*_{b(r) b'(r) }W_{a(r) a'(r)}\notag\\ &\times[P_{\alpha(r)}]_{b(r) a(r)}[P_{\alpha(r+1)}]_{a'(r) b'(r)},
\label{eq:C1decomposition}
\end{align}
and we now turn to the Haar random problem.

\subsubsection{Paths in RMF systems}
First consider the amplitude $C_{\alpha(0)\alpha(1)}\ldots C_{\alpha(t-1)\alpha(0)}$ for a single operator contributing to the autocorrelation function. Inserting Eq.~\eqref{eq:C1decomposition} into this expression, we see that some terms in the sum have non-negative contributions for all Floquet operators $W$. These are the ones where $a(\sigma(r))=b(r)$ and $a'(\sigma(r))=b'(r)$ for $r=0,\ldots,(t-1)$, with $\sigma$ a permutation of $t$ elements. There are $t!$ such permutations, and for each of them we find that the corresponding contribution to the amplitude involves a factor $\prod_{r=0}^{t-1} |W_{a(r) a'(r)}|^2$. 

\begin{figure}[h!]
\subfloat[\label{fig:path_cyclic1}]{\includegraphics[width=1cm]{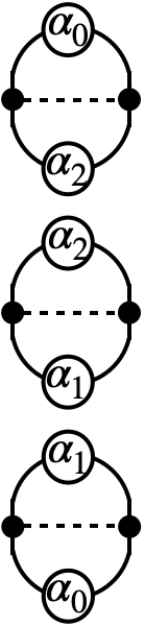}
}
\hfill
\subfloat[\label{fig:path_cyclic2}]{\includegraphics[width=1cm]{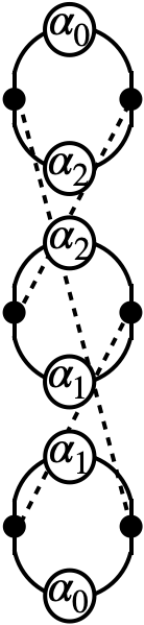}
}
\hfill
\subfloat[\label{fig:path_cyclic3}]{\includegraphics[width=1cm]{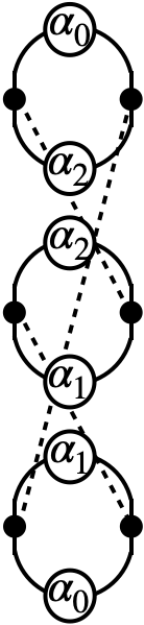}
}
\caption{Cyclic pairings in $C_{\alpha(0)\alpha(1)}\cdots C_{\alpha(t-1)\alpha(0)} $ for $t=3$.}
\end{figure}

In an unstructured RMF system we will now see that $t-1$ of the $t$ cyclic permutations dominate for $1 < t \ll D$. It is helpful to isolate the cyclic permutations, $\sigma(r)=r+s$, with $s=1 \ldots (t-1)$ and addition of indices defined modulo $t$. The cyclic permutation with $s=0$ can be verified to give a vanishing contribution as a consequence of the fact that all operators in the path are traceless (see Fig.~\ref{fig:path_cyclic1}-\ref{fig:path_cyclic3} for the diagrammatic representation of the cyclic pairings for $t=3$). We will additionally write
\begin{align}
    |W_{a(r) a'(r)}|^2 = D^{-1} + \ldots,
\label{eq:modsquareW}
\end{align}
where the ellipsis denotes a fluctuation with average zero that is of the same order as $D^{-1}$. Making the restriction that $t$ is prime for simplicity (we will relax this below), and summing over states $a(r), a'(r)$, then gives
\begin{align}
   & C_{\alpha(0)\alpha(1)}\ldots C_{\alpha(t-1)\alpha(0)} \label{eq:autocorrpaths2} \\&= D^{-2t}\sum_{s=1}^{t-1}|\mathrm{Tr}(P_{\alpha(0)}P_{\alpha(s)}\cdots P_{\alpha([t-1]s)})|^2 +...\notag
\end{align}
In this expression the ellipsis denotes (i) contributions from non-cyclic pairings of indices and (ii) fluctuations inherited from Eq.~\eqref{eq:modsquareW}. The appearance of a single squared trace is a consequence of the fact that $s$ and $t$ have no common factors. It is clear that the contribution to the operator path amplitude displayed on the right-hand side of Eq.~\eqref{eq:autocorrpaths2} is non-negative for all unitary evolution operators $W$. If we average $W$ over the Haar distribution of $D \times D$ unitary matrices, we find that only the displayed term survives at large $D$. For simplicity, we will restrict to such an average in the remainder of this subsection.

Observe now that $|\mathrm{Tr}(P_{\alpha(0)}P_{\alpha(s)}\cdots P_{\alpha([t-1]s)})|^2=D^2$ only if \begin{align}
P_{\alpha(0)}P_{\alpha(s)}\cdots P_{\alpha([t-1]s)}=\pm I \label{eq:selectionrule}
\end{align}
and is otherwise zero. We will refer to as a selection rule on operator paths. Note that, if a given operator path satisfies this condition for one value of $s$, it also satisfies it for all other values of $s$, simply because $P_{\alpha}P_{\beta}=\pm P_{\beta}P_{\alpha}$ for our operator basis of Pauli strings. Additionally, if a given operator path has nonzero amplitude, within this model of Haar-random matrices any reordering of that path also has nonzero amplitude.

The implication is that operator paths satisfying Eq.~\eqref{eq:selectionrule}, which make up a fraction of just $D^{-2}$ of all possible operator paths, have averaged amplitude
\begin{align}
    \overline{C_{\alpha(0)\alpha(1)}\ldots C_{\alpha(t-1)\alpha(0)}} = D^{-2(t-1)}(t-1),
\label{eq:autocorrpaths3}
\end{align}
while all others have zero averaged amplitude. By these arguments we have shown that there is a ramp at the level of individual operator path amplitudes, and that this ramp can be understood as arising from a restricted set of paths.

The analysis is similar for non-prime $t$. For cyclic pairings of state indices $b(r)=a(r+s)$ and $b'(r)=a'(r+s)$ we found a contribution proportional to $|\mathrm{Tr}(P_{\alpha(0)}P_{\alpha(s)}\cdots P_{\alpha([t-1]s)})|^2$ whenever $s$ and $t$ share no common factors [Eq.~\eqref{eq:autocorrpaths2}]. If $s$ and $t$ share a common factor, then instead of a single squared trace we find a product of them. For example, summing over $a(r)$ and $b(r)$ for $t=6$ and $s=2$ we find the product of $|\text{Tr}[P_{\alpha(0)}P_{\alpha(2)}P_{\alpha(4)}]|^2$ and $|\text{Tr}[P_{\alpha(1)}P_{\alpha(3)}P_{\alpha(5)}]|^2$. This object is zero for all but a fraction $D^{-4}$ of paths, but when it is nonzero we find a contribution to the amplitude $D^{-2(t-2)}$, c.f. Eq.~\eqref{eq:autocorrpaths3}. More generally, if we denote by $m$ and $n$ the smallest integers satisfying $ms=nt$, we find a contribution
\begin{align}
    \prod_{r=0}^{t/m-1}|\text{Tr}[P_{\alpha(r)}P_{\alpha(r+s)}\ldots P_{\alpha(r+(m-1)s)}]|^2, \label{eq:selectionrule2}
\end{align}
and summing over $s$ we find symmetry factors analogous to the factor $t-1$ in Eq.~\eqref{eq:autocorrpaths3}. For example if for $t=6$ the $s=2$ contribution to the amplitude is nonzero, the property $P_{\alpha}P_{\beta}=\pm P_{\beta}P_{\alpha}$ implies that the $s=4$ contribution is also nonzero. 

By summing over non-negative contributions to operator amplitudes, such as on the right-hand side of Eq.~\eqref{eq:autocorrpaths3}, we will now recover 
\begin{align}
    \overline{C_{\alpha(0) \alpha(0)}(t)} =D^{-2} (t-1) \label{eq:auto_rmt2}
\end{align}
which is the leading contribution to the autocorrelation function for $t \ll D$. The implication is that the late-time ramp in autocorrelation functions comes from operator paths satisfying the above selection rules. 

To see how Eq.~\eqref{eq:auto_rmt2} arises, let us sum the right-hand side of Eq.~\eqref{eq:autocorrpaths2} over the $D^{2(t-1)}$ paths which begin and end at $P_{\alpha(0)}$. Out of these, $D^{2(t-2)}$ satisfy Eq.~\eqref{eq:selectionrule}, and so for prime $t$ we find an overall contribution $D^{-2}(t-1)$ from Eq.~\eqref{eq:autocorrpaths3}. For $t$ that is not prime the result is the same at large $D$, but it is simpler to consider each cyclic permutation $s=1 \ldots (t-1)$ independently. For $s$ and $t$ sharing no common factors the $D^{2(t-2)}$ paths satisfying Eq.~\eqref{eq:selectionrule} make an overall contribution $D^{-2}$ as above, whereas if there exist integers $n < s$ and $m < t$ for which $ms=nt$ we find contributions of the form Eq.~\eqref{eq:selectionrule2}. The number of operator paths for which this expression is nonzero is $D^{2t(1-1/m)-2}$, but for each path we find a contribution $D^{2t(1/m-1)}$, and therefore each of the $t-1$ allowed values of $s$ gives an overall contribution $D^{-2}$. Their sum is the term displayed on the right-hand side of Eq.~\eqref{eq:auto_rmt2}.

To recover the ramps in the PSFF and SFF we need only sum over autocorrelation functions. Writing $K_A(t)=1+\sum_{P_\alpha \in \mathcal{P}_A}C_{\alpha \alpha}(t)$, we immediately find $\overline{K_A(t)}=1 + (D_A/D)^2(t-1)$ and $\overline{K(t)}=t$ for $1 \leq t \ll D$. All of these ramps therefore have their origin in operator paths satisfying the selection rule in Eq.~\eqref{eq:selectionrule}, and its generalization which is implicit in Eq.~\eqref{eq:selectionrule2}.

\subsubsection{Paths in the RPM}

We can adapt aspects of this analysis to the RPM in the $q \to \infty$ limit, although first some general comments are in order. Most importantly, the operator paths that can have nonzero amplitudes are severely restricted by locality. For example, in a system with local interactions, and over a single time step, an operator cannot change the distance between its left- and right-hand end-points by more than a number of sites of order unity. Subject to this constraint, we must ask which kinds of paths necessarily make real, positive contributions to autcorrelations functions. 
%(for all but a measure zero set of disorder realisations). 

Remarkably, in the RPM at $q \to \infty$, the paths which dominate autocorrelation functions are `localised'. By this we mean that the operators $P_{\alpha(r)}$ in the path all have the same support. In fact, this can be seen already from our calculations of averaged autocorrelation functions: $\overline{C_{\alpha \alpha}(t)}$ is independent of the system size at large $q$, and depends only on the support of the operator $P_{\alpha}$. The mechanism by which this arises is quite general, and later we will show that a related feature is present in the dynamics of qubit chains.

It is simplest to consider single-site operators, and so here we will restrict ourselves to this example. For a closed operator path $P_{\alpha(0)},\ldots,P_{\alpha(t-1)}$ in which all $P_{\alpha(r)}$ are nontrivial on just a single site, we can write $C_{\alpha(r+1)\alpha(r)}$ in Eq.~\eqref{eq:C1decomposition} as
\begin{align}
   &C_{\alpha(r+1)\alpha(r)} =  q^{-3} [P_{\alpha(r+1)}]_{b(r) a(r)} [P_{\alpha(r)}]_{a'(r) b'(r)}  \label{eq:C1decompositionRPM}\\ &\times V_{a(r) a'(r)} V^*_{b(r) b'(r)}  \sum_{c,c'=0}^{q-1} e^{i[\varphi_{c a(r)}-\varphi_{c b(r)}+\varphi'_{a(r) c'} - \varphi_{b(r) c'}]}\notag
\end{align}
where $V$ is a Haar-random unitary acting at the same site as the single-site operators $P_{\alpha(r)}$, and the diagonal matrices $\varphi$ and $\varphi$, whose diagonal entries we denote $\varphi_{c a(r)}$ and $\varphi'_{a(r) c'}$, describe interactions between this site and its left- and right-hand neighbours, respectively. Here $a(r),a'(r),b(r),b'(r)=0\ldots (q-1)$ are state indices for the single site at which the operators $P_{\alpha(r)}$ act, while $c$ and $c'$ are state indices for this site's neighbours. The overall prefactor $q^{-3}$ has arisen from contracting the circuit away from the three-site region centered on our operator path.

At large $q$, observe that we have $a(r) \neq b(r)$ and $b_r \neq b_r^*$ for the vast majority of terms. Summing over $c$ and $c'$, Eq.~\eqref{eq:C1decompositionRPM} then reduces to
\begin{align}
     C_{\alpha_{r+1}\alpha_r} =  q^{-1}e^{-2\varepsilon} &[P_{\alpha(r+1)}]_{b(r)a(r)} [P_{\alpha(r)}]_{a'(r) b'(r)}\notag\\& \times V_{a(r) a'(r)} V^*_{b(r) b'(r)} + \ldots \label{eq:C1decompositionRPM2}
\end{align}
Restricting to localised operator paths has at this stage left us with a single-site problem. The factors $e^{-2\epsilon}$ at each time step describe the dephasing induced by interactions with neighbouring sites. We can now follow essentially the same line of reasoning as in the previous subsection. 

That is, we first insert Eq.~\eqref{eq:C1decompositionRPM2} into the expression Eq.~\eqref{eq:paths} and ask about contributions from terms where the single-site indices are paired cyclically as $b(r)=a(r+s)$ and $b'(r)=a'(r+s)$. For these terms we find contributions $\prod_{r=0}^{t-1}|V_{a(r)a'(r)}|^2$ and, mirroring Eq.~\eqref{eq:modsquareW}, we will separate out non-negative parts
\begin{align}
    |V_{a(r) a'(r)}|^2 = q^{-1} + \ldots. \label{eq:modsquareV}
\end{align}
Summing over state indices then generates a non-negative contribution to the operator path. For prime $t$ this is
\begin{align}
    &C_{\alpha(0)\alpha_1}\ldots C_{\alpha(t-1)\alpha(0)} \label{eq:autocorrpaths2RPM} \\&= q^{-2t} e^{-2\varepsilon t}\sum_{s=1}^{t-1}|\mathrm{Tr}(P_{\alpha(0)}P_{\alpha_s}\cdots P_{\alpha([t-1]s)})|^2 +...\notag
\end{align}
We therefore have a single-site analogue of the selection rule in Eq.~\eqref{eq:selectionrule}, and the contributions displayed in Eq.~\eqref{eq:autocorrpaths2} will turn out to dominate the averaged autocorrelation functions of local operators at large $q$.

If $P_{\alpha(0)}\ldots P_{\alpha(t-1)} = \pm I$, we find that the expression displayed above is the dominant contribution to the averaged amplitude
\begin{align}
    \overline{C_{\alpha(0) \alpha(1)}\ldots C_{\alpha(t-1)\alpha(0)}} = q^{-2(t-1)} e^{-2\varepsilon t} (t-1)
\end{align}
at large $q$. This can be generalized to non-prime $t$ along identical lines to the previous section. Again, we find a ramp at the level of an individual path's amplitude, although in contrast to the previous subsection we have here restricted to an extremely small subset of the possible operator paths. The locality of the interactions compensates for this, and causes these paths to have relatively large amplitudes.  
Summing over paths (for prime or non-prime $t$) we pick up an overall factor $q^{2(t-2)}$ which leads us to
\begin{align}
    \overline{C_{\alpha(0) \alpha(0)}(t)} = q^{-2}e^{-2\varepsilon t}(t-1). \label{eq:auto_rpm2}
\end{align}
Remarkably, considering only the localised paths has given us the dominant contribution to the averaged autocorrelation function of the RPM. 

The above approach can be generalized to the autocorrelation functions of more complicated operators, and one then finds that all operators $P_{\alpha(r)}$ in the contributing paths $P_{\alpha(0)},\ldots,P_{\alpha(t-1)}$ have the same support. It is important to note, however, that our large-$q$ analysis does not expose the late-time ramp in $C_{\alpha(0) \alpha(0)}(t)$, although an approximate description of this feature can be formulated along similar lines to Sec.~\ref{sec:finiteq}.

Equation~\eqref{eq:auto_rpm2} is the central result of this section: the operator paths which contribute to ensemble-averaged autocorrelation functions of local operators in the RPM are themselves localised. This feature arises from the combination of (i) local interactions and (ii) the correlations between amplitudes of different steps of the operator path which are imposed by time-translation symmetry. 
%In the next section we will highlight the difference between this behaviour and that observed via the OTOC.

\subsection{Squared amplitudes}\label{sec:modsquareamplitudes}
Here we discuss the operator paths contributing to squared correlation functions, and to the OTOC. First note that in this limit the OTOC is a weighted sum over squared correlation functions; starting from Eq.~\eqref{eq:otoc} it can be verified that in the RPM at large $q$,
\begin{align}
    \overline{\mathcal{C}(x,t)} = 1 - \sum_{P_\alpha \in \mathcal{P}} \bra P_{\alpha} \mathcal{O}(x,0) P_{\alpha} \mathcal{O}(x,0)\ket \overline{C^2_{\mathcal{O}\alpha}(t)}.
\end{align}
We can therefore restrict our attention to averages of squared correlation functions. 

The key observation is that, even with time-translation symmetry, in the large-$q$ limit this does not play a role in the average. To see this let us first write
\begin{align}
    C_{\alpha \beta}(t) = \sum_{P_\gamma \in \mathcal{P}} C_{\alpha \gamma}(t') C_{\gamma \beta}(t-t'),
\end{align}
for $t' \neq 0,t$, and then square this expression. The square involves a sum over two sets of operators $P_\gamma,P_{\gamma'} \in \mathcal{P}$, and so we have both diagonal $\gamma = \gamma'$ and off-diagonal $\gamma \neq \gamma'$ contributions. The sum over diagonal contributions factorises at large $q$,
\begin{align}
    &\sum_{P_\gamma \in \mathcal{P}} \overline{[C_{\alpha \gamma}(t')]^2 [C_{\gamma \beta}(t-t')]^2} \\ &= \sum_{P_\gamma \in \mathcal{P}} \overline{[C_{\alpha \gamma}(t')]^2}\,\,\overline{[C_{\gamma \beta}(t-t')]^2}. \notag
\end{align}
It can then be shown that the off-diagonal contribution $\gamma \neq \gamma'$ is subleading relative to the above diagonal contribution above. Therefore,
\begin{align}
    \overline{[C_{\alpha \beta}(t)]^2} = \sum_{P_\gamma \in \mathcal{P}} \overline{[C_{\alpha \gamma}(t')]^2}\,\,\overline{[ C_{\gamma \beta}(t-t')]^2}.
\end{align}
This decomposition can be repeated until we are left with a product of average squared correlation functions for individual time steps,
\begin{equation}
  \overline{C^2_{\alpha\beta}(t)}=\sum_{P_{\alpha(1)},\cdots,P_{\alpha(t-1)}\in\mathcal{P}}\overline{C^2_{\alpha\alpha(1)}}\, \overline{C^2_{\alpha(1)\alpha(2)}}\cdots \overline{C^2_{\alpha(t-1)\beta}}.\label{eq:decom_weight}
\end{equation}
The fact that the averages on the right-hand side of Eq.~\eqref{eq:decom_weight} can be performed independently for each time step shows that averages of squared autocorrelation functions, as well as of the OTOC, are insensitive to time-translation symmetry at large $q$. We can therefore understand their behaviour from Ref.~\cite{Nahum_Real_2022} and Refs.~\cite{Nahum_Operator_2018,Keyserlingk_Operator_2018}, respectively. For example, the average squared autocorrelation function $\overline{C^2_{\alpha \alpha}(t)}$ is dominated by operator paths whose extent grows as $t^{1/2}$ \cite{Nahum_Real_2022}, to be contrasted with the localised operator paths contributing to $\overline{C_{\alpha \alpha}(t)}$ which we have discussed above. 

\subsection{Numerical results}\label{sec:numerics_path}
In this section we have so far shown that, at large $q$ in the RPM, the kinds of operator paths contributing to $\overline{C_{\alpha \alpha}(t)}$ and $\overline{C^2_{\alpha \alpha}(t)}$ are very different. We now demonstrate numerically that this difference carries over to the behavior of a Floquet system with $q=2$.

The first object we consider is a deconstructed sum over autocorrelation functions, and the second is a sum over squared correlation functions. To motivate the first object, let us write an autocorrelation function as
\begin{align}\label{eq:split}
    C_{\alpha \alpha}(t) = \sum_{P_\beta\in\mathcal{P}} C_{\alpha \beta}(\lceil t/2 \rceil) C_{\beta \alpha}(\lfloor t/2 \rfloor),
\end{align}
which is a sum over all paths from $P_{\alpha}$ to $P_{\beta}$ over the first half of the interval $t$, and over all paths back from $P_{\beta}$ to $P_{\alpha}$ in the second half. For odd $t$ we split the intervals as $\lceil t/2 \rceil = (t+1)/2$ and $\lfloor t/2 \rfloor = (t-1)/2$. Let us define the contribution to $C_{\alpha \alpha}(t)$ from operator paths that, halfway through the evolution, have support within a region $B$,
\begin{align}
    C^{B}_{\alpha \alpha}(t) = \sum_{P_\beta \in \mathcal{P}_B} C_{\alpha \beta}(\lceil t/2 \rceil) C_{\beta \alpha}(\lfloor t/2 \rfloor).
\end{align}
If we now define a sequence of regions $B_0,B_1,\ldots$, with $B_j$ strictly contained within $B_{j+1}$ we can write the contribution to $C_{\alpha \alpha}(t)$ from operator paths that have support outside of $B_j$ but not outside of $B_{j+1}$ as $C^{B_{j+1}}_{\alpha \alpha}(t)-C^{B_{j}}_{\alpha \alpha}(t)$. To make contact with our discussions of the PSFF, it will be convenient to sum over autocorrelation functions for sets of operators $P_{\alpha}$ that are nontrivial within a region $A$. In analogy with $C^{B}_{\alpha\alpha}(t)$ we define
\begin{align}
    K^B_A(t) = \sum_{P_\alpha \in \mathcal{P}_A} \sum_{P_\beta \in \mathcal{P}_B} C_{\alpha \beta}(\lceil t/2 \rceil) C_{\beta \alpha}(\lfloor t/2 \rfloor), \label{eq:KAB}
\end{align}
which is the sum over all amplitudes of operators paths that start in $A$, have support entirely within $B$ at the halfway point, and then return to $A$. Having defined a sequence of regions $B_j$ as above, with $B_0$ a region of zero sites for which $K_A^{B_0}(t)=0$, and with the largest region corresponding to entire system, we can write the PSFF as
\begin{align}
    K_A(t) = \sum_{j=0} [ K_A^{B_j}(t) - K_A^{B_{j-1}}(t)]. \label{eq:KA_KAB}
\end{align}
At large $q$ in the RPM we have $K_A(t)=K_A^A(t)$, i.e. all operator paths contributing to $K_A(t)$ are confined to the region $A$.

The second object we consider measures operator weight via squared correlation functions. To construct this weight we write
\begin{align}
    P_{\alpha}(\lceil t/2 \rceil) = \sum_{P_\beta\in\mathcal{P}} C_{\alpha \beta}(\lceil t/2 \rceil)P_{\beta},
\end{align}
and observe that $D^{-1}\text{Tr} P^2_{\alpha}(\lceil t/2 \rceil)=1$, which corresponds to the normalization condition on the sum over squared correlation functions $\sum_{P_\beta}C_{\alpha\beta}^2(\lceil t/2 \rceil)=1$. We can then ask how much of this weight comes from operators that have support only within $B$, i.e. $\sum_{P_\beta \in \mathcal{P}_B}C_{\alpha\beta}^2(\lceil t/2 \rceil)$. To compare with $K^B_A(t)$, we will sum this object over all operators within the region $A$,
\begin{align}
   N^B_A(t) = \sum_{P_\alpha \in \mathcal{P}_A} \sum_{P_\beta \in \mathcal{P}_B}C_{\alpha\beta}^2(\lceil t/2 \rceil).
\end{align}
This is the total weight of the operator strings which evolve from $A$ to $B$ over the interval $\lceil t/2 \rceil$, as opposed to the total amplitude of operator strings which follow this route and ultimately return to $A$ at time $t$. In analogy with Eq.~\eqref{eq:KA_KAB} we can write the total weight of operators which start in $A$ as
\begin{align}
    N_A(t) =  \sum_{j=0} [ N_A^{B_j}(t) - N_A^{B_{j-1}}(t)].
\end{align}
Since we have a normalisation condition $N_A(t)=D_A^2-1$, it will be convenient to study the objects $[ N_A^{B_j}(t) - N_A^{B_{j-1}}(t)]/(D_A^2-1)$ below. The sum of these objects over regions $B_j$ is then unity. Our analysis of the RPM at large $q$ (along with earlier studies of the OTOC in RUCs) shows that the value of $B$ for which $N^B_A(t)$ is maximized should grow ballistically with time.

\begin{figure}
\includegraphics[width=0.47\textwidth]{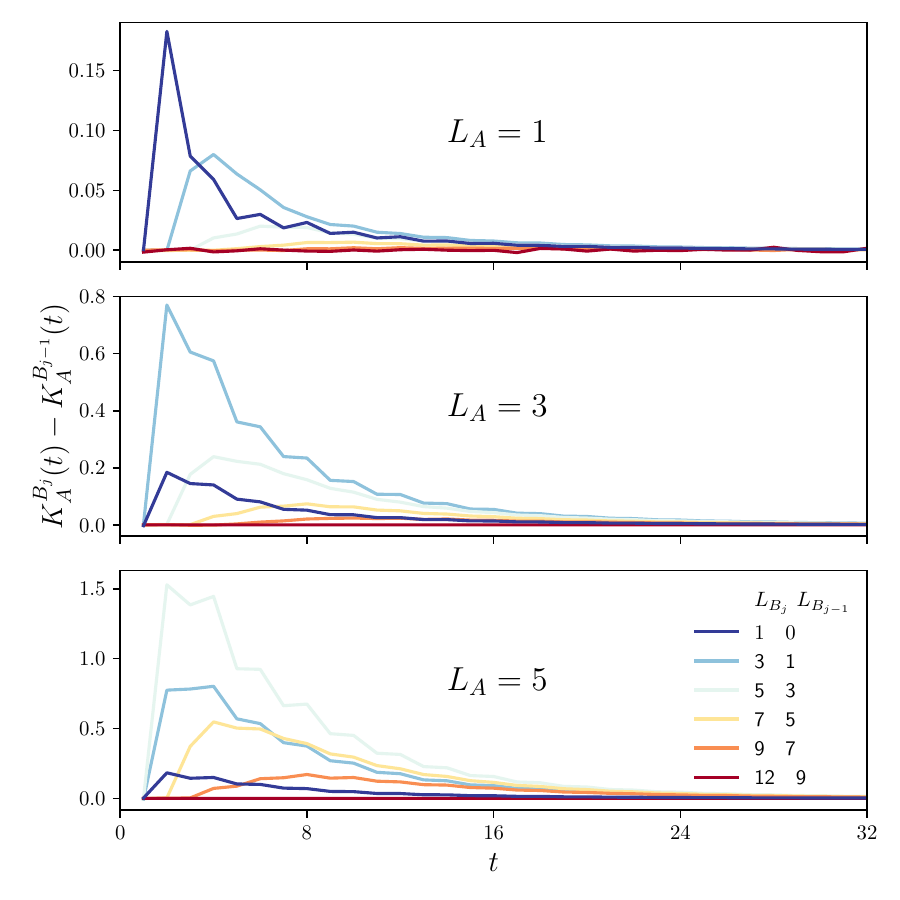}
\caption{Contributions $\overline{K^B_A(t)}$ to the average PSFF from operator paths of varying support. We consider regions $B_j$ of $L_{B_j}=2j-1$ sites for $j=1,2,3,4$ centred on the region $A$ of $L_A$ sites (see text in centres of panels). The data shown is $K^{B_j}_A(t)-K^{B_{j-1}}_A(t)$, corresponding to operator paths which at the halfway point have support outside of $B_{j-1}$ but within $B_j$. Here $L=12$.}
\label{fig:splitPSFF}
\end{figure}

\begin{figure}
\includegraphics[width=0.47\textwidth]{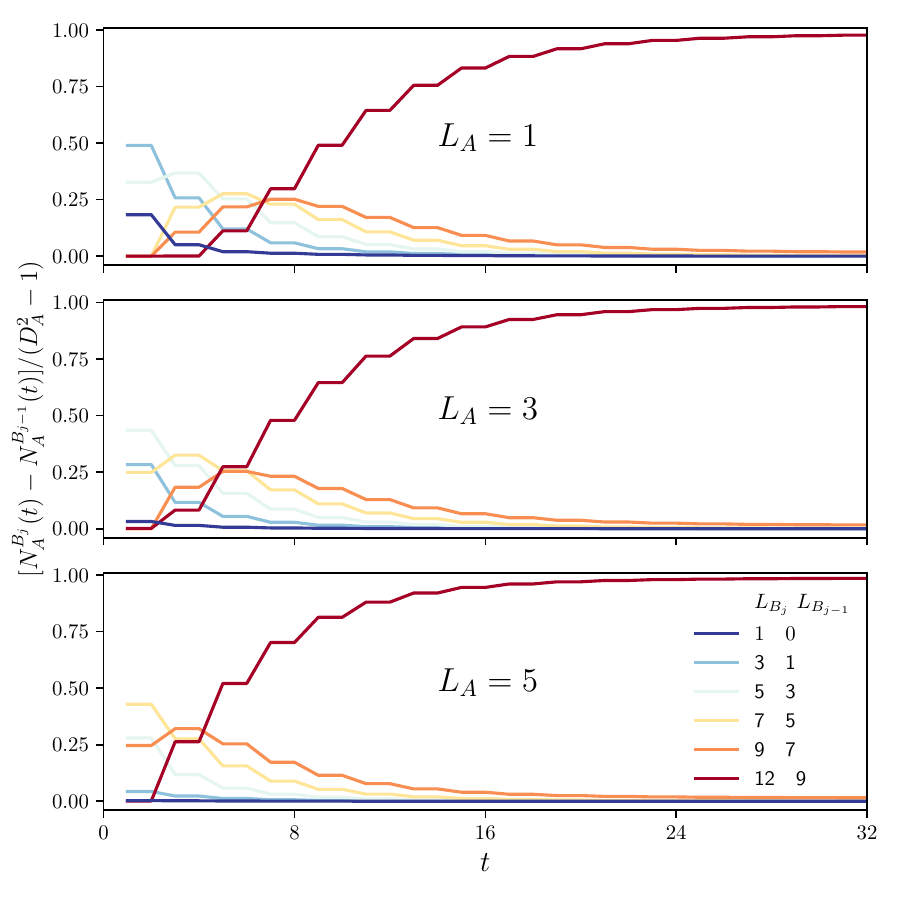}
\caption{Average contributions $(D_A^2-1)^{-1} \overline{N^B_A(t)}$ to the operator weight for operators which are initially within $A$ and at time $\lceil t/2 \rceil$ are in $B$. As in Fig.~\ref{fig:splitPSFF} we calculate differences between the contributions from different regions $B_j$, with $B_j$ strictly within $B_{j+1}$. Here $L=12$.}
\label{fig:sumOPPUR}
\end{figure}

In Fig.~\ref{fig:splitPSFF} we calculate $K^{B_j}_A(t)-K^{B_{j-1}}_A(t)$ numerically. For each of the values of $L_A$ shown it is clear that, at early times, the largest contribution to $K_A(t)$ is from operators which have support $\sim L_A$ at time $\lceil t/2 \rceil$,, which is the behavior anticipated from the RPM at large $q$. This behavior is to be contrasted with $N^{B_j}_A(t)-N^{B_{j-1}}(t)$: in Fig.~\ref{fig:sumOPPUR} we show that, by this measure, the operators $P_{\alpha}$ develop weight on ever longer Pauli strings as time progresses. At late times the difference between the structure of operators contributing to autocorrelation functions, and to squared autocorrelation functions, is particularly striking: only in the latter case is the operator weight is in the longest operator strings (see $L_{B_j}=12$, $L_{B_{j-1}}=9$ data in Fig.~\ref{fig:sumOPPUR}).

\section{Individual systems}\label{sec:fluctuations}
Above we have studied ensemble-averaged dynamics, and so it is natural to ask how much of the above phenomenology carries over to the behaviour of individual Floquet many-body systems. In the following we show first that sample-to-sample fluctuations of autocorrelation functions obscure their second peak. We then show that this structure can be revealed in an individual system by considering averages over sets of autocorrelation functions such as the PSFF in place of average over an ensemble of systems. In particular, we will show that the sample-to-sample fluctuations of $K_A(t)$ are parametrically smaller than its late-time tail for $1 \ll L_A \ll L$.

\begin{figure}
\includegraphics[width=0.47\textwidth]{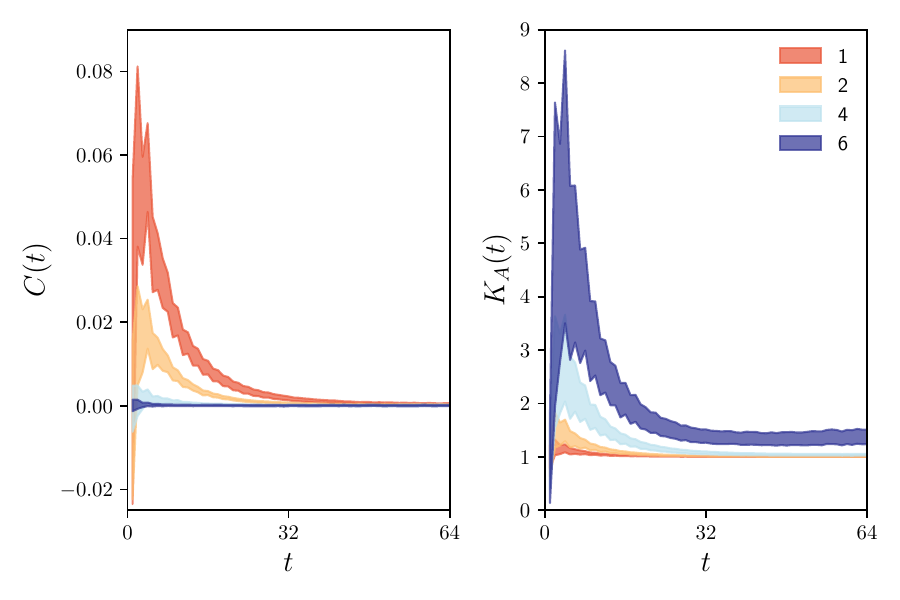}
\put(-233,145){(a)}
\put(-118,145){(b)}
\caption{(a) Autocorrelation functions of operators consisting of a single cluster ($n=1$) for various $a$ (legend); the coloured regions are centred on the average $\overline{C_{\alpha \alpha}(t)}$ and have vertical width equal to $10^{-1/2}$ times the root-mean-square fluctuation of $C_{\alpha \alpha}(t)$; they represent the effect of averaging over just $10$ random Floquet circuits. (b) Suppressing fluctuations through a sum over operators in individual Floquet systems. The coloured regions are centred on $\overline{K_A(t)}$ and have vertical width equal to the root-mean-square fluctuation of $K_A(t)$. The legend shows $L_A$. Here $L=10$.}
\label{fig:fluctuations}
\end{figure}

First consider the ensemble variance $\overline{C^2_{\alpha \alpha}(t)}-\overline{C_{\alpha \alpha}(t)}^2$ of the autocorrelation function 
in the large-$q$ limit of the RPM. For $a$-site single-cluster operators, it can be verified that at late times $\overline{C^2_{\alpha \alpha}(t)} \simeq q^{-2a}(t-1)e^{-4\varepsilon t}$, while $\overline{C_{\alpha \alpha}(t)} \simeq q^{-2a}(t-1)e^{-2\varepsilon t}$. The sample-to-sample fluctuations are therefore of order $q^{-a}$, and so are large compared with the ensemble average at large $q$. One might consider a time average, but since for single-site operators the autocorrelation function has nontrivial structure only over a time interval $1 \leq t \lesssim t_{\rm Th, \alpha}$, for an individual Floquet system there are too few time steps to suppress the fluctuations. Although at finite $q$ the ratio of the variance to the squared mean is of course finite, a mild ensemble average is nevertheless required to observe the second peak in autocorrelation functions; in Fig.~\ref{fig:fluctuations}\color{red}a \color{black} we illustrate the effects of averaging over $10$ random Floquet systems.

As an alternative to this ensemble average, we will sum $C_{\alpha \alpha}(t)$ over operators $P_{\alpha}$ supported within a region $A$, i.e. we consider the PSFF. To be concrete, we will calculate the second moment $\overline{K_A^2(t)}$ in the RPM at large $q$, and as above we use a transfer matrix technique. In calculating on-site Haar averages, denote by $a(r)$ and $b(r)$ indices appearing in the first factor of $K_A(t)$, as in Sec.~\ref{sect:rpm_derivation}, and by $c(r)$ and $d(r)$ indices in the second factor. Note that $a(r)$ and $c(r)$ appear in products of Floquet operators $W^t$, while $b(r)$ and $d(r)$ appear in the conjugates. The index $r=0,\ldots,t$, and for sites in $A$ we have the trace structure $a(0)=a(t)$, whereas for sites in $\bar{A}$ we have $a(0)=b(0)$, $a(t)=b(t)$, $c(0)=d(0)$, and $c(t)=d(t)$. We only need to account for cyclic Gaussian pairings, but there are now two sets of such pairings in region $A$. The first set, which we denote $p=0$, corresponds to $a(r)=b(r+s)$ and $c(r)=d(r+\tilde s)$ for $s,\tilde s=0,\ldots,(t-1)$. The second, which we denote $p=1$, corresponds to $a(r)=d(r+s)$, $b(r)=c(r+\tilde s)$. In region $\bar{A}$, the leading order pairings are simply $a(r)=b(r)$ and $c(r)=d(r)$.

We identify each pairing $(p,s,\tilde s)$ with one element of a set of $2t^2$ orthonormal basis vectors, and define a transfer matrix $T^{(2)}$ with elements
\begin{align}
    T^{(2)}_{ps\tilde s, p's'\tilde s'} &= \delta_{pp'}T_{ss'}T_{\tilde s \tilde s'} +[1-\delta_{pp'}]e^{-2\varepsilon t}.
\end{align}
Here $T$ is the transfer matrix defined in Eq.~\eqref{T}, which generates the average SFF \cite{Chan_Spectral_2018} as well as the PSFF above. We note that $T^{(2)}$ also generates the second moment of the SFF \cite{Garratt_Local_2021,Chan_Spectral_2021}. In terms of $T^{(2)}$ we have the ensemble variance
\begin{align}
    \overline{K_A^2(t)}- \overline{K_A(t)}^2=[(T^{(2)})^{L_A+1}]_{000,000} - [T^{L_A+1}]^2_{00},
\end{align}
where the indices $(p,s,\tilde s)=(0,0,0)$ correspond to the pairing in region $\bar{A}$. At late times the variance goes to zero, and it decays to this value as
\begin{align}
    \overline{K_A^2(t)}- \overline{K_A(t)}^2 = \frac{1}{2} L_A(L_A+1)t^2 e^{-4\varepsilon t}+\ldots,
\end{align}
which corresponds to a domain with $p \neq 0$ and any $s,\tilde s=0,\ldots,(t-1)$. For $L_A \gg 1$, the fluctuations around the average are therefore of order $L_A t e^{-2\varepsilon t}$.

We should compare the fluctuations with the average PSFF itself, and at late times we have shown that $\overline{K_A(t)} \simeq 1+\frac{1}{2}L_A(L_A+1)(t-1)e^{-2\varepsilon t}$. For large $L_A$, the sample-to-sample fluctuations in $K_A(t)$ are therefore parametrically smaller than its late-time tail, with the signal-to-noise ratio proportional to $L_A$ for $L_A \gg 1$. The implication is that structure in the PSFF arising from locality can be observed in individual systems. 

In Fig.~\ref{fig:fluctuations}\color{red}b \color{black} we verify this numerically for the $q=2$ RSM. There we show the root-mean-square fluctuations of $K_A(t)$ compared with $\overline{K_A(t)}$, and observe a clear separation between the different values of $L_A$ displayed. 
%In closing, we note that from its definition the PSFF is non-negative in individual systems, and this implies that the autocorrelation functions (which can take either sign) are anticorrelated. Exploring these correlations in detail is an interesting direction for future work.

\section{Summary and discussion}\label{sec:summary}
In summary, the work that we have presented is focussed on comparing behaviour in Floquet models with that in RUCs, in order to identify consequences of time-translation symmetry. It probes operator dynamics in two ways, through correlation functions and through OTOCs.
%and our focus has been on comparing behaviour in Floquet models with that in RUCs, in order to identify consequences of time-translation symmetry. 

%There are fundamental differences between the dynamics of operators as viewed through correlation functions and OTOCs. This is clear even in RUCs having gates which are Haar-random in both space and time: there the average of the OTOC can be understood through the hydrodynamic picture of operator spreading \cite{Nahum_Operator_2018,Keyserlingk_Operator_2018}, while averages of correlation functions vanish. 
It is already clear from previous results for RUCs with gates that are Haar-random in space and time that correlation functions and OTOCs reveal fundamentally different aspects of operator dynamics: in such models the average of the OTOC can be understood through the hydrodynamic picture of operator spreading \cite{Nahum_Operator_2018,Keyserlingk_Operator_2018}, while averages of correlation functions vanish. 
Complementing those results, we have considered Floquet quantum circuits without conservation laws, with gates that are Haar-random in space but periodic in time. A key outcome has been to show that in these systems time-translation symmetry of the evolution operator leads to observable and 
distinctive features in autocorrelation functions. In contrast, we have demonstrated that the structure of the OTOC is insensitive to time-translation invariance, at least at large $q$ in the RPM.

%In thiswork we have studied the effects of having a fixed time-evolution operator (i.e. the effects of time-translation invariance) on these objects. To simplify the analysis, we focused on Floquet quantum circuits having Haar-random local unitary operations and no conservation laws. A key result has been to show that, in locally-interacting spin chains, time-translation symmetry of the evolution operator leads to observable and 
%nontrivial
%distinctive 
%features in autocorrelation functions. In contrast, we demonstrated that the structure of the OTOC is insensitive to time-translation invariance, at least at large $q$ in the RPM.

This difference can be understood by representing correlation functions and OTOCs in terms of sums over paths in the space of operators. While correlation functions can be expressed as sums over amplitudes of operator paths [see Eq.~\eqref{eq:paths}], OTOCs involve sums over (non-negative) `probabilities' [see Eq.~\eqref{eq:pathnorm}]. In RUCs the amplitudes of operator paths have zero mean, and so averaged correlation functions vanish. Introducing time-translation symmetry generates correlations between the amplitudes of different segments of operator paths, and these give rise to nonzero averages of autocorrelation functions. In the local systems we have considered, this perspective has also revealed a striking difference between the morphologies of operator paths which contribute to ensemble-averaged autocorrelation functions relative to those dominating the OTOC. In particular, we find that the operator paths contributing to autocorrelation functions are much narrower in space; see Figs.~\ref{fig:path_dominant} and \ref{fig:path_subdominant}. We note that the morphologies of operators paths contributing to correlation functions were also recently discussed in Ref.~\cite{Nahum_Real_2022}. 

A further result has been to show how the spectral statistics of many-body Floquet systems can be understood in terms of the dynamics of operators. The SFF can be expressed as a sum over autocorrelation functions of all operators \cite{Gharibyan_Onset_2018}, and a local analogue (the PSFF) has been defined as the restriction of this sum to operators with support only within a subregion \cite{Joshi_Probing_2022}. In brief, at late times autocorrelation functions match their behavior in RMF systems, and their sum reproduces the famous ramp in the SFF. At the earliest times $t < t_\mathrm{d}$ autocorrelation functions relax. Strikingly, in some models this relaxation is followed by an increase for $t > t_\mathrm{d}$, which can be understood as arising from local unitary scrambling in the region of the operator's support (a residue of a `local ramp'). This local unitary scrambling competes with dephasing induced by interactions with neighbouring regions, and the result of this competition is a finite-time peak in the autocorrelation function. The decay following the peak ultimately gives way to late-time ramp and plateau, whose amplitudes are exponentially small in the system size. 

These findings were supported by analytical as well as numerical calculations. Our analytical investigations were centered on the RPM at large $q$ \cite{Chan_Spectral_2018}, as well as a spinful variant of it which we have introduced (see Appendix~\ref{sec:spinful}). In these models we calculated OTOCs, autocorrelation functions, and PSFFs. We also provided a heuristic argument to understand the finite-$q$ physics of PSFFs, although evidence for the generality of our analytic results in describing finite-$q$ systems can already be seen in Ref.~\cite{Garratt_Local_2021}, which studied spectral statistics numerically at $q=2$. Here we have demonstrated that calculations in the large-$q$ RPM also provide an accurate depiction of autocorrelation functions at $q=2$ (focusing on the ergodic phase of the model in Ref.~\cite{Garratt_ManyBody_2021}). 

Some aspects of autocorrelation functions remain to be understood. The local diagonal approximation has provided an approximate description of the emergence of the ramp $(t-1)/(q^{2L}-1)$ expected from RMT. However, our numerical calculations indicate that the gradient of the ramp may depend on the structure of the operator, and this feature is absent in RMT.

It is worth commenting on the difference between our results and those which would be found in ergodic dual unitary circuits (having brickwork geometry). In that setting autocorrelation functions of site-local operators are exactly zero for times $t>0$ in the limit of large $L$ \cite{Bertini_Exact_2019,Claeys_Maximum_2020,Gutkin_Exact_2020}, while in this regime the SFF matches the RMT result \cite{Bertini_Exact_2018,Bertini_Random_2021}. Representing the SFF as a sum over autocorrelation functions, these two properties are consistent with the `perfect dephaser' property of dual unitary circuits \cite{Lerose_Influence_2021}; the effective dephasing time for the autocorrelation function of a local operator is zero (i.e. it is as if the effective domain wall tension $\varepsilon \to \infty$). However, autocorrelation functions of multi-site operators in dual-unitary circuits do not necessarily vanish, and in fact some of them must be non-zero in order to generate the anticipated behaviour of the SFF from their sum. It would be interesting to understand how this arises. 

Moving away from the ergodic phase of Floquet circuits, it is helpful to consider first the extreme limit of decoupled qubits evolving under local Haar-random fields (i.e. $q=2$ and $\varepsilon=0$ in the RPM, or $J=0$ in RSM). There it is straightforward to show that $\overline{C_{\alpha \alpha}(t \geq 2)} = (1/3)^a$, with $a$ is the number of non-identity Pauli operators in the string $P_{\alpha}$, and $\overline{K(t\geq 2)}=2^L$, corresponding to Poisson spectral statistics. Weakly-coupled random spin chains are expected to be many-body localised  \cite{Nandkishore_Many_2015,Abanin_Many_2019}, and in that setting corrections to the above behaviour of autocorrelation functions can be understood in the energy domain through the theory of many-body resonances \cite{Gopalakrishnan_Low_2015,Crowley_Constructive_2022,Garratt_Local_2021b,Garratt_Resonant_2022,Long_Phenomenology_2023}; the connection to spectral statistics has been discussed in Ref.~\cite{Garratt_Local_2021b}. Moving to the time domain, it would be interesting to understand how that picture manifests itself in the structure of the operator paths contributing to correlation functions.

Discussions of operator paths are relevant to questions about the complexity of simulating quantum dynamics on classical computers (and hence also for discussions of quantum advantage \cite{Kim_Evidence_2023,Kechedzhi_Effective_2023}). For example, in a one-dimensional system of qubits, a generic operator having support on $\ell$ contiguous sites can be represented using matrix-product operator bond dimension $\chi \leq 2^{\ell}$. When studying an infinite-temperature autocorrelation function of a local operator, it is therefore important to understand how rapidly it grows in time. One might expect that the operator paths relevant to autocorrelation functions have $\ell \simeq 2 v_B t$ at their widest, although the results of Ref.~\cite{Nahum_Real_2022} suggest that the dominant paths may be confined to a parametrically smaller region of space with $\ell \sim t^{1/2}$. Our work suggests that the contributions to evolving operators which survive the ensemble average are confined to a region with $\ell \sim t^0$.

In closing, we note that the operator paths relevant to the simulation of diffusion have recently been investigated in ~\cite{Keyserlingk_Operator_2022,Rakovszky_Dissipation_2022}. Building on these ideas, as well as previous work on the spectral statistics of Floquet systems with conservation laws \cite{Friedman_Spectral_2019,Roy_2020_Random,Moudgalya_2021_Spectral}, it would be interesting to understand the connection between dynamics and spectra in Hamiltonian systems. This would provide an alternative to ideas based on interference effects \cite{Kos_ManyBody_2018,Chan_Spectral_2018,Garratt_Local_2021,Liao_Universal_2022}.

\section*{Acknowledgements}
We thank Remy Dubertrand, Michele Fava, Max McGinley, and Adam Nahum for useful discussions. T.Y. acknowledges hospitality at University of Tokyo where part of the manuscript was written. S.J.G. was supported by the Gordon and Betty Moore Foundation and by the U.S. Department of Energy, Office of Science, Office of High Energy Physics, under QuantISED Award DE-SC0019380. J.T.C. was supported in part by UK Engineering and Physics Sciences Research Council Grant no. EP/X030881/1.

\appendix
\section{Spinful RPM}\label{sec:spinful}
The RPM as defined in Sec.~\ref{sect:models} has the property that $\overline{C_{\alpha \alpha}(t=1)}=0$ for all operators $P_\alpha$. This feature is shared with other models in which the ensemble of Floquet operators is left and/or right invariant under independent unitary rotations at each site, but is not expected to be generic. For that reason it is interesting to construct a solvable model without such an invariance. We do so here by introducing into the RPM defined previously an extra degree of freedom, which has dynamics that does not consist simply of Haar-distributed rotations. We call the model with this extra degree of freedom the spinful RPM.

We define the spinful RPM by extending the local Hilbert space to have dimension $2q$, identifying this space as the tensor product of the original $q$-dimensional space (the `colour' degrees of freedom) $\mathbb{C}^q$ and the $2$-dimensional space of a `spin' degree of freedom $\mathbb{C}^2$. The single-site unitary operations (now $2q \times 2q$ matrices) mix the two spin states on a timescale set by a parameter $\theta$ (when $\theta=0$ they reduce to two independent copies of $q\times q$ Haar-distributed unitary matrices), and this mixing introduces new features in autocorrelation functions and the OTOC. To be more precise, the on-site unitaries consisting of the first layer of the Floquet operator $W_1=\bigotimes_{x=1}^LU_x$ are promoted to $2q\times 2q$ matrices
\begin{equation}
    U_x=(R\otimes I_q)\begin{pmatrix}
        V_{1,x} & 0\\
        0 & V_{2,x}
    \end{pmatrix}
\end{equation}
with
\begin{equation}
R = \begin{pmatrix}
\cos\theta & \sin\theta\\ -\sin\theta & \cos\theta
\end{pmatrix}
\end{equation}
where $\theta\in[0,\pi/4)$ and $V_{1,x}, V_{2,x}$ are independent $q\times q$ matrices that are Haar random. On the other hand, even in the spinful RPM the second layer $W_2$ still acts on the computational basis $\mathbb{C}^q\otimes\mathbb{C}^2$ diagonally.
\subsection{PSFF of the spinful RPM}\label{sec:rpm_spin_psff}
The PSFF for the spinful RPM can be computed in the same way as shown in Sec.~\ref{sect:diagrammatics} for the spinless case.   To compute Haar averages diagrammatically, let us first focus on a single site $x$ in $A$. The on-site diagram consists of two orbits, one a product of $U_x$ and the other a product of $U_x^*$. We need to evaluate
\begin{align}
    &\Big<[U_x]_{\mathtt{s}(0),a(0)}^{\mathtt{s}(1),a(1)}[U_x]_{\mathtt{s}(1),a(1)}^{\mathtt{s}(2),a(2)}\cdots[U_x]_{\mathtt{s}(t-1),a(t-1)}^{\mathtt{s}(0),a(0)} \n
    &\times[U^*_x]_{\mathtt{r}(0),b(0)}^{\mathtt{r}(1),b(1)}[U^*_x]_{\mathtt{r}(1),b(1)}^{\mathtt{r}(2),b(2)}\cdots[U^*_x]_{\mathtt{r}(t-1),b(t-1)}^{\mathtt{r}(0),b(0)}\Big>,
\end{align}
where indices $\mathtt{s}(u), \mathtt{r}(u)=1,2$ are associated with the spin degree of freedom, while $a(u), b(u)=1,\cdots,q$ are for the colour degree of freedom for $u=0,\cdots,t-1$. It turns out that the leading pairings upon Haar averaging are again $t$ cyclic pairings, which can be parameterised by $\underline{a}(u)=\underline{b}(u+s)$ for fixed $u$ and $s=0,\cdots,t-1$, where $\underline{a}=(\mathtt{s},a)$ (see Fig.~\ref{fig:psff_gauss_diag1}-\ref{fig:psff_gauss_diag3} in the case of $t=3$). Note that the parameterization for the spin degree of freedom $\mathtt{s}(u)=\mathtt{r}(u+s)$ comes from the fact that $V_1$ and $V_2$ are independent unitaries. This means that the phase factor due to having different local pairings on neighboring sites is still the same in this case. Finally, we also have to account for the contribution from the spin degree of freedom, which happens to be decoupled from the rest. Namely, for each site $x$ on $A$ we have
\begin{align} 
&\prod_{u=1}^t\sum_{\mathtt{s}_x(u)=1,2}R_{\mathtt{s}_x(u),\mathtt{s}_x(u+1)}R_{\mathtt{s}_x(u+s_x),\mathtt{s}_x(u+1+s_x)}\n
&=\mathrm{Tr}\,\Tilde{R}^t=1+\cos^t2\theta,
\end{align}
where $\Tilde{R}_{\mathtt{s}\mathtt{r}}=(R_{\mathtt{s}\mathtt{r}})^2$ for any $s_x$. Thus $K_A(t)$ can be recast into a product of two factors
\begin{equation}\label{eq:psff_rpm_exact}
    K_A(t)=K^\mathrm{s}_A(t)K^\mathrm{c}_A(t),
\end{equation}
where $K^\mathrm{s}_A(t)=(1+\cos^t2\theta)^{L_A}$ and
\begin{equation}
    K^\mathrm{c}_A(t)=\frac{1}{t}\left[\lambda_0^{L_A+1}+(t-1)\lambda^{L_A+1}\right]
\end{equation}
with $\lambda_0=1+(t-1)e^{-\varepsilon t}$ and $\lambda=1-e^{-\varepsilon t}$.

Two comments are in order. First, the PSFF no longer drops to $1$ at $t_\mathrm{d}=1$ but rather takes a nontrivial value $\overline{K_A(1)}=(1+\cos2\theta)^{L_A}$. We thus observe that the additional spin degree freedom in the model introduces nontrivial $\overline{K_A(t_\mathrm{d}=1)}\neq1$.
Another observation is that, in the spiful RPM, there are two sources that influence the Thouless time $t_{\mathrm{Th},A}$ of the PSFF. The first one is the domain-walls that are encoded in $K^\mathrm{c}_A(t)$ and another one is the extra on-site spin structure accounted for by $K^\mathrm{s}_A(t)$. Since $K^\mathrm{s}_A(t)$ decays over the timescale $-\ln L_A/\log \cos2\theta$, which again grows with $\ln L_A$, the Thouless time of the PSFF of the spinful RPM is given by
\begin{equation}
    t_{\mathrm{Th},A}=\max\{\varepsilon^{-1}\ln L_A/\varepsilon,-(\ln \cos2\theta)^{-1}\ln L_A\}
\end{equation}

\subsection{Autocorrelation functions of the spinful RPM}\label{sec:rpm_spin_corr}
\begin{figure}[h!]
\centering
\includegraphics[width=8cm]{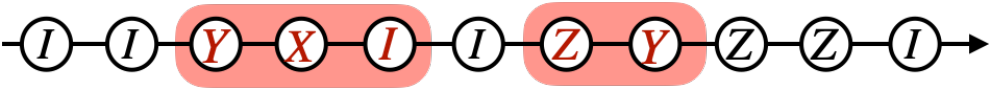}
\caption{An operator string where $L=11$, $n=2$, $b=2$, and $a=5$ (with $\mathtt{a}_0=\mathtt{a}_1=\mathtt{a}_3=1$ and $\mathtt{a}_2=2$). The letter in each on-site operator corresponds to the spin component of the operator $\sigma_\mu$, and those in red colour act nontrivially on the colour degree of freedom.  The weight of the operator string is $a+b=7$. } 
\label{fig:operator_string}
\end{figure}

As in Sec.~\ref{sec:overview_corr},
 let us start with accounting for the spin degree of freedom by parameterising an on-site operator on site $x$ as $\tilde{P}_{\mu_x,\alpha_x}=\sigma_{\mu_x}\otimes P_{\alpha_x}$ where  $\sigma_\mu$ are Pauli matrices ($\sigma_0=I_2, \sigma_1=X,\sigma_2=Y,\sigma_3=Z$) and normalised as $\mathrm{Tr}(\sigma_\mu\sigma_\nu)=2\delta_{\mu\nu}$. The on-site non-identity operators  $\tilde{P}_{\mu_x,\alpha_x}$ can be classified into two classes, which we call class I and class II, according to whether they act nontrivially on the colour degree of freedom (class I) or not (class II). It turns out that the only on-site operator in class II that produces a nontrivial contribution is $Z\otimes I_q$, as $X$ and $Y$ are nondiagonal. Let us consider the operator string $\tilde{P}_{\mu,\alpha}$ that consists of $n$ clusters of class I operators, each of which contains $a_m$ class I on-site operators for $m=1,\cdots,n$. Every pair of neighboring clusters is connected by a string of on-site operators that are either class II or the identity. We denote the total number of class II operators in the operator string $\tilde{P}_{\mu,\alpha}$ by $b$. This means that the total number of single-site identity operators is $L-b-a$ where $a=\sum_{m=1}^n a_m$, and the weight of $\tilde{P}_{\mu,\alpha}$ is $a+b$. Further, let us denote the total number of class I operators with $\sigma_\mu$ in the entire operator string by $\mathtt{a}_\mu$.
 %$a_{m,\mu}$ denote the number of class I operators with $\sigma^\mu$ in the cluster labeled by $m$. 
 %Likewise, by $a_\mu=\sum_{m=1}^na_{m,\mu}$, we mean . 
 See Fig.~\ref{fig:operator_string} for an example of an operator string.

Next we compute the Haar averages of single-site diagrams at $q\to\infty$. There are three types of single-site diagrams: those with class I operators, class II operators, and diagrams without local operators. Let us start with the first type. We can repeat the same argument as in the RPM without the spin degree of freedom, and observe that the leading pairings are again given by the cyclic ones with the factor $q^{-1}$ except the one associated with the identity permutation. The extra spin space, however, generates a factor $f_\mu(\theta)$, which turns out to be different depending on which Pauli matrix $\sigma_\mu$ is used in the on-site operator. A simple calculation shows that the factor is $f_\mu(\theta)=\sum_{\mathtt{s},\mathtt{r}}[R^\mathrm{T}\sigma_\mu R]_{\mathtt{s}\mathtt{r}}[\Tilde{R}^{t-2}\tilde{\sigma}_\mu]_{\mathtt{s}\mathtt{r}}R_{\mathtt{r}\mathtt{s}}$ for any local pairing, where $\tilde{\sigma}_{\mathtt{s}\mathtt{r}}=R_{\mathtt{s}\mathtt{r}}\sigma_{\mathtt{s}\mathtt{r}}$. More explicitly, $f_\mu(\theta)$ for each $\mu$ reads
\begin{equation}
    f_\mu(\theta)=\begin{dcases*}
    \cos^2\theta(1+\cos^{t-2}2\theta) & $\mu=0$ \\
    \sin^2\theta(1-\cos^{t-2}2\theta-2\cos^{t-1}2\theta) & $\mu=1$ \\
      \sin^2\theta(1+\cos^{t-2}2\theta) & $\mu=2$ \\
       \cos^2\theta(1-\cos^{t-2}2\theta+2\cos^{t-1}2\theta) & $\mu=3$
    \end{dcases*}.
\end{equation}
Note that $\sum_\mu f_\mu(\theta)=2(1+\cos^t2\theta)$, as expected. Therefore any $s
\neq0$ pairing that contains a class I operator $\sigma_{\mu_x}\otimes\Tilde{P}_{\alpha_x}$ yields the factor $q^{-1}f_{\mu_x}(\theta)$ at site $x$. This means that the contributions from the spin and the colour degrees of freedom are again decoupled as in the PSFF. The other two cases can be treated together as their actions on the colour degree of freedom are trivial. First, we notice that there is only one relevant local paring at $q\to\infty$ after Haar averaging, which is the one associated with the identity permutation that carries the weight $q$. This implies that the diagrams that include $\sigma_1$ or $\sigma_2$ have vanishing contributions at large $q$.  On the other hand, non-zero contributions are produced when $\mu=0,3$ with the factor  $g_\mu=\sum_{\mathtt{s},\mathtt{r}}[R^\mathrm{T}\sigma_\mu R]_{\mathtt{s}\mathtt{s}}\tilde{R}^{t-1}_{\mathtt{s}\mathtt{r}}[\sigma_\mu]_{\mathtt{r}\mathtt{r}}$. A straightforward calculation gives $g_0=2$ and $g_3=2\cos^t2\theta$. Again, they sum to $2(1+\cos^t2\theta)$ as they should.

Now we are in the position to evaluate autocorrelation functions. Combining the above results, we find that $\overline{C_{\alpha\alpha}(t)}$ behaves at large $q$ as
\begin{align}\label{eq:noconserve_corr}
\overline{C_{\alpha\alpha}(t)} 
   &=(2q)^{-L}q^{L-a}q^{-a}g_0^{L-b-a}g_3^b \n
   &\quad\times\prod_{\mu=0}^3f_\mu^{\mathtt{a}_\mu}(\theta)\prod_{m=1}^n\left[(TD_\mathrm{c})^{a_m}T\right]_{00} \n
        &=C^\mathrm{s}_\alpha(t)C^\mathrm{c}_\alpha(t)
\end{align}
where $C^\mathrm{s}_\alpha(t)=2^{-a}\cos^{bt}2\theta\prod_{\mu=0}^3f_\mu^{ §\mathtt{a}_\mu}(\theta)$ and $C^\mathrm{c}_\alpha(t)=q^{-2a}e^{-2n\varepsilon t}(t-1)^n\left[1+(t-2)e^{-\varepsilon t}\right]^{a-n}$
Note that $C^\mathrm{c}_\alpha(t)$ is nothing but $\overline{C_{\alpha\alpha}(t)}$ in the RPM without the spin degree of freedom Eq.~\eqref{eq:corr_rpm_exact2}. Since $C^\mathrm{s}_\alpha(t)$ is a function that decays monotonically in time, the contribution from the spin degree of freedom turns out to influence only the physics after the peak, which comes from $C^\mathrm{c}_\alpha(t)$. To be more precise, using the fact that asymptotically $C^\mathrm{s}_\alpha(t)$ goes as $C^\mathrm{s}_\alpha(t)\simeq  2^{-a}\cos^{bt}2\theta$ provided that $\theta\neq0$, the late time asymptotics of the auto-correlation functions is given by
\begin{equation}
    (2q^2)^{-a}\cos^{bt}2\theta e^{-2n\varepsilon t}(t-1)^n.
\end{equation}
This suggests that the operator strings that contribute to the decay of the PSFF are (i) those that are made of one cluster of class I operators and (ii) those that have only one cluster of class II operators with $b=1$. 

Another important observation is that the decay time $t_\mathrm{d}$ becomes nontrivial when the operator string is made of class II operators only, i.e. $a=n=0$, in which case autocorrelation functions become $\overline{C_{\alpha\alpha}(t)}=\cos^{bt}2\theta$. Invoking the sum rule Eq.~\eqref{eq:defPSFF}, it is thus these operators which give rise to the nontrivial value of the PSFF at $t=1$ in Eq.~\eqref{eq:psff_rpm_exact}.

Finally, let us also make a brief comment on the off-diagonal correlation functions $\overline{\langle C_{\alpha\beta}(t)\rangle}$. Following the same argument as above, it is readily seen that they vanish unless $\tilde{P}_{\mu,\alpha}$ and $\tilde{P}_{\mu',\beta}$ act on the colour degrees of freedom in the same way, i.e. $P_\alpha=P_\beta$. The nontrivial contribution comes from the Pauli matrices, and in particular, if $\tilde{P}_\alpha$ and $\tilde{P}_\beta$ contain $\sigma_\mu$ and $\sigma_{\mu''}$ on the same site, they amount to the factor $f_{\mu\mu'}=\sum_{\mathtt{s},\mathtt{r}}(R^\mathrm{T}\sigma_\mu R)_{\mathtt{s}\mathtt{r}}(\Tilde{R}^{t-2}\tilde{\sigma}_{\mu'})_{\mathtt{s}\mathtt{r}}R_{\mathtt{r}\mathtt{s}}$. Note $f_{\mu\mu}=f_\mu$ and the only nonvanishing nondiagonal matrix elements $f_{\mu\nu}$ are $f_{02}=-f_{20}=-\ii\sin2\theta(1-\cos^{t-2}2\theta)/2$ and $f_{13}=f_{31}=-\sin2\theta(1+\cos^{t-2}2\theta+2\cos^{t-1})/2$. The net factor coming from the Pauli matrix on each site can be obtained by multiplying $f_{\mu\nu}$.

\subsection{OTOC of the RPM with the spin sector}\label{sec:otoc_derivation_spin}
As in the computation for the autocorrelation functions as well as the PSFF, most of the techniques we developed in calculating the large-$q$ OTOC in Sec.~\ref{sec:otoc_derivation_nospin} carry over even in the presence of the spin degree of freedom. We parameterise the local operator as $\mathcal{O}=\sigma_\mu\otimes P_\alpha$, where both $\sigma_\mu$ and $P_\alpha$ act on the same site. Note that, as in the spinless case, the result does not depend on the choice of $P_\alpha$ after Haar averaging.

The task again boils down to computing the on-site diagrams so let us start with the sites that are neither at $0$ nor $x$.
\subsubsection{Diagrams without local operators}
\begin{figure}[h!]
\subfloat[\label{fig:floquet_op1}]{\includegraphics[width=3cm]{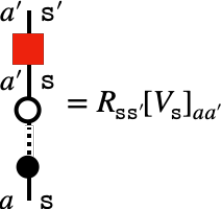}
}
\hfill
\subfloat[\label{fig:floquet_op2}]{\includegraphics[width=3cm]{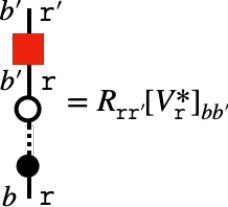} 
}
\hfill
\subfloat[\label{fig:square1}]{\includegraphics[width=3.cm]{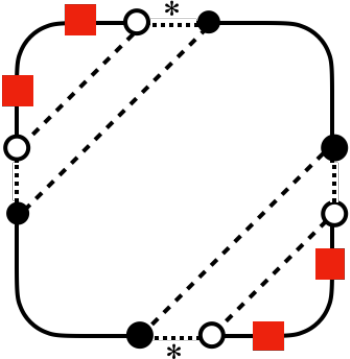}
}
\hfill
\subfloat[\label{fig:square2}]{\includegraphics[width=3.0cm]{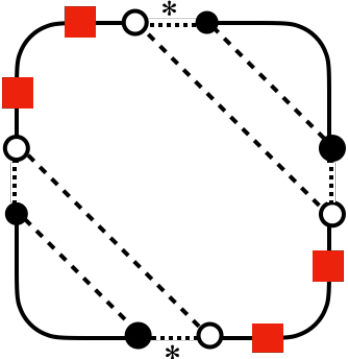}}
\hfill
\subfloat[\label{fig:square3}]{\includegraphics[width=3.0cm]{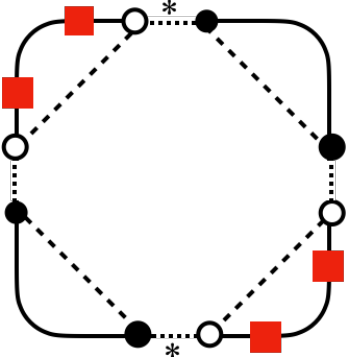}
}
\hfill
\subfloat[\label{fig:square4}]{\includegraphics[width=3.0cm]{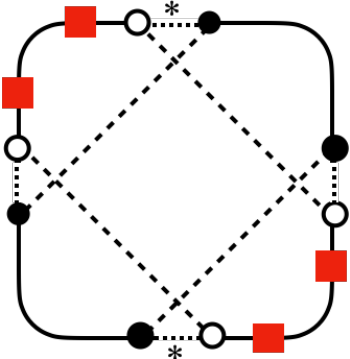} 
}
\caption{(a,b): On-site unitaries with the extra spin structure. (c,d,e,f): Local pairings for $t=1$ with the factors (c,d) $2q$, (e): $-2q$, and (f): $-2q^{-2}$ at large $q$. A red square indicates the matrix $R$.}
\end{figure}

The main change due to the additional on-site spin degree of freedom is that now each on-site unitary is represented by a pair of black and white dots together with a red square. While the dots act on the colour degree of freedom as before, red squares correspond to the matrix $R$ that acts on the spin degree of freedom (see Fig.~\ref{fig:floquet_op1} and \ref{fig:floquet_op2}). With this rule it is straightforward to see that, upon Haar averaging, the same leading pairings  as in the OTOC of the spinless RPM labeled by $p=0,\cdots,2t$ contribute, but in this case we also need to account for a possible factor stemming from $R$ (see Fig.~\ref{fig:square1}-\ref{fig:square4} for the possible local pairings at $t=1$. As in the spinless case, only Fig.~\ref{fig:square1}, \ref{fig:square2}, and \ref{fig:square3} contribute at large $q$). It turns out that this factor is trivial due to the trace structure of the diagram, and it only gives the factor $2$. Therefore, barring the sign, each contributing local pairing carries the factor $2q$. The situation is, however, slightly subtler when the diagram contains the operator $\mathcal{O}$, which we turn to next.

\subsubsection{Diagrams with local operators}
In contrast to the case for sites on which no local operator acts, at sites with local operators Haar averaging generates different local pairings depending on whether the local operator $\mathcal{O}$ is class I or class II. While in the former the situation is the same as in the spinless RPM, the case where the operator is class II is a little more involved. Haar averaging in fact gives 
 a similar result to that of on-site diagrams on sites without local operators. This is because the operator acts on the colour degree of freedom trivially when the operator is class II. Hence we have $2t+1$ contributing local pairings with the same transfer matrix $S$ for these diagrams as previously. The difference, however, is that since the operator $\mathcal{O}$ still acts on the spin space nontrivially, every local pairing carries the same factor $q\eta_\mu(\theta)$ where $\eta_\mu(\theta)=\sum_\mathtt{s}[(R^\mathrm{T}\sigma_\mu R)_{\mathtt{s}\mathtt{s}}]^2$ except the one corresponding to $p=0$ pairing (on site $0$) and $p=2t$ pairing (on site $x$). The factor $\eta_\mu(\theta)$ can be easily computed as $\eta_0(\theta)=2, \eta_1(\theta)=2\sin^2\theta, \eta_2(\theta)=0, \eta_3(\theta)=2\cos^2\theta$. 
 
\subsubsection{Computation of the OTOC}
Since there is essentially no effect of having the spin degree of freedom when the local operator is a class I operator, the OTOC for a class I operator $\mathcal{O}$ is exactly the same as the one we already computed Eq.~\eqref{eq:otoc_nospin}. When the operator is class II instead, the OTOC can be expressed as a trace
\begin{equation}\label{eq:otoc_class2_1}
    \overline{\mathcal{C}(x,t)}=1-\frac{1}{4}\mathrm{Tr}(SG_1\hat{S}^{|x|-1}SG_2\hat{S}^{L-|x|-1}),
\end{equation}
where $G_1$ and $G_2$ are $(2t+1)\times(2t+1)$ diagonal matrices whose diagonal entries are $[G_1]_{pp}=2\delta_{0p}+(1-\delta_{0p})\eta_\mu(\theta)$ and $[G_2]_{pp}=2\delta_{tp}+(1-\delta_{tp})\eta_\mu(\theta)$. It turns out that we can evaluate \eqref{eq:otoc_class2_1} in a way that is similar to the case of class I operators, which yields
\begin{equation}
    \overline{\mathcal{C}(x,t)}=\left(1-\frac{1}{2}\eta_\mu(\theta)\right)^2\sum_{m=0}^{t-|x|-1}\begin{pmatrix}
        t-1\\
        m
    \end{pmatrix}\rho^m(1-\rho)^{t-m-1},
\end{equation}
Thus at the coarse-grained scale (i.e. large $x$ and $t$), it  behaves as
\begin{equation}\label{eq:otoc_class2_2}
   \overline{\mathcal{C}(x,t)}=\left(1-\frac{1}{2}\eta_\mu(\theta)\right)^2\Phi\left(\frac{v_Bt-|x|}{\sqrt{2Dt}}\right).
\end{equation}
Even though the OTOC now depends on $\theta$ explicitly, it is striking that the way it does is rather simple and the additional spin space in the RPM only induces the time-independent multiplicative factor fixed by $\eta_\mu(\theta)$.

\subsection{Spinful RPM with multiple layers of Haar unitaries}\label{sec:rpm_more_layers}
So far we have only considered the (spinful) RPM with a single layer of Haar-distributed unitaries. It is a simple matter to generalise the model to have multiple layers of such unitaries, i.e. powers of $U_x$. The effect of such a change turns out merely to replace $t$ with $nt$ where $n$ is the number of layers. To see it, for simplicity, let us consider the spineless RPM with $n$ layers of Haar-distributed unitaries, which we shall call $n$-RPM, and evaluate autocorrelation functions $\overline{C^{[n]}_{\alpha\alpha}(t)}$ where the superscript emphasise that the correlators are for the $n$-RPM.

The onsite diagrams in this case have $nt$ unitaries $U_x$ on the left and their conjugate $U_x^*$ on the right (see e.g. Fig.~\ref{fig:psff_gauss_diag1}), and after every $n$ unitaries, the states on two neighbouring sites are paired by Gaussian coupling. As in the usual case $n=1$ the leading pairings at large $q$ are again given by the cyclic pairings, which give rise to the $nt\times nt$ transfer matrix $T_n=(1-e^{-\varepsilon nt})I+e^{-\varepsilon nt} \mathtt{E}_n$ where $\mathtt{E}_n$ is the $nt\times nt$ matrix of ones. This indicates that autocorrelation functions in the $n$-RPM simply read $\overline{C^{[n]}_{\alpha\alpha}(t)}=\overline{C^{[1]}_{\alpha\alpha}(nt)}$, where $\overline{C^{[1]}_{\alpha\alpha}(nt)}=\overline{C_{\alpha\alpha}(nt)}$ is given by Eq.~\eqref{eq:corr_rpm_exact}. The consequence of having more layers of Haar unitaries is thus simply to shift the position of the intermediate-time peak to earlier time. Namely, the timescale associated with the height of the peak $2+\varepsilon^{-1}$ (see Sec.~\ref{sect:correlation}) is rescaled to $(2+\varepsilon^{-1})/n$ in the $n$-RPM. Therefore, if the coupling strength $\varepsilon$ is large enough so that $(2+\varepsilon^{-1})/n\sim 1$, then the initial dip disappears, otherwise the minimum remains intact albeit with the asymmetric narrower intermediate-peak. A similar change also occurs to the PSFF and also in the spinful RPM and it is in agreement with the observed change in the $q=2$ random field Heisenberg model with $U_x$ being the square of a Haar-random unitary as reported in Figs.~\ref{fig:Cvariations} and \ref{fig:Kvariations}.

\section{Numerical methods}\label{sec:numerics}
To evaluate numerically the various quantities considered in this work, we rely on a matrix-free Monte Carlo method. First we describe the numerical method used to determine the PSFF. The scheme requires storing and operating on a complex vector of dimension $q^{L_A+L}=D^2_A D_{\bar{A}}$, and so is less memory-intensive than working with the full Floquet operator for $L_A < L$. As usual, we denote the two subregions in the system by $A$ and $\bar{A}$, with $L_{\bar{A}}=L-L_A$. We also introduce a supplementary subsystem $R$ consisting of $L_A$ sites. We initialise $A$ and $R$ in a maximally entangled state $|\Phi_{AR}\ket$ which we can view as a tensor product of $L_A$ Bell pairs, each having one end in $A$ and the other end in $R$. The region $\bar{A}$ is initialised in a Haar-random state $|\psi\ket_{\bar{A}}$. The full initial state is
\begin{align}
	|\Phi_{AR}\ket \otimes |\psi_{\bar{A}}\ket = \Big( D_A^{-1/2} \sum_{i=0}^{D_A-1} |i\ket_R \otimes |i\ket_A\Big) \otimes |\psi\ket_{\bar{A}}. \notag
\end{align}
We then evolve the system $A \bar{A}$ under the Floquet operator $W$. After $t$ time steps the state is
\begin{align}
	&1_R \otimes W^t \big(|\Phi_{AR}\ket \otimes |\psi_{\bar{A}}\ket\big)\label{eq:WtPhiARpsiAbar} \\&= D_A^{-1/2} \sum_{i=0}^{D_A-1} |i\ket_R \otimes W^t \big( |i\ket_A \otimes |\psi\ket_{\bar{A}}\big).\notag
\end{align}
Contracting this state with $\bra \Phi_{AR}|$ generates a pure state on $\overline{A}$,
\begin{align}
	D_A \bra \Phi_{AR}|1_R \otimes W^t \big(|\Phi_{AR}\rangle \otimes |\psi_{\bar{A}}\rangle\big) = \big[ \text{Tr}_A W^t \big]|\psi\rangle_{\bar{A}}. \notag
\end{align}
Finally we calculate the squared norm of the vector on the right-hand side. Averaging this over $|\psi\ket$ we generate
\begin{align}
	&\text{E}_{\psi}\Big[ \bra \psi|_{\bar{A}}\big[ \text{Tr}_A W^{-t} \big]\big[ \text{Tr}_A W^t \big]|\psi\ket_{\bar{A}}\Big] \\&= D_{\bar{A}}^{-1}\text{Tr}_{\bar{A}}\big[ \text{Tr}_A W^t  \text{Tr}_A W^{-t}\big] = K_A(t). \notag
\end{align}
Note that there are a number of different Monte Carlo methods that could be designed to determine the PSFF; the advantage of this one is that we directly contract the final-time indices in $\bar{A}$ between the forward and backward evolution, and this drastically reduces the statistical fluctuations. The variance over $|\psi\ket$ is
\begin{align}
	&\text{Var}_{\psi}\Big[ \bra \psi|_{\bar{A}}\big[ \text{Tr}_A W^{-t} \big]\big[ \text{Tr}_A W^t \big]|\psi\ket_{\bar{A}}\Big] \\&= D_{\bar{A}}^{-2} \text{Tr}_{\bar{A}}\big[ \text{Tr}_A W^t  \text{Tr}_A W^{-t} \text{Tr}_A W^t  \text{Tr}_A W^{-t}\big].\notag
\end{align}
For $W$ a Haar-random $D \times D$ matrix, the average over the right-hand side can in the large-$D$ limit be evaluated as $2D_{\bar{A}}^{-1}$. This is the contribution from the Gaussian diagram with equal-time pairing. This implies that for any given realization of the circuit, the fluctuations of $\bra \psi|_{\bar{A}}\big[ \text{Tr}_A W^{-t} \big]\big[ \text{Tr}_A W^t \big]|\psi\ket_{\bar{A}}$ around its average (i.e. the PSFF $K_A(t)$) are exponentially small in $L_{\bar{A}}$. 

The method used to determine the autocorrelation functions for operators with support within $A$ is closely related. After creating the state in Eq.~\eqref{eq:WtPhiARpsiAbar} we construct the reduced density matrix for the region $AR$, and then evaluate the expectation value of e.g. $P_{\alpha,R} \otimes P_{\alpha,A}$. Here we can view $P_{\alpha,R}$ as the same operator as $P_{\alpha,A}$ but acting at the initial time. Averaging the result of the calculation over $|\psi_{\bar{A}}\ket$ we find $D_{\bar{A}} \text{Tr}[P_{\alpha}(t)P_{\alpha}(0)]=D_A C_{\alpha \alpha}(t)$. Because the ensembles of circuits we consider enjoy a statistical invariance under single-site unitary operations it is sufficient to calculate autocorrelation functions for operators that are diagonal in the comptutational basis. For this reason we need only compute the diagonal elements of the density matrix for $AR$, and hence the memory required for the final step of this calculation is only $q^{2L_A} =D_A^2$.

We now describe our scheme for evaluating $K^B_A(t)$ and $N^B_A(t)$. To calculate $K^B_A(t)$ we first create two states $|\Psi_f\ket = |\Phi_{AR}\ket \otimes |\psi_{\bar{A},f}\ket$ and $|\Psi_b\ket= |\Phi_{AR}\ket \otimes |\psi_{\bar{A},b}\ket$, with $|\psi_{\bar{A},f}\ket$ and $|\psi_{\bar{A},b}\ket$ independent random states. Recall that $|\Phi_{AR}\rangle$ is a maximally entangled state of $L_A$ qubits in $A$ and $L_R=L_A$ qubits in $R$. The states $|\Psi_f\ket$ and $|\Psi_b\ket$ are evolved forward in time by $\lceil t/2 \rceil$, and backward in time by $\lfloor t/2 \rfloor$, respectively. At this step we have generated 
\begin{align}
|\Psi_f(\lceil t/2 \rceil)\ket &= (1_R \otimes W^{\lceil t/2 \rceil}) |\Psi_f\ket. \\
|\Psi_b(\lfloor t/2 \rfloor)\ket &= (1_R \otimes W^{-\lfloor t/2\rfloor}) |\Psi_b\ket.\notag
\end{align}
We now divide the $L$ sites of the system into subsystems $B$ and $\bar{B}$, consisting of $L_B$ and $L_{\bar{B}} = L-L_B$ sites, and contract indices of $|\Psi_f(\lceil t/2 \rceil)\ket$ and $\bra\Psi_b(\lfloor t/2 \rfloor)|$ in the regions $R$ and $B$. The result is a set of uncontracted indices in $\bar{B}$. Computing the squared norm of the resulting vector, and averaging over $|\psi_{\bar{A},f}\ket$ and $|\psi_{\bar{A},b}\ket$, we find $(D_A/D_B)K^B_A(t)$. In the modulus-squared object, we can understand the above contraction of indices in $R$ as generating the sum over all operators $\alpha \in A$ which defines the PSFF. Similarly, the contraction of indices in $B$ generates a resolution of the identity in the operator space for this region (corresponding to the sum over $\beta$ in Eq.~\eqref{eq:KAB}). The contraction in $\bar{B}$, on the other hand, has the effect of eliminating operator paths which, at time $\lceil t/2 \rceil$, have nontrivial support in $\bar{B}$.

To calculate $N^B_A(t)$, which in contrast with $K^B_A(t)$ is a sum over probabilities of operator paths, we construct a pair of states $|\Psi_f(\lceil t/2 \rceil)\ket$ and $|\Psi_{f'}(\lceil t/2 \rceil)\ket$, having identical structure but independent realisations of the randomness (i.e. a statistically independent initial state of $\bar{A}$). We then contract indices of $|\Psi_f(\lceil t/2 \rceil)\ket$ and $\bra\Psi_{f'}(\lceil t/2 \rceil)|$ in the regions $R$ and $B$, and compute the squared norm of the resulting state. After averaging over initial states of $\bar{A}$, we obtain $N^B_A(t)$.

\bibliography{autocorr_refs.bib}

\begin{thebibliography}{59}%
\makeatletter
\providecommand \@ifxundefined [1]{%
 \@ifx{#1\undefined}
}%
\providecommand \@ifnum [1]{%
 \ifnum #1\expandafter \@firstoftwo
 \else \expandafter \@secondoftwo
 \fi
}%
\providecommand \@ifx [1]{%
 \ifx #1\expandafter \@firstoftwo
 \else \expandafter \@secondoftwo
 \fi
}%
\providecommand \natexlab [1]{#1}%
\providecommand \enquote  [1]{``#1''}%
\providecommand \bibnamefont  [1]{#1}%
\providecommand \bibfnamefont [1]{#1}%
\providecommand \citenamefont [1]{#1}%
\providecommand \href@noop [0]{\@secondoftwo}%
\providecommand \href [0]{\begingroup \@sanitize@url \@href}%
\providecommand \@href[1]{\@@startlink{#1}\@@href}%
\providecommand \@@href[1]{\endgroup#1\@@endlink}%
\providecommand \@sanitize@url [0]{\catcode `\\12\catcode `\$12\catcode `\&12\catcode `\#12\catcode `\^12\catcode `\_12\catcode `\%12\relax}%
\providecommand \@@startlink[1]{}%
\providecommand \@@endlink[0]{}%
\providecommand \url  [0]{\begingroup\@sanitize@url \@url }%
\providecommand \@url [1]{\endgroup\@href {#1}{\urlprefix }}%
\providecommand \urlprefix  [0]{URL }%
\providecommand \Eprint [0]{\href }%
\providecommand \doibase [0]{https://doi.org/}%
\providecommand \selectlanguage [0]{\@gobble}%
\providecommand \bibinfo  [0]{\@secondoftwo}%
\providecommand \bibfield  [0]{\@secondoftwo}%
\providecommand \translation [1]{[#1]}%
\providecommand \BibitemOpen [0]{}%
\providecommand \bibitemStop [0]{}%
\providecommand \bibitemNoStop [0]{.\EOS\space}%
\providecommand \EOS [0]{\spacefactor3000\relax}%
\providecommand \BibitemShut  [1]{\csname bibitem#1\endcsname}%
\let\auto@bib@innerbib\@empty
%</preamble>
\bibitem [{\citenamefont {D'Alessio}\ \emph {et~al.}(2016)\citenamefont {D'Alessio}, \citenamefont {Kafri}, \citenamefont {Polkovnikov},\ and\ \citenamefont {Rigol}}]{DAlessio_From_2016}%
  \BibitemOpen
  \bibfield  {author} {\bibinfo {author} {\bibfnamefont {L.}~\bibnamefont {D'Alessio}}, \bibinfo {author} {\bibfnamefont {Y.}~\bibnamefont {Kafri}}, \bibinfo {author} {\bibfnamefont {A.}~\bibnamefont {Polkovnikov}},\ and\ \bibinfo {author} {\bibfnamefont {M.}~\bibnamefont {Rigol}},\ }\bibfield  {title} {\bibinfo {title} {From quantum chaos and eigenstate thermalization to statistical mechanics and thermodynamics},\ }\href {https://doi.org/10.1080/00018732.2016.1198134} {\bibfield  {journal} {\bibinfo  {journal} {Adv. Phys.}\ }\textbf {\bibinfo {volume} {65}},\ \bibinfo {pages} {239} (\bibinfo {year} {2016})}\BibitemShut {NoStop}%
\bibitem [{\citenamefont {Maldacena}\ \emph {et~al.}(2016)\citenamefont {Maldacena}, \citenamefont {Shenker},\ and\ \citenamefont {Stanford}}]{Maldacena_Bound_2016}%
  \BibitemOpen
  \bibfield  {author} {\bibinfo {author} {\bibfnamefont {J.}~\bibnamefont {Maldacena}}, \bibinfo {author} {\bibfnamefont {S.~H.}\ \bibnamefont {Shenker}},\ and\ \bibinfo {author} {\bibfnamefont {D.}~\bibnamefont {Stanford}},\ }\bibfield  {title} {\bibinfo {title} {A bound on chaos},\ }\href {https://doi.org/10.1007/JHEP08(2016)106} {\bibfield  {journal} {\bibinfo  {journal} {J. High Energy Phys.}\ }\textbf {\bibinfo {volume} {2016}}\bibinfo  {number} { (8)},\ \bibinfo {pages} {1}}\BibitemShut {NoStop}%
\bibitem [{\citenamefont {Nahum}\ \emph {et~al.}(2017)\citenamefont {Nahum}, \citenamefont {Ruhman}, \citenamefont {Vijay},\ and\ \citenamefont {Haah}}]{Nahum_2017_Quantum}%
  \BibitemOpen
\bibfield  {number} {  }\bibfield  {author} {\bibinfo {author} {\bibfnamefont {A.}~\bibnamefont {Nahum}}, \bibinfo {author} {\bibfnamefont {J.}~\bibnamefont {Ruhman}}, \bibinfo {author} {\bibfnamefont {S.}~\bibnamefont {Vijay}},\ and\ \bibinfo {author} {\bibfnamefont {J.}~\bibnamefont {Haah}},\ }\bibfield  {title} {\bibinfo {title} {Quantum entanglement growth under random unitary dynamics},\ }\href {https://doi.org/10.1103/PhysRevX.7.031016} {\bibfield  {journal} {\bibinfo  {journal} {Phys. Rev. X}\ }\textbf {\bibinfo {volume} {7}},\ \bibinfo {pages} {031016} (\bibinfo {year} {2017})}\BibitemShut {NoStop}%
\bibitem [{\citenamefont {Nahum}\ \emph {et~al.}(2018)\citenamefont {Nahum}, \citenamefont {Vijay},\ and\ \citenamefont {Haah}}]{Nahum_Operator_2018}%
  \BibitemOpen
  \bibfield  {author} {\bibinfo {author} {\bibfnamefont {A.}~\bibnamefont {Nahum}}, \bibinfo {author} {\bibfnamefont {S.}~\bibnamefont {Vijay}},\ and\ \bibinfo {author} {\bibfnamefont {J.}~\bibnamefont {Haah}},\ }\bibfield  {title} {\bibinfo {title} {Operator spreading in random unitary circuits},\ }\href {https://doi.org/10.1103/PhysRevX.8.021014} {\bibfield  {journal} {\bibinfo  {journal} {Phys. Rev. X}\ }\textbf {\bibinfo {volume} {8}},\ \bibinfo {pages} {021014} (\bibinfo {year} {2018})}\BibitemShut {NoStop}%
\bibitem [{\citenamefont {von Keyserlingk}\ \emph {et~al.}(2018)\citenamefont {von Keyserlingk}, \citenamefont {Rakovszky}, \citenamefont {Pollmann},\ and\ \citenamefont {Sondhi}}]{Keyserlingk_Operator_2018}%
  \BibitemOpen
  \bibfield  {author} {\bibinfo {author} {\bibfnamefont {C.~W.}\ \bibnamefont {von Keyserlingk}}, \bibinfo {author} {\bibfnamefont {T.}~\bibnamefont {Rakovszky}}, \bibinfo {author} {\bibfnamefont {F.}~\bibnamefont {Pollmann}},\ and\ \bibinfo {author} {\bibfnamefont {S.~L.}\ \bibnamefont {Sondhi}},\ }\bibfield  {title} {\bibinfo {title} {Operator hydrodynamics, {OTOC}s, and entanglement growth in systems without conservation laws},\ }\href {https://doi.org/10.1103/PhysRevX.8.021013} {\bibfield  {journal} {\bibinfo  {journal} {Phys. Rev. X}\ }\textbf {\bibinfo {volume} {8}},\ \bibinfo {pages} {021013} (\bibinfo {year} {2018})}\BibitemShut {NoStop}%
\bibitem [{\citenamefont {Fisher}\ \emph {et~al.}(2023)\citenamefont {Fisher}, \citenamefont {Khemani}, \citenamefont {Nahum},\ and\ \citenamefont {Vijay}}]{Fisher_Random_2023}%
  \BibitemOpen
  \bibfield  {author} {\bibinfo {author} {\bibfnamefont {M.~P.~A.}\ \bibnamefont {Fisher}}, \bibinfo {author} {\bibfnamefont {V.}~\bibnamefont {Khemani}}, \bibinfo {author} {\bibfnamefont {A.}~\bibnamefont {Nahum}},\ and\ \bibinfo {author} {\bibfnamefont {S.}~\bibnamefont {Vijay}},\ }\bibfield  {title} {\bibinfo {title} {Random quantum circuits},\ }\href {https://doi.org/10.1146/annurev-conmatphys-031720-030658} {\bibfield  {journal} {\bibinfo  {journal} {Annu. Rev. Condens. Matter Phys.}\ }\textbf {\bibinfo {volume} {14}},\ \bibinfo {pages} {335} (\bibinfo {year} {2023})}\BibitemShut {NoStop}%
\bibitem [{\citenamefont {Rakovszky}\ \emph {et~al.}(2018)\citenamefont {Rakovszky}, \citenamefont {Pollmann},\ and\ \citenamefont {von Keyserlingk}}]{Rakovszky_Diffusive_2018}%
  \BibitemOpen
  \bibfield  {author} {\bibinfo {author} {\bibfnamefont {T.}~\bibnamefont {Rakovszky}}, \bibinfo {author} {\bibfnamefont {F.}~\bibnamefont {Pollmann}},\ and\ \bibinfo {author} {\bibfnamefont {C.~W.}\ \bibnamefont {von Keyserlingk}},\ }\bibfield  {title} {\bibinfo {title} {Diffusive hydrodynamics of out-of-time-ordered correlators with charge conservation},\ }\href {https://doi.org/10.1103/PhysRevX.8.031058} {\bibfield  {journal} {\bibinfo  {journal} {Phys. Rev. X}\ }\textbf {\bibinfo {volume} {8}},\ \bibinfo {pages} {031058} (\bibinfo {year} {2018})}\BibitemShut {NoStop}%
\bibitem [{\citenamefont {Khemani}\ \emph {et~al.}(2018)\citenamefont {Khemani}, \citenamefont {Vishwanath},\ and\ \citenamefont {Huse}}]{Khemani_Operator_2018}%
  \BibitemOpen
  \bibfield  {author} {\bibinfo {author} {\bibfnamefont {V.}~\bibnamefont {Khemani}}, \bibinfo {author} {\bibfnamefont {A.}~\bibnamefont {Vishwanath}},\ and\ \bibinfo {author} {\bibfnamefont {D.~A.}\ \bibnamefont {Huse}},\ }\bibfield  {title} {\bibinfo {title} {Operator spreading and the emergence of dissipative hydrodynamics under unitary evolution with conservation laws},\ }\href {https://doi.org/10.1103/PhysRevX.8.031057} {\bibfield  {journal} {\bibinfo  {journal} {Phys. Rev. X}\ }\textbf {\bibinfo {volume} {8}},\ \bibinfo {pages} {031057} (\bibinfo {year} {2018})}\BibitemShut {NoStop}%
\bibitem [{\citenamefont {Chan}\ \emph {et~al.}(2018{\natexlab{a}})\citenamefont {Chan}, \citenamefont {De~Luca},\ and\ \citenamefont {Chalker}}]{Chan_Solution_2018}%
  \BibitemOpen
  \bibfield  {author} {\bibinfo {author} {\bibfnamefont {A.}~\bibnamefont {Chan}}, \bibinfo {author} {\bibfnamefont {A.}~\bibnamefont {De~Luca}},\ and\ \bibinfo {author} {\bibfnamefont {J.~T.}\ \bibnamefont {Chalker}},\ }\bibfield  {title} {\bibinfo {title} {Solution of a minimal model for many-body quantum chaos},\ }\href {https://doi.org/10.1103/PhysRevX.8.041019} {\bibfield  {journal} {\bibinfo  {journal} {Phys. Rev. X}\ }\textbf {\bibinfo {volume} {8}},\ \bibinfo {pages} {041019} (\bibinfo {year} {2018}{\natexlab{a}})}\BibitemShut {NoStop}%
\bibitem [{\citenamefont {Bertini}\ \emph {et~al.}(2018)\citenamefont {Bertini}, \citenamefont {Kos},\ and\ \citenamefont {Prosen}}]{Bertini_Exact_2018}%
  \BibitemOpen
  \bibfield  {author} {\bibinfo {author} {\bibfnamefont {B.}~\bibnamefont {Bertini}}, \bibinfo {author} {\bibfnamefont {P.}~\bibnamefont {Kos}},\ and\ \bibinfo {author} {\bibfnamefont {T.}~\bibnamefont {Prosen}},\ }\bibfield  {title} {\bibinfo {title} {Exact spectral form factor in a minimal model of many-body quantum chaos},\ }\href {https://doi.org/10.1103/PhysRevLett.121.264101} {\bibfield  {journal} {\bibinfo  {journal} {Phys. Rev. Lett.}\ }\textbf {\bibinfo {volume} {121}},\ \bibinfo {pages} {264101} (\bibinfo {year} {2018})}\BibitemShut {NoStop}%
\bibitem [{\citenamefont {Mehta}(2004)}]{Mehta_Random_2004}%
  \BibitemOpen
  \bibfield  {author} {\bibinfo {author} {\bibfnamefont {M.~L.}\ \bibnamefont {Mehta}},\ }\href@noop {} {\emph {\bibinfo {title} {Random Matrices}}},\ \bibinfo {edition} {3rd}\ ed.,\ Pure and Applied Mathematics\ (\bibinfo  {publisher} {Academic Press},\ \bibinfo {year} {2004})\BibitemShut {NoStop}%
\bibitem [{\citenamefont {Schiulaz}\ \emph {et~al.}(2019)\citenamefont {Schiulaz}, \citenamefont {Torres-Herrera},\ and\ \citenamefont {Santos}}]{Schiulaz_19}%
  \BibitemOpen
  \bibfield  {author} {\bibinfo {author} {\bibfnamefont {M.}~\bibnamefont {Schiulaz}}, \bibinfo {author} {\bibfnamefont {E.~J.}\ \bibnamefont {Torres-Herrera}},\ and\ \bibinfo {author} {\bibfnamefont {L.~F.}\ \bibnamefont {Santos}},\ }\bibfield  {title} {\bibinfo {title} {Thouless and relaxation time scales in many-body quantum systems},\ }\href {https://doi.org/10.1103/PhysRevB.99.174313} {\bibfield  {journal} {\bibinfo  {journal} {Phys. Rev. B}\ }\textbf {\bibinfo {volume} {99}},\ \bibinfo {pages} {174313} (\bibinfo {year} {2019})}\BibitemShut {NoStop}%
\bibitem [{\citenamefont {Garratt}\ and\ \citenamefont {Chalker}(2021{\natexlab{a}})}]{Garratt_Local_2021}%
  \BibitemOpen
  \bibfield  {author} {\bibinfo {author} {\bibfnamefont {S.~J.}\ \bibnamefont {Garratt}}\ and\ \bibinfo {author} {\bibfnamefont {J.~T.}\ \bibnamefont {Chalker}},\ }\bibfield  {title} {\bibinfo {title} {Local pairing of {F}eynman histories in many-body {F}loquet models},\ }\href {https://doi.org/10.1103/PhysRevX.11.021051} {\bibfield  {journal} {\bibinfo  {journal} {Phys. Rev. X}\ }\textbf {\bibinfo {volume} {11}},\ \bibinfo {pages} {021051} (\bibinfo {year} {2021}{\natexlab{a}})}\BibitemShut {NoStop}%
\bibitem [{\citenamefont {Joshi}\ \emph {et~al.}(2022)\citenamefont {Joshi}, \citenamefont {Elben}, \citenamefont {Vikram}, \citenamefont {Vermersch}, \citenamefont {Galitski},\ and\ \citenamefont {Zoller}}]{Joshi_Probing_2022}%
  \BibitemOpen
  \bibfield  {author} {\bibinfo {author} {\bibfnamefont {L.~K.}\ \bibnamefont {Joshi}}, \bibinfo {author} {\bibfnamefont {A.}~\bibnamefont {Elben}}, \bibinfo {author} {\bibfnamefont {A.}~\bibnamefont {Vikram}}, \bibinfo {author} {\bibfnamefont {B.}~\bibnamefont {Vermersch}}, \bibinfo {author} {\bibfnamefont {V.}~\bibnamefont {Galitski}},\ and\ \bibinfo {author} {\bibfnamefont {P.}~\bibnamefont {Zoller}},\ }\bibfield  {title} {\bibinfo {title} {Probing many-body quantum chaos with quantum simulators},\ }\href {https://doi.org/10.1103/PhysRevX.12.011018} {\bibfield  {journal} {\bibinfo  {journal} {Phys. Rev. X}\ }\textbf {\bibinfo {volume} {12}},\ \bibinfo {pages} {011018} (\bibinfo {year} {2022})}\BibitemShut {NoStop}%
\bibitem [{\citenamefont {Prigodin}\ \emph {et~al.}(1994)\citenamefont {Prigodin}, \citenamefont {Altshuler}, \citenamefont {Efetov},\ and\ \citenamefont {Iida}}]{Prigodin_94}%
  \BibitemOpen
  \bibfield  {author} {\bibinfo {author} {\bibfnamefont {V.~N.}\ \bibnamefont {Prigodin}}, \bibinfo {author} {\bibfnamefont {B.~L.}\ \bibnamefont {Altshuler}}, \bibinfo {author} {\bibfnamefont {K.~B.}\ \bibnamefont {Efetov}},\ and\ \bibinfo {author} {\bibfnamefont {S.}~\bibnamefont {Iida}},\ }\bibfield  {title} {\bibinfo {title} {Mesoscopic dynamical echo in quantum dots},\ }\href {https://doi.org/10.1103/PhysRevLett.72.546} {\bibfield  {journal} {\bibinfo  {journal} {Phys. Rev. Lett.}\ }\textbf {\bibinfo {volume} {72}},\ \bibinfo {pages} {546} (\bibinfo {year} {1994})}\BibitemShut {NoStop}%
\bibitem [{\citenamefont {Lerma-Hern\'andez}\ \emph {et~al.}(2019)\citenamefont {Lerma-Hern\'andez}, \citenamefont {Villase\~nor}, \citenamefont {Bastarrachea-Magnani}, \citenamefont {Torres-Herrera}, \citenamefont {Santos},\ and\ \citenamefont {Hirsch}}]{Santos_19}%
  \BibitemOpen
  \bibfield  {author} {\bibinfo {author} {\bibfnamefont {S.}~\bibnamefont {Lerma-Hern\'andez}}, \bibinfo {author} {\bibfnamefont {D.}~\bibnamefont {Villase\~nor}}, \bibinfo {author} {\bibfnamefont {M.~A.}\ \bibnamefont {Bastarrachea-Magnani}}, \bibinfo {author} {\bibfnamefont {E.~J.}\ \bibnamefont {Torres-Herrera}}, \bibinfo {author} {\bibfnamefont {L.~F.}\ \bibnamefont {Santos}},\ and\ \bibinfo {author} {\bibfnamefont {J.~G.}\ \bibnamefont {Hirsch}},\ }\bibfield  {title} {\bibinfo {title} {Dynamical signatures of quantum chaos and relaxation time scales in a spin-boson system},\ }\href {https://doi.org/10.1103/PhysRevE.100.012218} {\bibfield  {journal} {\bibinfo  {journal} {Phys. Rev. E}\ }\textbf {\bibinfo {volume} {100}},\ \bibinfo {pages} {012218} (\bibinfo {year} {2019})}\BibitemShut {NoStop}%
\bibitem [{\citenamefont {Chan}\ \emph {et~al.}(2018{\natexlab{b}})\citenamefont {Chan}, \citenamefont {De~Luca},\ and\ \citenamefont {Chalker}}]{Chan_Spectral_2018}%
  \BibitemOpen
  \bibfield  {author} {\bibinfo {author} {\bibfnamefont {A.}~\bibnamefont {Chan}}, \bibinfo {author} {\bibfnamefont {A.}~\bibnamefont {De~Luca}},\ and\ \bibinfo {author} {\bibfnamefont {J.~T.}\ \bibnamefont {Chalker}},\ }\bibfield  {title} {\bibinfo {title} {Spectral statistics in spatially extended chaotic quantum many-body systems},\ }\href {https://doi.org/10.1103/PhysRevLett.121.060601} {\bibfield  {journal} {\bibinfo  {journal} {Phys. Rev. Lett.}\ }\textbf {\bibinfo {volume} {121}},\ \bibinfo {pages} {060601} (\bibinfo {year} {2018}{\natexlab{b}})}\BibitemShut {NoStop}%
\bibitem [{\citenamefont {Garratt}\ and\ \citenamefont {Chalker}(2021{\natexlab{b}})}]{Garratt_ManyBody_2021}%
  \BibitemOpen
  \bibfield  {author} {\bibinfo {author} {\bibfnamefont {S.~J.}\ \bibnamefont {Garratt}}\ and\ \bibinfo {author} {\bibfnamefont {J.~T.}\ \bibnamefont {Chalker}},\ }\bibfield  {title} {\bibinfo {title} {Many-body delocalization as symmetry breaking},\ }\href {https://doi.org/10.1103/PhysRevLett.127.026802} {\bibfield  {journal} {\bibinfo  {journal} {Phys. Rev. Lett.}\ }\textbf {\bibinfo {volume} {127}},\ \bibinfo {pages} {026802} (\bibinfo {year} {2021}{\natexlab{b}})}\BibitemShut {NoStop}%
\bibitem [{\citenamefont {Prange}(1997)}]{Prange_Spectral_1997}%
  \BibitemOpen
  \bibfield  {author} {\bibinfo {author} {\bibfnamefont {R.~E.}\ \bibnamefont {Prange}},\ }\bibfield  {title} {\bibinfo {title} {The spectral form factor is not self-averaging},\ }\href {https://doi.org/10.1103/PhysRevLett.78.2280} {\bibfield  {journal} {\bibinfo  {journal} {Phys. Rev. Lett.}\ }\textbf {\bibinfo {volume} {78}},\ \bibinfo {pages} {2280} (\bibinfo {year} {1997})}\BibitemShut {NoStop}%
\bibitem [{\citenamefont {Kunz}(1999)}]{Kunz_Probability_1999}%
  \BibitemOpen
  \bibfield  {author} {\bibinfo {author} {\bibfnamefont {H.}~\bibnamefont {Kunz}},\ }\bibfield  {title} {\bibinfo {title} {The probability distribution of the spectral form factor in random matrix theory},\ }\href {https://doi.org/10.1088/0305-4470/32/11/011} {\bibfield  {journal} {\bibinfo  {journal} {J. Phys. A: Math. Gen.}\ }\textbf {\bibinfo {volume} {32}},\ \bibinfo {pages} {2171} (\bibinfo {year} {1999})}\BibitemShut {NoStop}%
\bibitem [{\citenamefont {Thouless}(1977)}]{Thouless_77}%
  \BibitemOpen
  \bibfield  {author} {\bibinfo {author} {\bibfnamefont {D.~J.}\ \bibnamefont {Thouless}},\ }\bibfield  {title} {\bibinfo {title} {Maximum metallic resistance in thin wires},\ }\href {https://doi.org/10.1103/PhysRevLett.39.1167} {\bibfield  {journal} {\bibinfo  {journal} {Phys. Rev. Lett.}\ }\textbf {\bibinfo {volume} {39}},\ \bibinfo {pages} {1167} (\bibinfo {year} {1977})}\BibitemShut {NoStop}%
\bibitem [{\citenamefont {Altshuler}\ and\ \citenamefont {Shklovskii}(1986)}]{Altshuler_86}%
  \BibitemOpen
  \bibfield  {author} {\bibinfo {author} {\bibfnamefont {B.~L.}\ \bibnamefont {Altshuler}}\ and\ \bibinfo {author} {\bibfnamefont {B.~I.}\ \bibnamefont {Shklovskii}},\ }\bibfield  {title} {\bibinfo {title} {Repulsion of energy levels and conductivity of small metal samples},\ }\href {http://www.jetp.ac.ru/cgi-bin/dn/e_064_01_0127.pdf} {\bibfield  {journal} {\bibinfo  {journal} {JETP}\ }\textbf {\bibinfo {volume} {64}},\ \bibinfo {pages} {127} (\bibinfo {year} {1986})}\BibitemShut {NoStop}%
\bibitem [{\citenamefont {Bertrand}\ and\ \citenamefont {Garc\'{\i}a-Garc\'{\i}a}(2016)}]{Bertrand_Anomalous_2016}%
  \BibitemOpen
  \bibfield  {author} {\bibinfo {author} {\bibfnamefont {C.~L.}\ \bibnamefont {Bertrand}}\ and\ \bibinfo {author} {\bibfnamefont {A.~M.}\ \bibnamefont {Garc\'{\i}a-Garc\'{\i}a}},\ }\bibfield  {title} {\bibinfo {title} {Anomalous {T}houless energy and critical statistics on the metallic side of the many-body localization transition},\ }\href {https://doi.org/10.1103/PhysRevB.94.144201} {\bibfield  {journal} {\bibinfo  {journal} {Phys. Rev. B}\ }\textbf {\bibinfo {volume} {94}},\ \bibinfo {pages} {144201} (\bibinfo {year} {2016})}\BibitemShut {NoStop}%
\bibitem [{\citenamefont {Friedman}\ \emph {et~al.}(2019)\citenamefont {Friedman}, \citenamefont {Chan}, \citenamefont {De~Luca},\ and\ \citenamefont {Chalker}}]{Friedman_Spectral_2019}%
  \BibitemOpen
  \bibfield  {author} {\bibinfo {author} {\bibfnamefont {A.~J.}\ \bibnamefont {Friedman}}, \bibinfo {author} {\bibfnamefont {A.}~\bibnamefont {Chan}}, \bibinfo {author} {\bibfnamefont {A.}~\bibnamefont {De~Luca}},\ and\ \bibinfo {author} {\bibfnamefont {J.~T.}\ \bibnamefont {Chalker}},\ }\bibfield  {title} {\bibinfo {title} {Spectral statistics and many-body quantum chaos with conserved charge},\ }\href {https://doi.org/10.1103/PhysRevLett.123.210603} {\bibfield  {journal} {\bibinfo  {journal} {Phys. Rev. Lett.}\ }\textbf {\bibinfo {volume} {123}},\ \bibinfo {pages} {210603} (\bibinfo {year} {2019})}\BibitemShut {NoStop}%
\bibitem [{\citenamefont {Sierant}\ \emph {et~al.}(2020)\citenamefont {Sierant}, \citenamefont {Delande},\ and\ \citenamefont {Zakrzewski}}]{Sierant_Thouless_2020}%
  \BibitemOpen
  \bibfield  {author} {\bibinfo {author} {\bibfnamefont {P.}~\bibnamefont {Sierant}}, \bibinfo {author} {\bibfnamefont {D.}~\bibnamefont {Delande}},\ and\ \bibinfo {author} {\bibfnamefont {J.}~\bibnamefont {Zakrzewski}},\ }\bibfield  {title} {\bibinfo {title} {{T}houless time analysis of {A}nderson and many-body localization transitions},\ }\href {https://doi.org/10.1103/PhysRevLett.124.186601} {\bibfield  {journal} {\bibinfo  {journal} {Phys. Rev. Lett.}\ }\textbf {\bibinfo {volume} {124}},\ \bibinfo {pages} {186601} (\bibinfo {year} {2020})}\BibitemShut {NoStop}%
\bibitem [{\citenamefont {Sonner}\ \emph {et~al.}(2021)\citenamefont {Sonner}, \citenamefont {Serbyn}, \citenamefont {Papi{\'c}},\ and\ \citenamefont {Abanin}}]{Sonner_Thouless_2021}%
  \BibitemOpen
  \bibfield  {author} {\bibinfo {author} {\bibfnamefont {M.}~\bibnamefont {Sonner}}, \bibinfo {author} {\bibfnamefont {M.}~\bibnamefont {Serbyn}}, \bibinfo {author} {\bibfnamefont {Z.}~\bibnamefont {Papi{\'c}}},\ and\ \bibinfo {author} {\bibfnamefont {D.~A.}\ \bibnamefont {Abanin}},\ }\bibfield  {title} {\bibinfo {title} {{T}houless energy across the many-body localization transition in {F}loquet systems},\ }\href {https://doi.org/10.1103/PhysRevB.104.L081112} {\bibfield  {journal} {\bibinfo  {journal} {Phys. Rev. B}\ }\textbf {\bibinfo {volume} {104}},\ \bibinfo {pages} {L081112} (\bibinfo {year} {2021})}\BibitemShut {NoStop}%
\bibitem [{\citenamefont {Serbyn}\ \emph {et~al.}(2017)\citenamefont {Serbyn}, \citenamefont {Papi{\'c}},\ and\ \citenamefont {Abanin}}]{Serbyn_Thouless_2017}%
  \BibitemOpen
  \bibfield  {author} {\bibinfo {author} {\bibfnamefont {M.}~\bibnamefont {Serbyn}}, \bibinfo {author} {\bibfnamefont {Z.}~\bibnamefont {Papi{\'c}}},\ and\ \bibinfo {author} {\bibfnamefont {D.~A.}\ \bibnamefont {Abanin}},\ }\bibfield  {title} {\bibinfo {title} {{T}houless energy and multifractality across the many-body localization transition},\ }\href {https://doi.org/10.1103/PhysRevB.96.104201} {\bibfield  {journal} {\bibinfo  {journal} {Phys. Rev. B}\ }\textbf {\bibinfo {volume} {96}},\ \bibinfo {pages} {104201} (\bibinfo {year} {2017})}\BibitemShut {NoStop}%
\bibitem [{\citenamefont {Cotler}\ \emph {et~al.}(2017)\citenamefont {Cotler}, \citenamefont {Hunter-Jones}, \citenamefont {Liu},\ and\ \citenamefont {Yoshida}}]{Cotler_Chaos_2017}%
  \BibitemOpen
  \bibfield  {author} {\bibinfo {author} {\bibfnamefont {J.}~\bibnamefont {Cotler}}, \bibinfo {author} {\bibfnamefont {N.}~\bibnamefont {Hunter-Jones}}, \bibinfo {author} {\bibfnamefont {J.}~\bibnamefont {Liu}},\ and\ \bibinfo {author} {\bibfnamefont {B.}~\bibnamefont {Yoshida}},\ }\bibfield  {title} {\bibinfo {title} {Chaos, complexity, and random matrices},\ }\href {https://doi.org/10.1007/JHEP11(2017)048} {\bibfield  {journal} {\bibinfo  {journal} {J. High Energy Phys.}\ }\textbf {\bibinfo {volume} {2017}}\bibinfo  {number} { (11)},\ \bibinfo {pages} {1}}\BibitemShut {NoStop}%
\bibitem [{\citenamefont {Gharibyan}\ \emph {et~al.}(2018)\citenamefont {Gharibyan}, \citenamefont {Hanada}, \citenamefont {Shenker},\ and\ \citenamefont {Tezuka}}]{Gharibyan_Onset_2018}%
  \BibitemOpen
\bibfield  {number} {  }\bibfield  {author} {\bibinfo {author} {\bibfnamefont {H.}~\bibnamefont {Gharibyan}}, \bibinfo {author} {\bibfnamefont {M.}~\bibnamefont {Hanada}}, \bibinfo {author} {\bibfnamefont {S.~H.}\ \bibnamefont {Shenker}},\ and\ \bibinfo {author} {\bibfnamefont {M.}~\bibnamefont {Tezuka}},\ }\bibfield  {title} {\bibinfo {title} {Onset of random matrix behavior in scrambling systems},\ }\href {https://doi.org/10.1007/JHEP07(2018)124} {\bibfield  {journal} {\bibinfo  {journal} {J. High Energy Phys.}\ }\textbf {\bibinfo {volume} {2018}}\bibinfo  {number} { (7)},\ \bibinfo {pages} {124}}\BibitemShut {NoStop}%
\bibitem [{\citenamefont {Braun}\ \emph {et~al.}(2020)\citenamefont {Braun}, \citenamefont {Waltner}, \citenamefont {Akila}, \citenamefont {Gutkin},\ and\ \citenamefont {Guhr}}]{Braun_Transition_2020}%
  \BibitemOpen
\bibfield  {number} {  }\bibfield  {author} {\bibinfo {author} {\bibfnamefont {P.}~\bibnamefont {Braun}}, \bibinfo {author} {\bibfnamefont {D.}~\bibnamefont {Waltner}}, \bibinfo {author} {\bibfnamefont {M.}~\bibnamefont {Akila}}, \bibinfo {author} {\bibfnamefont {B.}~\bibnamefont {Gutkin}},\ and\ \bibinfo {author} {\bibfnamefont {T.}~\bibnamefont {Guhr}},\ }\bibfield  {title} {\bibinfo {title} {Transition from quantum chaos to localization in spin chains},\ }\href {https://doi.org/10.1103/PhysRevE.101.052201} {\bibfield  {journal} {\bibinfo  {journal} {Phys. Rev. E}\ }\textbf {\bibinfo {volume} {101}},\ \bibinfo {pages} {052201} (\bibinfo {year} {2020})}\BibitemShut {NoStop}%
\bibitem [{\citenamefont {Nahum}\ \emph {et~al.}(2022)\citenamefont {Nahum}, \citenamefont {Roy}, \citenamefont {Vijay},\ and\ \citenamefont {Zhou}}]{Nahum_Real_2022}%
  \BibitemOpen
  \bibfield  {author} {\bibinfo {author} {\bibfnamefont {A.}~\bibnamefont {Nahum}}, \bibinfo {author} {\bibfnamefont {S.}~\bibnamefont {Roy}}, \bibinfo {author} {\bibfnamefont {S.}~\bibnamefont {Vijay}},\ and\ \bibinfo {author} {\bibfnamefont {T.}~\bibnamefont {Zhou}},\ }\bibfield  {title} {\bibinfo {title} {Real-time correlators in chaotic quantum many-body systems},\ }\href {https://doi.org/10.1103/PhysRevB.106.224310} {\bibfield  {journal} {\bibinfo  {journal} {Phys. Rev. B}\ }\textbf {\bibinfo {volume} {106}},\ \bibinfo {pages} {224310} (\bibinfo {year} {2022})}\BibitemShut {NoStop}%
\bibitem [{\citenamefont {Rakovszky}\ \emph {et~al.}(2022)\citenamefont {Rakovszky}, \citenamefont {von Keyserlingk},\ and\ \citenamefont {Pollmann}}]{Rakovszky_Dissipation_2022}%
  \BibitemOpen
  \bibfield  {author} {\bibinfo {author} {\bibfnamefont {T.}~\bibnamefont {Rakovszky}}, \bibinfo {author} {\bibfnamefont {C.~W.}\ \bibnamefont {von Keyserlingk}},\ and\ \bibinfo {author} {\bibfnamefont {F.}~\bibnamefont {Pollmann}},\ }\bibfield  {title} {\bibinfo {title} {Dissipation-assisted operator evolution method for capturing hydrodynamic transport},\ }\href {https://doi.org/10.1103/PhysRevB.105.075131} {\bibfield  {journal} {\bibinfo  {journal} {Phys. Rev. B}\ }\textbf {\bibinfo {volume} {105}},\ \bibinfo {pages} {075131} (\bibinfo {year} {2022})}\BibitemShut {NoStop}%
\bibitem [{\citenamefont {von Keyserlingk}\ \emph {et~al.}(2022)\citenamefont {von Keyserlingk}, \citenamefont {Pollmann},\ and\ \citenamefont {Rakovszky}}]{Keyserlingk_Operator_2022}%
  \BibitemOpen
  \bibfield  {author} {\bibinfo {author} {\bibfnamefont {C.}~\bibnamefont {von Keyserlingk}}, \bibinfo {author} {\bibfnamefont {F.}~\bibnamefont {Pollmann}},\ and\ \bibinfo {author} {\bibfnamefont {T.}~\bibnamefont {Rakovszky}},\ }\bibfield  {title} {\bibinfo {title} {Operator backflow and the classical simulation of quantum transport},\ }\href {https://doi.org/10.1103/PhysRevB.105.245101} {\bibfield  {journal} {\bibinfo  {journal} {Phys. Rev. B}\ }\textbf {\bibinfo {volume} {105}},\ \bibinfo {pages} {245101} (\bibinfo {year} {2022})}\BibitemShut {NoStop}%
\bibitem [{\citenamefont {Swingle}\ \emph {et~al.}(2016)\citenamefont {Swingle}, \citenamefont {Bentsen}, \citenamefont {Schleier-Smith},\ and\ \citenamefont {Hayden}}]{Swingle_Measuring_2016}%
  \BibitemOpen
  \bibfield  {author} {\bibinfo {author} {\bibfnamefont {B.}~\bibnamefont {Swingle}}, \bibinfo {author} {\bibfnamefont {G.}~\bibnamefont {Bentsen}}, \bibinfo {author} {\bibfnamefont {M.}~\bibnamefont {Schleier-Smith}},\ and\ \bibinfo {author} {\bibfnamefont {P.}~\bibnamefont {Hayden}},\ }\bibfield  {title} {\bibinfo {title} {Measuring the scrambling of quantum information},\ }\href {https://doi.org/10.1103/PhysRevA.94.040302} {\bibfield  {journal} {\bibinfo  {journal} {Phys. Rev. A}\ }\textbf {\bibinfo {volume} {94}},\ \bibinfo {pages} {040302(R)} (\bibinfo {year} {2016})}\BibitemShut {NoStop}%
\bibitem [{\citenamefont {Vermersch}\ \emph {et~al.}(2019)\citenamefont {Vermersch}, \citenamefont {Elben}, \citenamefont {Sieberer}, \citenamefont {Yao},\ and\ \citenamefont {Zoller}}]{Vermersch_Probing_2019}%
  \BibitemOpen
  \bibfield  {author} {\bibinfo {author} {\bibfnamefont {B.}~\bibnamefont {Vermersch}}, \bibinfo {author} {\bibfnamefont {A.}~\bibnamefont {Elben}}, \bibinfo {author} {\bibfnamefont {L.~M.}\ \bibnamefont {Sieberer}}, \bibinfo {author} {\bibfnamefont {N.~Y.}\ \bibnamefont {Yao}},\ and\ \bibinfo {author} {\bibfnamefont {P.}~\bibnamefont {Zoller}},\ }\bibfield  {title} {\bibinfo {title} {Probing scrambling using statistical correlations between randomized measurements},\ }\href {https://doi.org/10.1103/PhysRevX.9.021061} {\bibfield  {journal} {\bibinfo  {journal} {Phys. Rev. X}\ }\textbf {\bibinfo {volume} {9}},\ \bibinfo {pages} {021061} (\bibinfo {year} {2019})}\BibitemShut {NoStop}%
\bibitem [{\citenamefont {Garratt}\ \emph {et~al.}(2021)\citenamefont {Garratt}, \citenamefont {Roy},\ and\ \citenamefont {Chalker}}]{Garratt_Local_2021b}%
  \BibitemOpen
  \bibfield  {author} {\bibinfo {author} {\bibfnamefont {S.~J.}\ \bibnamefont {Garratt}}, \bibinfo {author} {\bibfnamefont {S.}~\bibnamefont {Roy}},\ and\ \bibinfo {author} {\bibfnamefont {J.~T.}\ \bibnamefont {Chalker}},\ }\bibfield  {title} {\bibinfo {title} {Local resonances and parametric level dynamics in the many-body localized phase},\ }\href {https://doi.org/10.1103/PhysRevB.104.184203} {\bibfield  {journal} {\bibinfo  {journal} {Phys. Rev. B}\ }\textbf {\bibinfo {volume} {104}},\ \bibinfo {pages} {184203} (\bibinfo {year} {2021})}\BibitemShut {NoStop}%
\bibitem [{\citenamefont {Lieb}\ and\ \citenamefont {Robinson}(1972)}]{Lieb_Finite_1972}%
  \BibitemOpen
  \bibfield  {author} {\bibinfo {author} {\bibfnamefont {E.~H.}\ \bibnamefont {Lieb}}\ and\ \bibinfo {author} {\bibfnamefont {D.~W.}\ \bibnamefont {Robinson}},\ }\bibfield  {title} {\bibinfo {title} {The finite group velocity of quantum spin systems},\ }\href {https://doi.org/10.1007/BF01645779} {\bibfield  {journal} {\bibinfo  {journal} {Commun. Math. Phys.}\ }\textbf {\bibinfo {volume} {28}},\ \bibinfo {pages} {251} (\bibinfo {year} {1972})}\BibitemShut {NoStop}%
\bibitem [{\citenamefont {Claeys}\ and\ \citenamefont {Lamacraft}(2020)}]{Claeys_Maximum_2020}%
  \BibitemOpen
  \bibfield  {author} {\bibinfo {author} {\bibfnamefont {P.~W.}\ \bibnamefont {Claeys}}\ and\ \bibinfo {author} {\bibfnamefont {A.}~\bibnamefont {Lamacraft}},\ }\bibfield  {title} {\bibinfo {title} {Maximum velocity quantum circuits},\ }\href {https://doi.org/10.1103/PhysRevResearch.2.033032} {\bibfield  {journal} {\bibinfo  {journal} {Phys. Rev. Res.}\ }\textbf {\bibinfo {volume} {2}},\ \bibinfo {pages} {033032} (\bibinfo {year} {2020})}\BibitemShut {NoStop}%
\bibitem [{\citenamefont {Rampp}\ \emph {et~al.}(2023)\citenamefont {Rampp}, \citenamefont {Moessner},\ and\ \citenamefont {Claeys}}]{Rampp_From_2023}%
  \BibitemOpen
  \bibfield  {author} {\bibinfo {author} {\bibfnamefont {M.~A.}\ \bibnamefont {Rampp}}, \bibinfo {author} {\bibfnamefont {R.}~\bibnamefont {Moessner}},\ and\ \bibinfo {author} {\bibfnamefont {P.~W.}\ \bibnamefont {Claeys}},\ }\bibfield  {title} {\bibinfo {title} {From dual unitarity to generic quantum operator spreading},\ }\href {https://doi.org/10.1103/PhysRevLett.130.130402} {\bibfield  {journal} {\bibinfo  {journal} {Phys. Rev. Lett.}\ }\textbf {\bibinfo {volume} {130}},\ \bibinfo {pages} {130402} (\bibinfo {year} {2023})}\BibitemShut {NoStop}%
\bibitem [{\citenamefont {McCulloch}\ and\ \citenamefont {von Keyserlingk}(2022)}]{McCulloch_Operator_2022}%
  \BibitemOpen
  \bibfield  {author} {\bibinfo {author} {\bibfnamefont {E.}~\bibnamefont {McCulloch}}\ and\ \bibinfo {author} {\bibfnamefont {C.~W.}\ \bibnamefont {von Keyserlingk}},\ }\bibfield  {title} {\bibinfo {title} {Operator spreading in the memory matrix formalism},\ }\href {https://doi.org/10.1088/1751-8121/ac7091} {\bibfield  {journal} {\bibinfo  {journal} {J. Phys. A: Math. Theor.}\ }\textbf {\bibinfo {volume} {55}},\ \bibinfo {pages} {274007} (\bibinfo {year} {2022})}\BibitemShut {NoStop}%
\bibitem [{\citenamefont {Brouwer}\ and\ \citenamefont {Beenakker}(1996)}]{Brouwer_Diagrammatic_1996}%
  \BibitemOpen
  \bibfield  {author} {\bibinfo {author} {\bibfnamefont {P.~W.}\ \bibnamefont {Brouwer}}\ and\ \bibinfo {author} {\bibfnamefont {C.~W.~J.}\ \bibnamefont {Beenakker}},\ }\bibfield  {title} {\bibinfo {title} {Diagrammatic method of integration over the unitary group, with applications to quantum transport in mesoscopic systems},\ }\href {https://doi.org/10.1063/1.531667} {\bibfield  {journal} {\bibinfo  {journal} {J. Math. Phys.}\ }\textbf {\bibinfo {volume} {37}},\ \bibinfo {pages} {4904} (\bibinfo {year} {1996})}\BibitemShut {NoStop}%
\bibitem [{\citenamefont {Lerose}\ \emph {et~al.}(2021)\citenamefont {Lerose}, \citenamefont {Sonner},\ and\ \citenamefont {Abanin}}]{Lerose_Influence_2021}%
  \BibitemOpen
  \bibfield  {author} {\bibinfo {author} {\bibfnamefont {A.}~\bibnamefont {Lerose}}, \bibinfo {author} {\bibfnamefont {M.}~\bibnamefont {Sonner}},\ and\ \bibinfo {author} {\bibfnamefont {D.~A.}\ \bibnamefont {Abanin}},\ }\bibfield  {title} {\bibinfo {title} {Influence matrix approach to many-body {F}loquet dynamics},\ }\href {https://doi.org/10.1103/PhysRevX.11.021040} {\bibfield  {journal} {\bibinfo  {journal} {Phys. Rev. X}\ }\textbf {\bibinfo {volume} {11}},\ \bibinfo {pages} {021040} (\bibinfo {year} {2021})}\BibitemShut {NoStop}%
\bibitem [{\citenamefont {Roy}\ and\ \citenamefont {Prosen}(2020)}]{Roy_2020_Random}%
  \BibitemOpen
  \bibfield  {author} {\bibinfo {author} {\bibfnamefont {D.}~\bibnamefont {Roy}}\ and\ \bibinfo {author} {\bibfnamefont {T.}~\bibnamefont {Prosen}},\ }\bibfield  {title} {\bibinfo {title} {Random matrix spectral form factor in kicked interacting fermionic chains},\ }\href {https://doi.org/10.1103/PhysRevE.102.060202} {\bibfield  {journal} {\bibinfo  {journal} {Phys. Rev. E}\ }\textbf {\bibinfo {volume} {102}},\ \bibinfo {pages} {060202(R)} (\bibinfo {year} {2020})}\BibitemShut {NoStop}%
\bibitem [{\citenamefont {Moudgalya}\ \emph {et~al.}(2021)\citenamefont {Moudgalya}, \citenamefont {Prem}, \citenamefont {Huse},\ and\ \citenamefont {Chan}}]{Moudgalya_2021_Spectral}%
  \BibitemOpen
  \bibfield  {author} {\bibinfo {author} {\bibfnamefont {S.}~\bibnamefont {Moudgalya}}, \bibinfo {author} {\bibfnamefont {A.}~\bibnamefont {Prem}}, \bibinfo {author} {\bibfnamefont {D.~A.}\ \bibnamefont {Huse}},\ and\ \bibinfo {author} {\bibfnamefont {A.}~\bibnamefont {Chan}},\ }\bibfield  {title} {\bibinfo {title} {Spectral statistics in constrained many-body quantum chaotic systems},\ }\href {https://doi.org/10.1103/PhysRevResearch.3.023176} {\bibfield  {journal} {\bibinfo  {journal} {Phys. Rev. Research}\ }\textbf {\bibinfo {volume} {3}},\ \bibinfo {pages} {023176} (\bibinfo {year} {2021})}\BibitemShut {NoStop}%
\bibitem [{\citenamefont {Parker}\ \emph {et~al.}(2019)\citenamefont {Parker}, \citenamefont {Cao}, \citenamefont {Avdoshkin}, \citenamefont {Scaffidi},\ and\ \citenamefont {Altman}}]{Parker_Universal_2019}%
  \BibitemOpen
  \bibfield  {author} {\bibinfo {author} {\bibfnamefont {D.~E.}\ \bibnamefont {Parker}}, \bibinfo {author} {\bibfnamefont {X.}~\bibnamefont {Cao}}, \bibinfo {author} {\bibfnamefont {A.}~\bibnamefont {Avdoshkin}}, \bibinfo {author} {\bibfnamefont {T.}~\bibnamefont {Scaffidi}},\ and\ \bibinfo {author} {\bibfnamefont {E.}~\bibnamefont {Altman}},\ }\bibfield  {title} {\bibinfo {title} {A universal operator growth hypothesis},\ }\href {https://doi.org/10.1103/PhysRevX.9.041017} {\bibfield  {journal} {\bibinfo  {journal} {Phys. Rev. X}\ }\textbf {\bibinfo {volume} {9}},\ \bibinfo {pages} {041017} (\bibinfo {year} {2019})}\BibitemShut {NoStop}%
\bibitem [{\citenamefont {Chan}\ \emph {et~al.}(2021)\citenamefont {Chan}, \citenamefont {De~Luca},\ and\ \citenamefont {Chalker}}]{Chan_Spectral_2021}%
  \BibitemOpen
  \bibfield  {author} {\bibinfo {author} {\bibfnamefont {A.}~\bibnamefont {Chan}}, \bibinfo {author} {\bibfnamefont {A.}~\bibnamefont {De~Luca}},\ and\ \bibinfo {author} {\bibfnamefont {J.~T.}\ \bibnamefont {Chalker}},\ }\bibfield  {title} {\bibinfo {title} {Spectral {L}yapunov exponents in chaotic and localized many-body quantum systems},\ }\href {https://doi.org/10.1103/PhysRevResearch.3.023118} {\bibfield  {journal} {\bibinfo  {journal} {Phys. Rev. Research}\ }\textbf {\bibinfo {volume} {3}},\ \bibinfo {pages} {023118} (\bibinfo {year} {2021})}\BibitemShut {NoStop}%
\bibitem [{\citenamefont {Bertini}\ \emph {et~al.}(2019)\citenamefont {Bertini}, \citenamefont {Kos},\ and\ \citenamefont {Prosen}}]{Bertini_Exact_2019}%
  \BibitemOpen
  \bibfield  {author} {\bibinfo {author} {\bibfnamefont {B.}~\bibnamefont {Bertini}}, \bibinfo {author} {\bibfnamefont {P.}~\bibnamefont {Kos}},\ and\ \bibinfo {author} {\bibfnamefont {T.}~\bibnamefont {Prosen}},\ }\bibfield  {title} {\bibinfo {title} {Exact correlation functions for dual-unitary lattice models in $1+1$ dimensions},\ }\href {https://doi.org/10.1103/PhysRevLett.123.210601} {\bibfield  {journal} {\bibinfo  {journal} {Phys. Rev. Lett.}\ }\textbf {\bibinfo {volume} {123}},\ \bibinfo {pages} {210601} (\bibinfo {year} {2019})}\BibitemShut {NoStop}%
\bibitem [{\citenamefont {Gutkin}\ \emph {et~al.}(2020)\citenamefont {Gutkin}, \citenamefont {Braun}, \citenamefont {Akila}, \citenamefont {Waltner},\ and\ \citenamefont {Guhr}}]{Gutkin_Exact_2020}%
  \BibitemOpen
  \bibfield  {author} {\bibinfo {author} {\bibfnamefont {B.}~\bibnamefont {Gutkin}}, \bibinfo {author} {\bibfnamefont {P.}~\bibnamefont {Braun}}, \bibinfo {author} {\bibfnamefont {M.}~\bibnamefont {Akila}}, \bibinfo {author} {\bibfnamefont {D.}~\bibnamefont {Waltner}},\ and\ \bibinfo {author} {\bibfnamefont {T.}~\bibnamefont {Guhr}},\ }\bibfield  {title} {\bibinfo {title} {Exact local correlations in kicked chains},\ }\href {https://doi.org/10.1103/PhysRevB.102.174307} {\bibfield  {journal} {\bibinfo  {journal} {Phys. Rev. B}\ }\textbf {\bibinfo {volume} {102}},\ \bibinfo {pages} {174307} (\bibinfo {year} {2020})}\BibitemShut {NoStop}%
\bibitem [{\citenamefont {Bertini}\ \emph {et~al.}(2021)\citenamefont {Bertini}, \citenamefont {Kos},\ and\ \citenamefont {Prosen}}]{Bertini_Random_2021}%
  \BibitemOpen
  \bibfield  {author} {\bibinfo {author} {\bibfnamefont {B.}~\bibnamefont {Bertini}}, \bibinfo {author} {\bibfnamefont {P.}~\bibnamefont {Kos}},\ and\ \bibinfo {author} {\bibfnamefont {T.}~\bibnamefont {Prosen}},\ }\bibfield  {title} {\bibinfo {title} {Random matrix spectral form factor of dual-unitary quantum circuits},\ }\href {https://doi.org/10.1007/s00220-021-04139-2} {\bibfield  {journal} {\bibinfo  {journal} {Commun. Math. Phys.}\ }\textbf {\bibinfo {volume} {387}},\ \bibinfo {pages} {597} (\bibinfo {year} {2021})}\BibitemShut {NoStop}%
\bibitem [{\citenamefont {Nandkishore}\ and\ \citenamefont {Huse}(2015)}]{Nandkishore_Many_2015}%
  \BibitemOpen
  \bibfield  {author} {\bibinfo {author} {\bibfnamefont {R.}~\bibnamefont {Nandkishore}}\ and\ \bibinfo {author} {\bibfnamefont {D.~A.}\ \bibnamefont {Huse}},\ }\bibfield  {title} {\bibinfo {title} {Many-body localization and thermalization in quantum statistical mechanics},\ }\href@noop {} {\bibfield  {journal} {\bibinfo  {journal} {Annu. Rev. Condens. Matter Phys.}\ }\textbf {\bibinfo {volume} {6}},\ \bibinfo {pages} {15} (\bibinfo {year} {2015})}\BibitemShut {NoStop}%
\bibitem [{\citenamefont {Abanin}\ \emph {et~al.}(2019)\citenamefont {Abanin}, \citenamefont {Altman}, \citenamefont {Bloch},\ and\ \citenamefont {Serbyn}}]{Abanin_Many_2019}%
  \BibitemOpen
  \bibfield  {author} {\bibinfo {author} {\bibfnamefont {D.~A.}\ \bibnamefont {Abanin}}, \bibinfo {author} {\bibfnamefont {E.}~\bibnamefont {Altman}}, \bibinfo {author} {\bibfnamefont {I.}~\bibnamefont {Bloch}},\ and\ \bibinfo {author} {\bibfnamefont {M.}~\bibnamefont {Serbyn}},\ }\bibfield  {title} {\bibinfo {title} {Colloquium: Many-body localization, thermalization, and entanglement},\ }\href {https://doi.org/10.1103/RevModPhys.91.021001} {\bibfield  {journal} {\bibinfo  {journal} {Rev. Mod. Phys.}\ }\textbf {\bibinfo {volume} {91}},\ \bibinfo {pages} {021001} (\bibinfo {year} {2019})}\BibitemShut {NoStop}%
\bibitem [{\citenamefont {Gopalakrishnan}\ \emph {et~al.}(2015)\citenamefont {Gopalakrishnan}, \citenamefont {M\"uller}, \citenamefont {Khemani}, \citenamefont {Knap}, \citenamefont {Demler},\ and\ \citenamefont {Huse}}]{Gopalakrishnan_Low_2015}%
  \BibitemOpen
  \bibfield  {author} {\bibinfo {author} {\bibfnamefont {S.}~\bibnamefont {Gopalakrishnan}}, \bibinfo {author} {\bibfnamefont {M.}~\bibnamefont {M\"uller}}, \bibinfo {author} {\bibfnamefont {V.}~\bibnamefont {Khemani}}, \bibinfo {author} {\bibfnamefont {M.}~\bibnamefont {Knap}}, \bibinfo {author} {\bibfnamefont {E.}~\bibnamefont {Demler}},\ and\ \bibinfo {author} {\bibfnamefont {D.~A.}\ \bibnamefont {Huse}},\ }\bibfield  {title} {\bibinfo {title} {Low-frequency conductivity in many-body localized systems},\ }\href {https://doi.org/10.1103/PhysRevB.92.104202} {\bibfield  {journal} {\bibinfo  {journal} {Phys. Rev. B}\ }\textbf {\bibinfo {volume} {92}},\ \bibinfo {pages} {104202} (\bibinfo {year} {2015})}\BibitemShut {NoStop}%
\bibitem [{\citenamefont {Crowley}\ and\ \citenamefont {Chandran}(2022)}]{Crowley_Constructive_2022}%
  \BibitemOpen
  \bibfield  {author} {\bibinfo {author} {\bibfnamefont {P.~J.~D.}\ \bibnamefont {Crowley}}\ and\ \bibinfo {author} {\bibfnamefont {A.}~\bibnamefont {Chandran}},\ }\bibfield  {title} {\bibinfo {title} {A constructive theory of the numerically accessible many-body localized to thermal crossover},\ }\href {https://doi.org/10.21468/SciPostPhys.12.6.201} {\bibfield  {journal} {\bibinfo  {journal} {SciPost Phys.}\ }\textbf {\bibinfo {volume} {12}},\ \bibinfo {pages} {201} (\bibinfo {year} {2022})}\BibitemShut {NoStop}%
\bibitem [{\citenamefont {Garratt}\ and\ \citenamefont {Roy}(2022)}]{Garratt_Resonant_2022}%
  \BibitemOpen
  \bibfield  {author} {\bibinfo {author} {\bibfnamefont {S.~J.}\ \bibnamefont {Garratt}}\ and\ \bibinfo {author} {\bibfnamefont {S.}~\bibnamefont {Roy}},\ }\bibfield  {title} {\bibinfo {title} {Resonant energy scales and local observables in the many-body localized phase},\ }\href {https://doi.org/10.1103/PhysRevB.106.054309} {\bibfield  {journal} {\bibinfo  {journal} {Phys. Rev. B}\ }\textbf {\bibinfo {volume} {106}},\ \bibinfo {pages} {054309} (\bibinfo {year} {2022})}\BibitemShut {NoStop}%
\bibitem [{\citenamefont {Long}\ \emph {et~al.}(2023)\citenamefont {Long}, \citenamefont {Crowley}, \citenamefont {Khemani},\ and\ \citenamefont {Chandran}}]{Long_Phenomenology_2023}%
  \BibitemOpen
  \bibfield  {author} {\bibinfo {author} {\bibfnamefont {D.~M.}\ \bibnamefont {Long}}, \bibinfo {author} {\bibfnamefont {P.~J.~D.}\ \bibnamefont {Crowley}}, \bibinfo {author} {\bibfnamefont {V.}~\bibnamefont {Khemani}},\ and\ \bibinfo {author} {\bibfnamefont {A.}~\bibnamefont {Chandran}},\ }\bibfield  {title} {\bibinfo {title} {Phenomenology of the prethermal many-body localized regime},\ }\href {https://doi.org/10.1103/PhysRevLett.131.106301} {\bibfield  {journal} {\bibinfo  {journal} {Phys. Rev. Lett.}\ }\textbf {\bibinfo {volume} {131}},\ \bibinfo {pages} {106301} (\bibinfo {year} {2023})}\BibitemShut {NoStop}%
\bibitem [{\citenamefont {Kim}\ \emph {et~al.}(2023)\citenamefont {Kim}, \citenamefont {Eddins}, \citenamefont {Anand}, \citenamefont {Wei}, \citenamefont {Van Den~Berg}, \citenamefont {Rosenblatt}, \citenamefont {Nayfeh}, \citenamefont {Wu}, \citenamefont {Zaletel}, \citenamefont {Temme} \emph {et~al.}}]{Kim_Evidence_2023}%
  \BibitemOpen
  \bibfield  {author} {\bibinfo {author} {\bibfnamefont {Y.}~\bibnamefont {Kim}}, \bibinfo {author} {\bibfnamefont {A.}~\bibnamefont {Eddins}}, \bibinfo {author} {\bibfnamefont {S.}~\bibnamefont {Anand}}, \bibinfo {author} {\bibfnamefont {K.~X.}\ \bibnamefont {Wei}}, \bibinfo {author} {\bibfnamefont {E.}~\bibnamefont {Van Den~Berg}}, \bibinfo {author} {\bibfnamefont {S.}~\bibnamefont {Rosenblatt}}, \bibinfo {author} {\bibfnamefont {H.}~\bibnamefont {Nayfeh}}, \bibinfo {author} {\bibfnamefont {Y.}~\bibnamefont {Wu}}, \bibinfo {author} {\bibfnamefont {M.}~\bibnamefont {Zaletel}}, \bibinfo {author} {\bibfnamefont {K.}~\bibnamefont {Temme}}, \emph {et~al.},\ }\bibfield  {title} {\bibinfo {title} {Evidence for the utility of quantum computing before fault tolerance},\ }\href {https://doi.org/10.1038/s41586-023-06096-3} {\bibfield  {journal} {\bibinfo  {journal} {Nature}\ }\textbf {\bibinfo {volume} {618}},\ \bibinfo {pages} {500} (\bibinfo {year} {2023})}\BibitemShut {NoStop}%
\bibitem [{\citenamefont {Kechedzhi}\ \emph {et~al.}(2023)\citenamefont {Kechedzhi}, \citenamefont {Isakov}, \citenamefont {Mandr{\`a}}, \citenamefont {Villalonga}, \citenamefont {Mi}, \citenamefont {Boixo},\ and\ \citenamefont {Smelyanskiy}}]{Kechedzhi_Effective_2023}%
  \BibitemOpen
  \bibfield  {author} {\bibinfo {author} {\bibfnamefont {K.}~\bibnamefont {Kechedzhi}}, \bibinfo {author} {\bibfnamefont {S.}~\bibnamefont {Isakov}}, \bibinfo {author} {\bibfnamefont {S.}~\bibnamefont {Mandr{\`a}}}, \bibinfo {author} {\bibfnamefont {B.}~\bibnamefont {Villalonga}}, \bibinfo {author} {\bibfnamefont {X.}~\bibnamefont {Mi}}, \bibinfo {author} {\bibfnamefont {S.}~\bibnamefont {Boixo}},\ and\ \bibinfo {author} {\bibfnamefont {V.}~\bibnamefont {Smelyanskiy}},\ }\bibfield  {title} {\bibinfo {title} {Effective quantum volume, fidelity and computational cost of noisy quantum processing experiments},\ }\href {https://doi.org/10.1016/j.future.2023.12.002} {\bibfield  {journal} {\bibinfo  {journal} {Future Gener. Comput. Syst.}\ ,\ } (\bibinfo {year} {2023})}\BibitemShut {NoStop}%
\bibitem [{\citenamefont {Kos}\ \emph {et~al.}(2018)\citenamefont {Kos}, \citenamefont {Ljubotina},\ and\ \citenamefont {Prosen}}]{Kos_ManyBody_2018}%
  \BibitemOpen
  \bibfield  {author} {\bibinfo {author} {\bibfnamefont {P.}~\bibnamefont {Kos}}, \bibinfo {author} {\bibfnamefont {M.}~\bibnamefont {Ljubotina}},\ and\ \bibinfo {author} {\bibfnamefont {T.}~\bibnamefont {Prosen}},\ }\bibfield  {title} {\bibinfo {title} {Many-body quantum chaos: Analytic connection to random matrix theory},\ }\href {https://doi.org/10.1103/PhysRevX.8.021062} {\bibfield  {journal} {\bibinfo  {journal} {Phys. Rev. X}\ }\textbf {\bibinfo {volume} {8}},\ \bibinfo {pages} {021062} (\bibinfo {year} {2018})}\BibitemShut {NoStop}%
\bibitem [{\citenamefont {Liao}\ and\ \citenamefont {Galitski}(2022)}]{Liao_Universal_2022}%
  \BibitemOpen
  \bibfield  {author} {\bibinfo {author} {\bibfnamefont {Y.}~\bibnamefont {Liao}}\ and\ \bibinfo {author} {\bibfnamefont {V.}~\bibnamefont {Galitski}},\ }\bibfield  {title} {\bibinfo {title} {Universal dephasing mechanism of many-body quantum chaos},\ }\href {https://doi.org/10.1103/PhysRevResearch.4.L012037} {\bibfield  {journal} {\bibinfo  {journal} {Phys. Rev. Research}\ }\textbf {\bibinfo {volume} {4}},\ \bibinfo {pages} {L012037} (\bibinfo {year} {2022})}\BibitemShut {NoStop}%
\end{thebibliography}%

\end{document}